\documentclass[usenatbib,twocolumn]{mn2e}
\usepackage[]{graphicx}
\usepackage[]{amssymb}
\usepackage{longtable}

\newcommand\apjl{ApJ}%
\newcommand\apjs{ApJS}%
\newcommand\ao{Appl.~Opt.}%
\newcommand\aap{A\&A}%

\title[Optically attenuation of GRBs in the \textit{Swift} era]{A detailed study of the optical attenuation of gamma-ray bursts in the \textit{Swift} era}   
\author[O.~M.~Littlejohns et al.]
{O.~M.~Littlejohns$^{1}$,\thanks{E-mail:    olittlej@asu.edu (OML)}
N.~R.~Butler$^{1}$,
A.~Cucchiara$^{2}$,
A.~M.~Watson$^{3}$,
O.~D.~Fox$^{4}$,
\and
W.~H.~Lee$^{3}$,
A.~S.~Kutyrev$^{5}$,
M.~G.~Richer$^{6}$,
C.~R.~Klein$^{4}$,
J.~X.~Prochaska$^{7}$,
\and
J.~S.~Bloom$^{4}$,
E.~Troja$^{5}$,
E.~Ramirez-Ruiz$^{7}$,
J.~A.~de~Diego$^{3}$,
L.~Georgiev$^{3}$,
\and
J.~Gonz\'{a}lez$^{3}$,
C.~G.~Rom\'{a}n-Z\'{u}\~{n}iga$^{6}$,
N.~Gehrels$^{5}$,
H.~Moseley$^{5}$
\\
\\
$^{1}$ School of Earth \& Space Exploration, Arizona State University,
    AZ 85287, USA \\
$^{2}$ NASA Postdoctoral Program Fellow, Goddard Space Flight Center, Greenbelt, MD 20771, USA \\
$^{3}$ Instituto de Astronom\'{i}a, Universidad Nacional Aut\'{o}noma de M\'{e}xico, Apartado Postal 70-264, 04510 M\'{e}xico, D. F., M\'{e}xico \\
$^{4}$ Astronomy Department, University of California, Berkeley,
    CA 94720-7450, USA \\
$^{5}$ NASA, Goddard Space Flight Center, Greenbelt, MD 20771, USA \\
$^{6}$ Instituto de Astronom\'{i}a, Universidad Nacional Aut\'{o}noma de M\'{e}xico, Apartado Postal 106, 22800 Ensenada, Baja California, M\'{e}xico \\
$^{7}$ Department of Astronomy and Astrophysics, UCO/Lick Observatory, University of California, 1156 High Street, Santa Cruz, CA 95064, USA \\
}
\begin{document}
\date{\today}
%\date{Accepted 2005 June 15. Received 1988 December 14; in original form 1988 October 11}

\pagerange{\pageref{firstpage}--\pageref{lastpage}} \pubyear{2014}

\maketitle

\label{firstpage}
\begin{abstract}
We present optical and  near-infrared (NIR) photometry of 28 gamma-ray
bursts  (GRBs) detected  by the  \textit{Swift} satellite  and rapidly
observed by  the Reionization and  Transients Infrared/Optical (RATIR)
camera.  We compare  the optical flux at fiducial times  of 5.5 and 11
hours after  the high-energy  trigger to that  in the X-ray  regime to
quantify optical darkness.  46$\pm$9 per cent (13/28) of all bursts in
our sample and  55$\pm$10 per cent (13/26) of  long GRBs are optically
dark,  which  is  statistically  consistently with  previous  studies.
Fitting RATIR optical and  NIR spectral energy distributions (SEDs) of
19 GRBs, most (6/7) optically  dark GRBs either occur at high-redshift
($z>4.5$) or have a high  dust content in their host galaxies ($A_{\rm
  V} >  0.3$).  Performing K-S tests,  we compare the  RATIR sample to
those   previously   presented   in   the  literature,   finding   our
distributions of redshift, optical  darkness, host dust extinction and
X-ray  derived column  density  to be  consistent.   The one  reported
discrepancy is with host galaxy dust content in the BAT6 sample, which
appears   inconsistent   with    our   sample   and   other   previous
literature.  Comparing  X-ray  derived  host  galaxy  hydrogen  column
densities to  host galaxy dust extinction,  we find that  GRBs tend to
occur in host galaxies with  a higher metal-to-dust ratio than our own
Galaxy, more akin to the  Large and Small Magellanic Clouds.  Finally,
to mitigate time evolution of optical darkness, we measure $\beta_{\rm
  OX,rest}$ at  a fixed rest frame time,  $t_{\rm rest}=1.5$~hours and
fixed rest frame energies in  the X-ray and optical regimes.  Choosing
to evaluate optical flux  at $\lambda_{\rm rest}=0.25~\mu$m, we remove
high-redshift  as a  source  of optical  darkness, demonstrating  that
optical darkness  must result from either  high-redshift, dust content
in the host  galaxy along the GRB sight line, or  a combination of the
two.
\end{abstract}


\section{Introduction}
\label{sec:intro}

Gamma-ray  bursts (GRBs)  are  powerful explosions  likely  to be  the
outcome of  the collapse of  massive stars \citep{2006ARA&A..44..507W}
or  of the  merger of  compact objects  in binaries  or  dense stellar
systems
\citep{2006NatPh...2..116G,2007PhR...442..166N,2007NJPh....9...17L,2010ApJ...720..953L}. The
central engine, accreting in the hypercritical neutrino-cooled regime,
produces       a      collimated       ultra-relativistic      outflow
\citep{1999ApJ...525..737R,2002ApJ...571..779P}, which converts energy
to radiation through  internal shocks \citep{1994ApJ...430L..93R}, and
external   shocks   with  the   surrounding   medium   (see   e.g.;
\citealt{2004RvMP...76.1143P}   and  references   therein),  producing
bright fluxes across the electromagnetic spectrum.  As such they allow
us   to  probe   the   Universe   at  a   wide   range  of   redshifts
\citep{2009Natur.461.1254T,2009Natur.461.1258S,2011ApJ...736....7C}
and  explore the properties  of host  galaxies and  intervening matter
along the GRB line of sight.\par

The  intrinsic  emission  from   GRBs  is  attributed  to  synchrotron
radiation  during  both  the  prompt  \citep{2014IJMPD..2330002Z}  and
afterglow \citep{2002ApJ...568..820G}  phases.  Broadband observations
across  the  electromagnetic spectrum  are  required  to confirm  this
mechanism  (e.g.; \citealt{2014ApJ...781...37P}).   Along the  line of
sight  to the  observer  photons  must pass  through  the host  galaxy
interstellar  medium  \citep{2009ApJ...691..182S},  the  intergalactic
medium  (IGM; \citealt{1965ApJ...142.1633G})  and the  Milky  Way (MW;
\citealt{1998ApJ...500..525S}).  Each of these environments affect the
observed spectral energy distribution  (SED) of the burst, potentially
leading  to the  GRB being  fainter  in some  wavelength regimes  than
expected under the standard GRB paradigm.\par

Broadband,  multiwavelength  observations of  GRBs  require rapid  and
precise  localisations,  which  are  now  routinely  provided  by  the
\textit{Swift}   satellite   \citep{2004ApJ...611.1005G}.    Extensive
ground-based  follow-up  of GRBs  is  enabled  by arcsecond  precision
measurements   made   by    the   on-board   X-ray   Telescope   (XRT;
\citealt{2005SSRv..120..165B})   and   Ultraviolet/Optical   Telescope
(UVOT;  \citealt{2005SSRv..120...95R}).    Included  in  the  numerous
facilities  now  routinely  observing  GRBs is  the  Reionization  and
Transients           Infrared/Optical          (RATIR)          camera
\citep{2012SPIE.8446E..10B}.   RATIR  has  an  automatic  response  to
\textit{Swift} triggers, allowing it to  observe a given field of view
within  minutes  of   an  alert  notice  of  a   new  gamma-ray  burst
\citep{2012SPIE.8444E..5LW,2012SPIE.8453E..2SK}.       With     almost
simultaneous coverage in  six filters in the optical  to near infrared
(NIR) regimes (from 5600 to 16000 \AA; \textit{riZYJH}), RATIR enables
the  modelling of  SEDs  using  templates for  the  IGM and  different
extinction        models       for        the        host       galaxy
\citep{2014AJ....148....2L}. Such modelling allows us to quantify host
galaxy   dust   extinction  and   estimate   a  photometric   redshift
\citep{2008A&A...490.1047C,2011A&A...526A.153K}.\par

The \textit{Swift}/XRT detects  emission associated with approximately
90 per  cent of  the GRBs detected  by the \textit{Swift}  Burst Alert
Telescope             (BAT;             \citealt{2005SSRv..120..143B})
\citep{2009MNRAS.397.1177E,2013ApJS..209...20G}. However, observations
with the  \textit{Swift}/UVOT and ground-based  telescopes detect only
about     40--60     per     cent     in    the     optical     regime
\citep{2010ApJ...720.1513K,2012A&A...545A..77R,2012ApJ...758...27L}. Some
of  the optical  non-detections are  consistent with  an extrapolation
from  the  X-ray  emission  using a  standard  unreddened  synchrotron
power-law spectrum.  However, approximately 25--50 per  cent require a
steeper                                                        spectrum
\citep{2008ApJ...686.1209M,2009ApJ...693.1484C,2009ApJS..185..526F,2011A&A...526A..30G,2012MNRAS.421.1265M},
and     these      are     called     ``optically      dark''     GRBs
\citep{2004ApJ...617L..21J,2005ApJ...624..868R,2009ApJ...699.1087V}.\par

There  are  two  explanations  attributed to  such  optical  darkness;
attenuation by material in the host galaxy \citep{2009ApJ...691..182S}
or    suppression    by   Ly-$\alpha$    absorption    in   the    IGM
\citep{2000ApJ...536....1L}.  For the latter  effect to reduce flux in
the optical regime,  a GRB must lie at high  redshift ($z \gtrsim 5$).
Quantifying both  of these effects is  important, as they  allow us to
study  galaxy evolution as  well as  that of  the star  formation rate
(SFR)  and   metallicity  of  galaxies  as  a   function  of  redshift
\citep{2012MNRAS.420..627S,2014arXiv1408.3578C}.\par

In  this  work we  first  describe the  data  obtained  from both  the
\textit{Swift}/XRT  and RATIR.   We then  briefly present  the success
RATIR has  had in rapid  follow-up of \textit{Swift} GRB  triggers. In
\S~\ref{sec:anal}  we  define  optical  darkness, and  identify  those
bursts which are considered  under-luminous in the optical regime when
compared to the X-ray observations. We present our optical and NIR SED
fitting results  in \S~\ref{sec:sed_fit}. With SED  templates in hand,
we comment upon how many GRBs  may occur at high redshift and upon the
dust  content of  these  GRB  host galaxies.  Finally,  we attempt  to
mitigate any temporal  evolution in optical darkness by  taking a rest
frame   defined  measure,   which  is   presented  and   discussed  in
\S~\ref{sec:rest_frame}.\par

\section{Data}
\label{sec:obs}

\subsection{X-ray data}
\label{sec:xraydata}

\textit{Swift}/XRT  count rate  light  curves were  obtained from  the
\textit{Swift}/XRT           light           curve          repository
\citep{2007A&A...469..379E,2009MNRAS.397.1177E}   hosted  at   the  UK
\textit{Swift}             Science             Data             Centre
(UKSSDC)\footnote{www.swift.ac.uk/xrt\_curves}.   Spectral information
for  the Windowed  Timing  (WT)  and Photon  Counting  (PC) modes  was
obtained from the pipeline detailed in \citet{2007ApJ...663..407B}. We
first convert  the \textit{Swift}/XRT count rate light  curves to flux
light curves across the  entire 0.3--10.0~keV energy band. Then, using
the  spectral models of  \citet{2007ApJ...663..407B}, we  convert this
full band flux to a flux density at 1~keV.\par

As   \textit{Swift}/XRT  and  RATIR   observations  were   not  always
simultaneous, we  fitted the XRT light curves  using the morphological
model of  \citet{2007ApJ...662.1093W}. To  do so, we  first identified
epochs of  flaring within  the XRT light  curve using  the methodology
described                 in                the                updated
documentation\footnote{http://www.swift.ac.uk//xrt\_live\_cat/docs.php\#lc}
of            the           UKSSDC            burst           analyser
\citep{2009MNRAS.397.1177E,2010A&A...519A.102E},   which  is  outlined
below.\par

Each     light    curve    was     initially    fitted     with    the
\citet{2007ApJ...662.1093W} model.   If XRT observations  began within
2~ks   of   the   initial   \textit{Swift}/BAT   trigger,   then   two
\citet{2007ApJ...662.1093W}  components  were  used.   Otherwise,  the
light curve was fitted using only a single component.\par

In     cases     where     the     rapid     decay     phase     (RDP)
\citep{2006ApJ...647.1213O,2006ApJ...642..389N} was observed, this was
used to constrain  the power-law index and plateau  time of the prompt
emission  tail component.  This  prompt component  was fitted  to data
prior  to  the end  of  the RDP.   A  second  afterglow component  was
initially fitted to the data after  the end of the RDP. A combined fit
was then performed  using both components and the  values derived from
the  preliminary modelling  of each  component individually.  In cases
where observations  began at  least 2~ks after  the \textit{Swift}/BAT
trigger time  the single component was  fitted to the  entire range of
data.\par

Once  a  model for  the  data set  had  been  produced, our  algorithm
searched   the  data   for  points   where  the   model  significantly
under-predicted  the  observed flux.  The  condition for  significance
within 3~ks of  the trigger time was 8$\sigma$  whilst bins after this
time  required 10$\sigma$  significance to  be marked  as  a candidate
flare.\par

If any candidate  flares were found, the most  significant was removed
from  the data  set  and the  data  were re-fitted.  This process  was
repeated iteratively  until no new  significant flares were  found. In
cases where  5 or more consecutive  bins were designated  a flare, the
significance  threshold was  reduced  to 6$\sigma$  and 8$\sigma$  for
flares peaking before and after 3~ks, respectively.\par

For  the  majority  of GRBs,  the  rise  of  the X-ray  afterglow  was
unobserved. This is  primarily due to the light  curve being dominated
by the RDP at this epoch.   We therefore fixed the afterglow rise time
to  $T_{\rm   rise}=100$  s,  with  the  exception   of  three  bursts
(GRB~130514B,  GRB~130603B and GRB~130606A),  in which  $T_{\rm rise}$
was  allowed to  float to  ensure a  good fit  was  obtained.\par

\subsection{RATIR data}
\label{sec:ratirdata}

RATIR is a six band simultaneous optical and NIR imager mounted on the
autonomous  1.5~m Harold  L.   Johnson Telescope  at the  Observatorio
Astron\'{o}mico  Nacional  on  Sierra  San Pedro  M\'{a}rtir  in  Baja
California, Mexico. Since commencing full operations in 2012 December,
RATIR  has been  responding  to GRB  triggers  from the  \textit{Swift}
satellite,  obtaining  simultaneous   photometry  in  the  \textit{r},
\textit{i},  \textit{Z}, \textit{Y},  \textit{J} and  \textit{H} bands
\citep{2012SPIE.8446E..10B,2012SPIE.8444E..5LW,2012SPIE.8453E..2SK,2012SPIE.8453E..1OF}.\par

RATIR  has  four detectors,  two  optical  and  two infrared  cameras,
allowing four images of a source to be taken simultaneously, either in
\textit{riZJ} or  \textit{riYH}. Both  of the infrared  detectors have
split filters so that, by  dithering sources across the field of view,
they can be observed in all six RATIR filters.  Individual frames from
the optical cameras have exposure times of 80~s, whilst those from the
NIR cameras are 67~s due to additional overheads.\par

The  images   are  reduced  in  near  real-time   using  an  automatic
pipeline. Bias  subtraction and  twilight flat division  are performed
using algorithms written in {\sc python}, image alignment is conducted
by astrometry.net \citep{2010AJ....139.1782L} and image co-addition is
achieved using {\sc swarp} \citep{2010ascl.soft10068B}.\par

We  use  {\sc  sextractor}  \citep{1996A&AS..117..393B}  to  calculate
photometry for  individual science  frames and mosaics  with apertures
ranging from 2 to 30 pixels in diameter, with an optical and NIR pixel
scales    of   0.32$^{\prime   \prime}$.pixel$^{-1}$    and   0.3$^{\prime
  \prime}$.pixel$^{-1}$,  respectively. Taking a  weighted average  of the
flux in  these apertures  for all  stars in a  field, we  construct an
annular point-spread-function  (PSF). Point source  photometry is then
optimised  by fitting  this PSF  to the  annular flux  values  of each
source.\par

To calibrate our  field photometry, we compare our  values to existing
catalogues,  including the  Sloan Digital  Sky Survey  Data  Release 9
(SDSS  DR9;   \citealt{2012ApJS..203...21A}).   The  RATIR   and  SDSS
\textit{r},   \textit{i}  and   \textit{Z}  bands   agree   to  within
$\lesssim$3  per  cent  \citep{2014Butlerprep}.   The  \textit{J}  and
\textit{H} bands  are calibrated  relative to the  Two Micron  All Sky
Survey  (2MASS; \citealt{2006AJ....131.1163S}).   We use  an empirical
relation  for  \textit{Y}  in   terms  of  \textit{J}  and  \textit{H}
magnitudes derived from the  United Kingdom Infrared Telescope (UKIRT)
Wide         Field         Camera         observations         (WFCAM;
\citealt{2009MNRAS.394..675H,2007A&A...467..777C}).    For  fields  of
view  without  SDSS  observations  we  use  the  United  States  Naval
Observatory
(USNO)-B1\footnote{http://tdc-www.harvard.edu/catalogs/ub1.html}
catalogue \citep{2003AJ....125..984M} to  calibrate the \textit{r} and
\textit{i} band  photometry.  In these  instances we use  an empirical
relation from WFCAM to calculate \textit{Z} band magnitudes.\par

We initially considered all GRBs observed by RATIR from 2013 January 1
to 2014 July  11 inclusive. In this time RATIR  observed the fields of
80 GRBs. 64 of these  were from \textit{Swift} on-board triggers, with
the  other 16 being  from \textit{Swift}  Target of  Opportunity (ToO)
requests.    A  breakdown   of  the   response  time   for   RATIR  to
\textit{Swift}/BAT    on-board   triggers    is    shown   in    Table
\ref{tab:response_times}.\par

\begin{table}
  \centering
  \caption{Delay   between    onset   of   RATIR    observations   and
    \textit{Swift}/BAT  trigger time  for  on-board \textit{Swift}/BAT
    triggers. The middle column shows  the fraction of GRBs within the
    time interval  that are  detected, whilst the  last column  is the
    percentage of  all on-board \textit{Swift}/BAT  triggers that were
    responded to within the indicated time interval.}
  \label{tab:response_times}
  \begin{tabular}{ccc}
    \hline
    \hline
    Time delay & GRBs & Percentage \\
    (hours) & (detected/observed) & of total \\
    \hline
    $<$ 0.5 & 5 / 10 (50\%) & 15.6\% \\
    0.5 -- 4 & 9 / 12 (75\%) & 18.8\% \\
    4 -- 8 & 4 / 7 (57\%) & 10.9\% \\
    8 -- 16 & 6 / 19 (32\%) & 29.7\% \\
    16 -- 24 & 0 / 10 (0\%) & 15.6\% \\
    $>$ 24 & 2 / 6 (33\%) & 9.4\% \\
    \hline
  \end{tabular}
\end{table}

In this study, we only consider those GRBs observed by RATIR within 10
hours of the initial high-energy trigger. We analysed the completeness
of the  RATIR sample as  a function of  delay between the  initial GRB
trigger  time and  the  beginning of  RATIR  observations. For  bursts
responded to within 10 hours  of the \textit{Swift} trigger, the RATIR
sample has  a detection rate of  approximately 50 per  cent.  After 10
hours  this fraction  rapidly reduces  (5/31), showing  that  at times
greater than 10 hours the RATIR sample is significantly less complete.
This     is    illustrated,     to    some     extent,     in    Table
\ref{tab:response_times}.   GRBs with  early  epoch observations  also
provide a  better data  set for later  modelling of the  optical light
curve.  This  limits our sample  to only 33  bursts, all of  which are
\textit{Swift}/BAT on-board triggers.  Three of these GRBs do not have
XRT observations (GRB~130626A, GRB~140118A and GRB~130215A), one burst
occurred during cloudy weather at the observatory (GRB~130122A), and a
further  GRB  was  observed  by  RATIR  in  only  the  \textit{Z}  and
\textit{Y}  bands  (GRB~130504A).   These  five  GRBs  were  therefore
removed from the sample.\par

To obtain  light curves for each of  the GRBs in our  sample, we first
concatenated all epochs of observation for each burst. We then removed
the effects of  Galactic foreground extinction using the  dust maps of
\citet{2011ApJ...737..103S}.   For those  bursts with  sufficient data
(14/28), we modelled the optical light curve with both a power-law and
broken                                                        power-law
\citep{2008ApJ...675..528L,2009MNRAS.395..490O,2009ApJ...693.1484C}
using {\sc  mpfit} \citep{2009ASPC..411..251M}. An F-test  was used to
determine if the temporal break was warranted, finding such a break to
be statistically  significant at the  3$\sigma$ level for only  2 GRBs
(GRB~130427A \& GRB~131030A).\par

\section{Analysis}
\label{sec:anal}

\begin{figure*}
  \begin{center}
    \includegraphics[width=8.5cm,height=8.5cm,clip,angle=0]{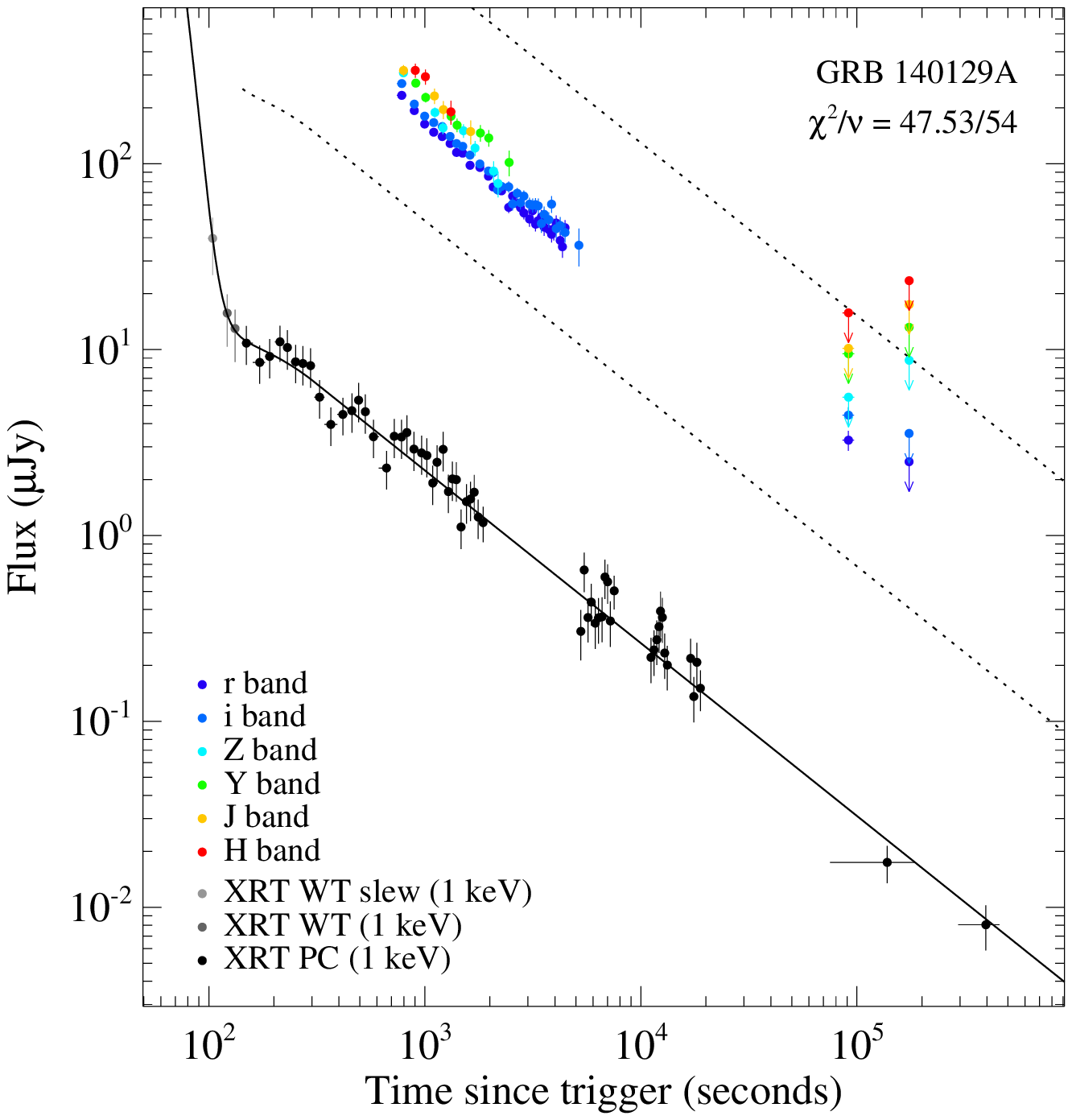}
    \quad
    \includegraphics[width=8.5cm,height=8.5cm,clip,angle=0]{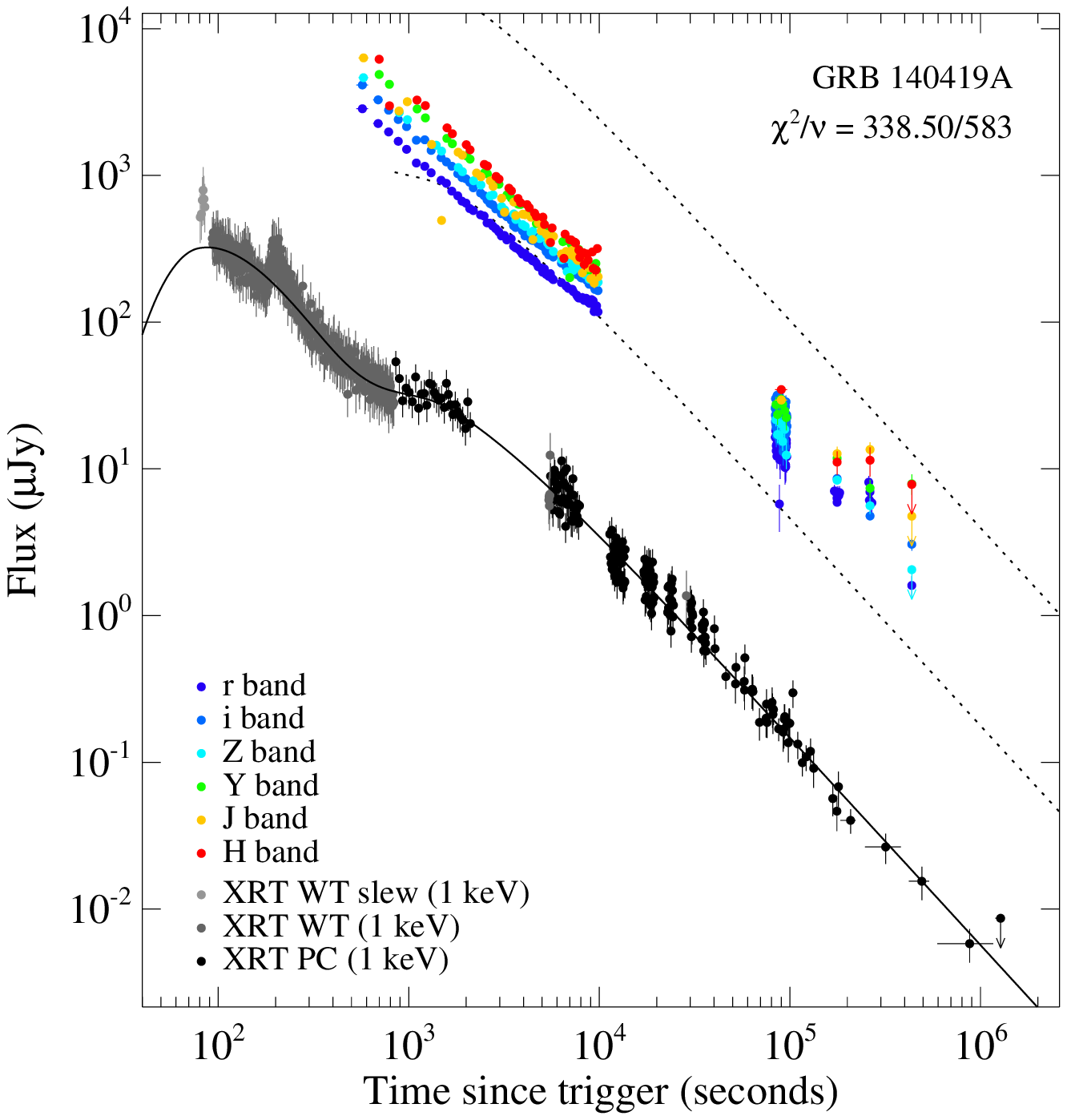}
  \end{center}
  \caption{Example light curves  in the joint RATIR-\textit{Swift}/XRT
    sample. Each  point is colour coded  as described in  the key. The
    solid black  line denotes the  best fit to  the \textit{Swift}/XRT
    light curve,  whilst the  upper and lower  black dashed  lines are
    extrapolations to  the \textit{r} band assuming  $\beta_{\rm OX} =
    \beta_{X}$  and   $\beta_{OX}=\beta_{X}-0.5$,  respectively.   The
    intrinsic  \textit{r} band  data is  expected  to  lie  on, or  be
    bounded by, these lines.}
  \label{fig:example_lcs}
\end{figure*}

\begin{table*}
  \centering
    \caption{GRB fields responded  to by RATIR within 10  hours of the
      initial \textit{Swift}  trigger.  The  delay refers to  the time
      between the initial \textit{Swift}/BAT  trigger and the start of
      RATIR  observations. The duration  of the  first RATIR  epoch of
      observations  is also  given,  along with  whether  the GRB  was
      detected  by RATIR  and observed  by the  \textit{Swift}/XRT. If
      there were  sufficient optical data  for modelling, a  bursts is
      denoted by  M, if optical detections  were made, but  too few to
      model,  a burst  is denoted  by D.   If only  upper  limits were
      obtained  by  RATIR,  the  burst  is denoted  by  UL.   Measured
      redshifts, X-ray spectral indices  for PC mode data, $\beta_{\rm
        X}$, $\beta_{\rm  OX}$ (at  both 5.5 and  11 hours)  and X-ray
      neutral hydrogen column density are also given, as well as other
      relevant  additional information  about each  GRB.   Bursts with
      X-ray  observations   and  a  redshift  also   have  a  reported
      rest-frame  $N_{\rm H}$  value  as obtained  from the  catalogue
      described in \citet{2007ApJ...663..407B}.}
    \label{tab:fast_sample}
    \begin{tabular}{ccccccccccc}
    \hline \hline
    GRB & Delay & Duration & RATIR & XRT & $z$ & $\beta_{\rm X}$ 
    & $\beta_{\rm OX}$ & $\beta_{\rm OX}$ & $N_{\rm H,rest}$ & Notes \\
    & (mins) & (mins) & & & & & (11 hours) & (5.5 hours) 
    & (10$^{21}$ cm$^{-2}$) & \\
    \hline
    130122A & 483.7 & 35 & ... & Y & ... & 0.84$_{-0.22}^{+0.30}$ & ... & ... 
    & ...  & Clouds \\
    130215A & 96.8 & 27 & M & N & 0.597$^{1}$ & ... & ... & ... & ... & ... \\
    130327A & 71.0 & 49 & D & Y & ... & 1.00$_{-0.30}^{+0.34}$ & 0.64$\pm$0.13 
    & 0.66$\pm$0.13 & ... & ... \\
    130418A & 494.9 & 184 & M & Y & 1.218$^{2}$ & 0.59$_{-0.18}^{+0.30}$ 
    & 1.01$\pm$0.09 & 1.07$\pm$0.08 & 0.0$_{-0.0}^{+5.5}$ & ... \\
    130420A & 136.7 & 21 & M & Y & 1.297$^{3}$ & 1.28$_{-0.11}^{+0.12}$ 
    & 0.65$\pm$0.06 & 0.68$\pm$0.06 & 3.9$_{-1.1}^{+1.2}$ & ... \\
    130427A & 16.9 & 64 & M & Y & 0.340$^{4}$ & 0.82$_{-0.04}^{+0.04}$ 
    & 0.60$\pm$0.03 & 0.61$\pm$0.02 & 1.0$_{-0.1}^{+0.1}$ & ... \\
    130502A & 569.4 & 43 & UL & Y & ... & 1.12$_{-0.32}^{+0.59}$ 
    & $<$0.50 & $<$0.51 & ... & ... \\
    130504A & 90.3 & 207 & D & Y & ... & 1.91$_{-0.28}^{+0.29}$ & ... 
    & ... & ... & ... \\
    130514A & 5.4 & 149 & UL & Y & 3.6$^{5}$ & 1.34$_{-0.32}^{+0.36}$ 
    & $<$0.22 & $<$0.20 & 37.4$_{-30.3}^{+36.9}$ & Photometric redshift \\
    130606A & 443.0 & 265 & M & Y & 5.913$^{6}$ & 0.87$_{-0.14}^{+0.15}$ 
    & 0.29$\pm$0.41 & 0.33$\pm$0.40 & 8.4$_{-8.4}^{+24.8}$ & ... \\
    130609A & 702.0 & 85 & UL & Y & ... & 1.95$_{-0.71}^{+0.80}$ & $<$0.14 
    & $<$0.15 & ... & ... \\
    130610A & 52.5 & 175 & M & Y & 2.092$^{7}$ & 1.08$_{-0.21}^{+0.24}$ 
    & 0.85$\pm$0.07 & 0.81$\pm$0.07 & 3.3$_{-3.3}^{+6.5}$ & ... \\
    130612A & 24.1 & 319 & M & Y & ... & 1.07$_{-0.26}^{+0.27}$ & 0.89$\pm$5.93 
    & 0.90$\pm$2.29 & ... & ... \\
    130626A & 3.0 & 13 & ... & N & ... & ... & ... & ... & ... 
    & $T_{\rm 90}=0.16\pm0.03$s \\
    130701A & 300.3 & 21 & D & Y & 1.155$^{8}$ & 1.08$_{-0.17}^{+0.19}$ 
    & 0.75$\pm$0.02 & 0.72$\pm$0.01 & 4.3$_{-2.4}^{+2.8}$ & ... \\
    130907A & 334.4 & 21 & M & Y & 1.238$^{9}$ & 0.96$_{-0.04}^{+0.04}$ 
    & 0.33$\pm$0.05 & 0.25$\pm$0.05 & 7.5$_{-0.6}^{+0.6}$ & ... \\
    130925A & 138.0 & 233 & M & Y & 0.347$^{10}$ & 2.44$_{-0.08}^{+0.08}$ 
    & $-$0.02$\pm$0.06 & $-$0.16$\pm$0.06 & 19.6$_{-0.8}^{+0.9}$ & ... \\
    131004A & 307.1 & 144 & UL & Y & 0.717$^{11}$ & 0.94$_{-0.25}^{+0.28}$ 
    & $<$1.08 & $<$0.90 & 5.1$_{-3.2}^{+3.7}$ & $T_{\rm 90}=1.54\pm0.33$s \\
    131030A & 294.4 & 260 & M & Y & 1.293$^{12}$ & 1.19$_{-0.11}^{+0.11}$ 
    & 0.73$\pm$0.02 & 0.71$\pm$0.02 & 4.6$_{-1.3}^{+1.5}$ & ... \\
    140114A & 9.4 & 43 & D & Y & ... & 0.99$_{-0.20}^{+0.41}$ & 0.26$\pm$0.06
    & 0.27$\pm$0.05 & ... & ... \\
    140118A & 39.6 & 4 & ... & N & ... & ... & ... & ... & ... & ... \\
    140129A & 12.4 & 56 & M & Y & ... & 1.00$_{-0.14}^{+0.15}$ & 0.66$\pm$0.06 
    & 0.66$\pm$0.05 & ... & ... \\
    140215A & 38.7 & 64 & M & Y & ... & 0.97$_{-0.13}^{+0.14}$ & 0.91$\pm$0.04 
    & 0.91$\pm$0.04 & ... & ... \\
    140311A & 524.8 & 125 & D & Y & 4.954$^{13}$ & 0.72$_{-0.15}^{+0.21}$ 
    & 0.57$\pm$0.25 & 0.53$\pm$0.24 & 0.0$_{-0.0}^{+37.0}$ & ... \\
    140318A & 294.6 & 211 & D & Y & 1.02$^{14}$ & 1.43$_{-0.59}^{+0.65}$ 
    & 0.79$\pm$0.55 & 0.82$\pm$0.47 & 8.0$_{-6.6}^{+7.3}$ & ... \\
    140331A & 21.2 & 91 & UL & Y & ... & 1.09$_{-0.15}^{+0.17}$ & $<$0.18 
    & $<$0.18 & ... & ... \\
    140419A & 8.8 & 97 & M & Y & 3.956$^{15}$ & 1.05$_{-0.07}^{+0.07}$ 
    & 0.61$\pm$0.01 & 0.58$\pm$0.01 & 11.2$_{-4.5}^{+4.9}$ & ... \\
    140518A & 36.3 & 43 & M & Y & 4.707$^{16}$ & 0.94$_{-0.12}^{+0.12}$ 
    & 0.34$\pm$0.11 & 0.22$\pm$0.10 & 0.0$_{-0.0}^{+60.5}$ & ... \\
    140614B & 6.5 & 43 & UL & Y & ... & 0.46$_{-0.18}^{+0.19}$ & $<$2.41 
    & $<$2.05 & ... & ... \\
    140622A & 1.3 & 64 & UL & Y & 0.959$^{17}$ & 1.60$_{-0.30}^{+0.60}$ 
    & $<$1.55 & $<$1.51 & 0.0$_{-0.0}^{+1.2}$ & $T_{\rm 90}=0.13\pm0.04$s \\
    140703A & 584.2 & 43 & M & Y & 3.14$^{18}$ & 0.98$_{-0.11}^{+0.11}$ 
    & 0.70$\pm$1.16 & 0.61$\pm$1.12 & 11.6$_{-6.9}^{+7.7}$ & ... \\
    140709A & 165.8 & 299 & D & Y & ... & 1.09$_{-0.15}^{+0.16}$ & 0.11$\pm$0.03
    & 0.10$\pm$0.03 & ... & ... \\
    140710A & 3.5 & 43 & D & Y & 0.558$^{19}$ & 0.92$_{-0.17}^{+0.29}$ 
    & 0.34$\pm$0.07 & 0.35$\pm$0.06 & 0.0$_{-0.0}^{+3.2}$ & ... \\
    \hline
  \end{tabular} \\
  \textit{References}: $^{1}$\citet{2013GCN..14207...1C}, $^{2}$\citet{2013GCN..14380...1D}, $^{3}$\citet{2013GCN..14437...1D}, $^{4}$\citet{2014ApJ...781...37P}, $^{5}$\citet{2013GCN..14634...1S}, $^{6}$\citet{2013ApJ...774...26C}, $^{7}$\citet{2013GCN..14848...1S}, $^{8}$\citet{2013GCN..14956...1X}, $^{9}$ \citet{2013GCN..15187...1D}, $^{10}$\citet{2013GCN..15249...1V}, $^{11}$\citet{2013GCN..15307...1C}, $^{12}$\citet{2013GCN..15407...1X}, $^{13}$\citet{2014GCN..15966...1C}, $^{14}$\citet{2014GCN..15988...1T}, $^{15}$\citet{2014GCN..16125...1T}, $^{16}$\citet{2014GCN..16301...1C}, $^{17}$\citet{2014GCN..16437...1H}, $^{18}$\citet{2014GCN..16505...1C} and $^{19}$\citet{2014GCN..16570...1T}.
\end{table*}

\subsection{Identifying dark GRBs}
\label{sec:darkgrbs}

Two criteria for dark GRBs  are traditionally used. The first is based
solely   on  the   X-ray  to   optical  spectral   index,  $\beta_{\rm
  OX}$. \citet{2004ApJ...617L..21J}  proposed that a  dark burst could
be classified  as one where  $\beta_{\rm OX} < 0.5$.   Following this,
\citet{2009ApJ...699.1087V}  used  a  large  sample of  41  GRBs  from
\citet{2008ApJ...689.1161G}  to  suggest  that  any  optical  darkness
criterion placed on $\beta_{\rm OX}$ should also account for the X-ray
spectral index, $\beta_{X}$. In  this alternative scenario, a dark GRB
is one that meets the condition that $\beta_{\rm OX} < \beta_{\rm X} -
0.5$.\par

The  latter criterion, as  proposed by  \citet{2009ApJ...699.1087V} is
motivated by a specific theory  of GRB emission.  As both internal and
external  shocks  are  expected  to  emit  via  synchrotron  radiation
\citep{2002ApJ...568..820G,2004IJMPA..19.2385Z},    there    are   two
expected scenarios for the nature of a GRB SED between the optical and
X-ray regimes  during late afterglow observations.  The  first is that
the  optical  emission  is  on  the  same  power-law  segment  of  the
synchrotron  SED as  the  X-rays.   In this  case,  $\beta_{\rm OX}  =
\beta_{\rm X}$,  giving an expected  upper bound to  $\beta_{\rm OX}$.
Alternatively, the  cooling break,  $\nu_{c}$, an expected  feature of
synchrotron emission, may be present  between the two regimes.  Such a
spectral break is characterised by  a steepening of power-law slope in
the  SED  above   the  break  frequency  by  $\Delta   \beta  =  0.5$.
$\beta_{\rm  OX}$ is an  average spectral  index over  the intervening
range, however the largest value $\nu_{c}$ can adopt is just below the
measured  X-ray regime.   As such,  $\beta_{\rm OX}$  therefore  has a
lower limit of  $\beta_{\rm OX} \geqslant \beta_{\rm X}  - 0.5$, which
leads   to   the  condition   for   optical   darkness  discussed   in
\citet{2009ApJ...699.1087V}.\par

In  Figure \ref{fig:example_lcs}  we  show flux  density light  curves
obtained from both \textit{Swift}/XRT  and RATIR for two bursts within
our sample. The grey scale points are the X-ray data, while the colour
points  denote RATIR  data, both  being described  in the  key  of the
figures. The dotted lines show the bounds in which the \textit{r} band
flux is expected. To calculate  these limits we extrapolated the X-ray
flux assuming  either $\beta_{\rm OX} = \beta_{\rm  X}$ or $\beta_{\rm
  OX}  =  \beta_{X}-0.5$.  The  former  condition  corresponds to  the
optical  regime lying  on the  same  power-law segment  as the  X-ray,
whilst the  latter assumes  a cooling break  at 0.3~keV.   This second
condition  predicts the  minimum flux  from the  intrinsic synchrotron
spectrum assuming there is no attenuation in the optical band.\par

To estimate  the optical  darkness, we take  a measure of  optical and
X-ray             flux             at             11             hours
\citep{2003ApJ...592.1018D,2004ApJ...617L..21J,2009ApJ...699.1087V,2011A&A...526A..30G}. At
such  a time  the emission  is  expected to  be in  an external  shock
dominated           phase          of           the          afterglow
\citep{2001ApJ...560L.167P,2003ApJ...590..379B,2007ApJ...668..400B},
with the prompt emission and  later X-ray flaring having ended and the
afterglow remaining significantly brighter than any host galaxy.\par

Half of  our rapidly  observed sample (14/28)  had sufficient  data to
allow  for the modelling  of the  optical light  curves with  either a
power-law or broken power-law.  In these instances, we interpolate the
optical flux,  $F_{\rm opt}$, at 11  hours from the  fitted model. For
those other GRBs  with few single detections or  upper limits, we take
the average temporal power-law decay  index from those modelled in our
sample and extrapolate from the available data to our fiducial time of
11 hours  after the initial  high-energy trigger to  calculate $F_{\rm
  opt}$.    \citet{2007ApJ...668..400B}   investigated  the   hardness
evolution of GRB X-ray afterglows  to understand when the internal and
external shocks  mechanisms dominate  the observed emission.   In that
work,  it  was found  that  X-ray  afterglows  were well  modelled  by
synchrotron   external   shock   emission   at  times   greater   than
$2\times10^{4}$~s  ($\approx$~5.5~hours)  after  the  initial  trigger
time. With this in mind we also estimated the optical and X-ray fluxes
for each  GRB in our sample at  5.5 hours. This allows  us to consider
the     time     evolution      of     $\beta_{\rm     OX}$     (e.g.;
\citealt{2012MNRAS.421.1265M}).\par

There are  seven GRBs  for which  there are only  upper limits  in the
RATIR  \textit{r} band observations. As  such it  is only  possible to
determine  upper  limits in  $\beta_{\rm  OX}$  for  these GRBs,  once
extrapolating the optical  upper limit to the fiducial  time. While we
use the mean fitted RATIR value  of temporal decay index to evolve the
upper limit  to constrain the optical  flux at this time,  there is an
inherent uncertainty  in this process.  Three of  these limits require
an  optically  dark  GRB  event.  The  remaining  three  (GRB~131004A,
GRB~140614B and GRB~140622A) do not allow us to classify the bursts as
dark, however, the calculated upper  limits in $\beta_{\rm OX}$ do not
preclude optical brightness.\par

In Table  \ref{tab:fast_sample} we show  the full sample of  GRBs with
RATIR  observations   beginning  within   10  hours  of   the  initial
high-energy  trigger. In this  table we  indicate whether  the optical
RATIR  light curve  data were  sufficient to  enable modelling  with a
power-law   or   broken    power-law.    We   also   present   whether
\textit{Swift}/XRT data were  available, reported redshifts and, where
measurable, the  calculated value of  $\beta_{\rm OX}$ at both  11 and
5.5~hours.  The  reported X-ray derived  column densities in  the rest
frame,   $N_{\rm   H,rest}$,    are   determined   by   an   automated
pipeline\footnote{http://butler.lab.asu.edu/swift}  which is described
in  detail  in  \citet{2007ApJ...663..407B}.  Each  \textit{Swift}/XRT
photon counting (PC) mode spectrum is fitted with a power-law spectrum
and   two   absorption  components,   corresponding   to  a   Galactic
\citep{2005A&A...440..775K} and extragalactic column.  We assume solar
metallicities      according      to      the     abundances      from
\citet{1989GeCoA..53..197A},    we     utilise    the    photoelectric
cross-section of \citet{1992ApJ...400..699B}  and the He cross-section
based on \citet{1998ApJ...496.1044Y}.\par

We note  a large value of  error in $\beta_{\rm  OX}$ for GRB~130612A.
This is due  to a large uncertainty in $F_{\rm  X}$ as calculated from
the   fitted   light   curve   at   5.5   and   11 hours   after   the
\textit{Swift}/BAT trigger.   Fundamentally this is a  result of large
error bars in  the X-ray flux density light curve  once converted to a
flux  at  1~keV, which  results  in  greater  uncertainty in  the  fit
parameters  used to derive  $F_{\rm X}$.  GRB~130504A was  detected by
RATIR, but was not observed in the \textit{r} band, and so $\beta_{\rm
  OX}$ was not calculated.\par

\begin{figure}
  \begin{center}
    \includegraphics[width=8.5cm,height=8.2cm,clip,angle=0]{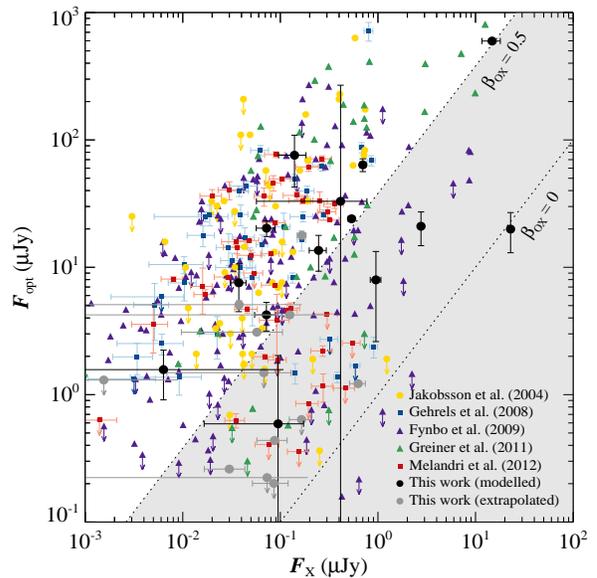}
    \caption{Optical  \textit{r}  band   flux,  $F_{\rm  opt}$,  as  a
      function of 1~keV X-ray flux, $F_{\rm X}$, at a fiducial time of
      11  hours.   Black points  correspond  to  $F_{\rm opt}$  values
      calculated from  fitted optical  models, grey points  with error
      bars  have $F_{\rm opt}$  extrapolated from  optical detections,
      using the mean temporal decay  index of the fitted RATIR sample,
      while grey  upper limits  are extrapolations from  optical upper
      limits as measured  by RATIR. The light grey  region denotes the
      parameter  space where  bursts  are considered  to be  optically
      dark. Also plotted  are values of $F_{\rm opt}$  and $F_{\rm X}$
      available  from  the   literature.   Plot  symbols  and  colours
      corresponding  to each  sample are  denoted  in the  key in  the
      bottom right corner of the panel.}
    \label{fig:fopt_vs_fx}
  \end{center}
\end{figure}

In  Figure \ref{fig:fopt_vs_fx},  we  show the  optical flux,  $F_{\rm
  opt}$, as  a function  of X-ray flux,  $F_{\rm X}$, at  the fiducial
time of  11 hours after the initial  \textit{Swift}/BAT trigger. Those
bursts with fully modelled optical light curves are plotted with black
points  and error  bars.   Those  with few  data  extrapolated to  the
fiducial  time   are  plotted  in   grey.   Also  plotted   on  Figure
\ref{fig:fopt_vs_fx}  are lines  denoting $\beta_{\rm  OX} =  0.5$ and
$\beta_{\rm OX}=0$.  Any GRB in the grey region of the parameter space
below and to the right of $\beta_{\rm OX} = 0.5$ is considered to be a
dark GRB by the condition of \citet{2004ApJ...617L..21J}.\par

As well  as the RATIR  sample, we plot  the data from  several samples
available               from               the              literature
\citep{2004ApJ...617L..21J,2008ApJ...689.1161G,2009ApJS..185..526F,2011A&A...526A..30G,2012MNRAS.421.1265M}. Of
the five samples shown  with ours, that of \citet{2011A&A...526A..30G}
has the  most similar  selection criteria to  ours. In this  work GRBs
detected by the Gamma-Ray burst Optical/Near-infrared Detector (GROND;
\citealt{2008PASP..120..405G}),  a  seven   channel  optical  and  NIR
imager,  within four hours  of the  high-energy trigger  are included.
Also, we  note that the sample of  \citet{2008ApJ...689.1161G} is that
with which  \citet{2009ApJ...699.1087V} define their  optical darkness
criterion.   \citet{2009ApJS..185..526F} define their  measurements of
optical  darkness at  earlier epochs  than the  11 hour  fiducial time
selected  in this work.   We also  show results  from the  BAT6 sample
\citep{2012ApJ...749...68S},  taken  from \citet{2012MNRAS.421.1265M}.
GRBs in this sample are  selected contingent on being bright enough in
the   hard   X-ray   regime   (15--350~keV),  as   measured   by   the
\textit{Swift}/BAT, to be detected if they were six times fainter.\par

From Figure \ref{fig:fopt_vs_fx}  it can be seen that  the majority of
optically dark bursts are generally also at the faint end of the total
distribution of optical fluxes at 11 hours.  There is also a hint of a
bifurcation  in the  population  separated by  $\Delta \beta_{\rm  OX}
\approx 0.5$,  which would be  the expected split in  the distribution
between those GRBs with or without a cooling break between the optical
and X-ray regimes at 11 hours.\par

\begin{figure}
  \begin{center}
    \includegraphics[width=8.5cm,height=8.2cm,clip,angle=0]{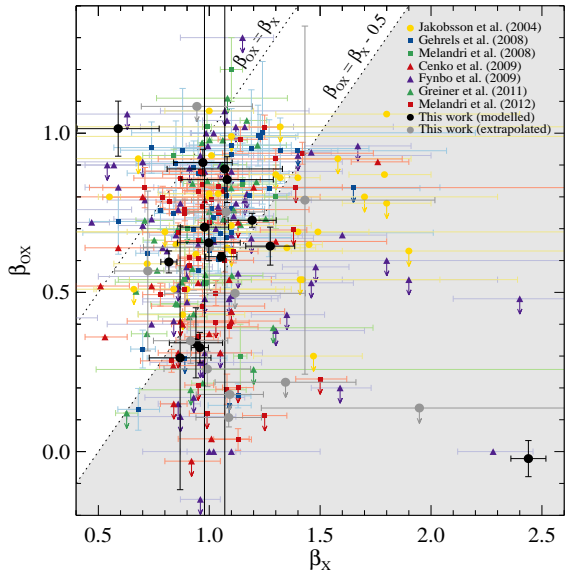}
    \caption{$\beta_{\rm OX}$  as a function  of $\beta_{\rm X}$  at a
      fiducial time  of 11 hours.  Black points  correspond to $F_{\rm
        opt}$  values interpolated  from fitted  optical  models. Grey
      points  with error  bars  have $F_{\rm  opt}$ extrapolated  from
      optical detections,  using the mean temporal decay  index of the
      fitted RATIR sample, while  grey upper limits are extrapolations
      from optical upper limits as  measured by RATIR.  The light grey
      region denotes  the parameter space where  bursts are considered
      to be  optically dark.  Also  plotted are values  of $\beta_{\rm
        OX}$ and $\beta_{\rm X}$  available from the literature.  Plot
      symbols and colours corresponding  to each sample are denoted in
      the key in the top right corner of the panel.}
    \label{fig:betaox_vs_betax}
  \end{center}
\end{figure}

In  Figure \ref{fig:betaox_vs_betax}  we  plot $\beta_{\rm  OX}$ as  a
function of $\beta_{\rm X}$, again for our sample and for those values
available               from               the              literature
\citep{2004ApJ...617L..21J,2008ApJ...689.1161G,2008ApJ...686.1209M,2009ApJ...693.1484C,2009ApJS..185..526F,2012MNRAS.421.1265M}. The
majority of  bursts in our sample  with well fitted  optical and X-ray
light  curves populate  the same  region of  parameter space  as those
analysed in previous samples. The two dashed lines on the panel denote
the range in  which GRBs well described in both  the optical and X-ray
regimes purely  by synchrotron emission should inhabit.   We find only
one  GRB with  a  detected optical  flux  suggesting that  $\beta_{\rm
  OX}>\beta_{\rm X}$,  GRB~130418A. This burst has one  of the softest
measured X-ray spectral indices, and is discussed in further detail in
\S~\ref{sec:rest_frame}.    GRB~140614B  and  GRB~140622A   both  have
optical upper  limits that correspond  to upper limits  in $\beta_{\rm
  OX}$,  which do  not  preclude this  possibility.   The position  of
GRB~140622A lies approximately  40$^{\prime \prime}$ from the 12$^{\rm
  th}$ magnitude star TYC~5783-1382-1, meaning that the photometry for
GRB~140622A   could  suffer  from   contamination  from   this  bright
object.\par

Figure \ref{fig:betaox_vs_betax}  shows that  the majority of  GRBs in
the  RATIR sample  inhabit regions  of the  parameter space  also well
populated by other samples. We  find 39$\pm$9 per cent (11/28) of GRBs
in   our  sample   are  identified   as  optically   dark   using  the
\citet{2004ApJ...617L..21J}   criterion,  whilst  46$\pm$9   per  cent
(13/28) are dark as  defined by \citet{2009ApJ...699.1087V}.  All GRBs
identified  as dark by  the \citet{2004ApJ...617L..21J}  criterion are
also identified  by that of  \citet{2009ApJ...699.1087V}.  GRB~130420A
and  GRB~140318A qualify  as optically  dark when  accounting  for the
value of  $\beta_{\rm X}$.  Both $\beta_{\rm OX}$  and $\beta_{\rm X}$
are reported in  Table \ref{tab:fast_sample}.\par

We  adopt   the  \citet{2009ApJ...699.1087V}  definition   of  optical
darkness.   As such, we  find that  46$\pm$9 per  cent (13/28)  of the
rapidly  observed RATIR  GRB sample  are optically  dark (GRB~130420A,
GRB~130502A,   GRB~130514A,  GRB~130606A,   GRB~130609A,  GRB~130907A,
GRB~130925A,   GRB~140114A,  GRB~140318A,   GRB~140331A,  GRB~140518A,
GRB~140709A  \& GRB~140710A).  This  selection remains  identical when
considering  $\beta_{\rm  OX}$  as  calculated  at either  11  or  5.5
hours. Comparing the $\beta_{\rm OX}$ values calculated at both epochs
reveals  only   four  GRBs  that   have  $\beta_{\rm  OX}$   that  are
inconsistent  at   the  1$\sigma$  level   (GRB~130701A,  GRB~130907A,
GRB~130925A \& GRB~140419A).  In each case, there is  a small increase
in $\beta_{\rm OX}$, indicating that the optical flux becomes slightly
less attenuated with  time. This is somewhat expected,  as the average
optical  decay power-law  index is  shallower than  that of  the X-ray
afterglows.\par

This fraction of optically dark GRBs is comparable to previous studies
such  as  \citet{2011A&A...526A..30G} and  \citet{2009ApJS..185..526F}
who find  the dark fraction of bursts  in their samples to  be 25-- 40
per      cent     and      25--42     per      cent,     respectively.
\citet{2009ApJ...693.1484C}   and   \citet{2008ApJ...686.1209M}   both
estimate a  dark burst fraction  of 50 per  cent, although it  must be
noted   that  these   studies   use  the   \citet{2004ApJ...617L..21J}
definition of optical darkness.   Most recently, work on the gamma-ray
selected BAT6  sample suggests  the dark fraction  in their  sample is
also 25--35 per cent \citep{2012MNRAS.421.1265M}.\par

If  we restrict  our sample  to  only long  GRBs ($T_{\rm  90} >  2$s;
\citealt{1993ApJ...413L.101K}),  this fraction increases  to 50$\pm$10
per cent  (13/26).  Those bursts  identified as short GRBs  have their
reported $T_{\rm 90}$ values  as measured by the \textit{Swift}/BAT in
the 15--350~keV  range in  Table \ref{tab:fast_sample}. The  two short
GRBs with  estimates of $\beta_{\rm OX}$  (GRB~131004A \& GRB~140622A)
have optical upper limits that  do not provide rigorous constraints on
$\beta_{\rm OX}$.\par

Typically,   GRBs  with  spectroscopic   redshifts  also   have  RATIR
\textit{r} band detections. The two exceptions to this are GRB~131004A
and  GRB~140622A,  both  of  which  are  short  GRBs.   To  measure  a
spectroscopic redshift requires a GRB to remain bright for the typical
response time  of large  spectrograph facilities, thus  increasing the
likelihood of a  RATIR optical detection. By limiting  our sample only
based on the condition of  rapid RATIR observation we have presented a
homogeneous  sample, which  limits any  brightness bias  introduced by
requiring a spectroscopic redshift. Considering only those GRBs with a
sufficient  optical  data  to   allow  light  curve  modelling  and  a
spectroscopic  redshift,  we  find   55$\pm$15  per  cent  (6/11)  are
optically    dark,    which    is    consistent    with    the    full
sample. Interestingly,  we find that GRB~140311A is  not classified as
optically dark,  despite having a spectroscopic  redshift of $z=4.954$
\citep{2014GCN..16301...1C}.    This   is    discussed    further   in
\S~\ref{sec:sed_fit}.\par

To statistically assess the  similarity between the RATIR and previous
distributions  of $\beta_{\rm OX}$  and the  dark burst  fractions, we
performed Kolmogorov-Smirnov  (K-S) tests.  This tests  two samples of
data  under the  null hypothesis  that the  two derive  from  the same
parent population  with significant low  probabilities indicating this
null hypothesis to  be inaccurate. A K-S test  compares the cumulative
distribution  functions  (CDFs)  of  the two  samples,  measuring  the
maximum distance between the two  CDFs, $D_{\rm KS}$, which is the K-S
statistic. Larger  values of $D_{\rm  KS}$ are indicative of  the CDFs
have a larger  maximum separation.  Performing this K-S  test yields a
probability $p_{\rm KS}=0.15$, indicating that the two populations are
not significantly  different.  Results from these K-S  tests are shown
in  Table   \ref{tab:ks_test_results},  which  also   details  similar
statistical comparisons  between parameters calculated  from RATIR SED
fitting.  From  these results,  we can see  that the  distributions of
redshift, $\beta_{\rm X}$ and $\beta_{\rm  OX}$ for our sample are not
statistically  significantly  different   from  any  other  individual
sample. We  also compiled a  total sample of all  previous literature,
taking  care to  only include  each GRB  once if  present  in multiple
samples,  finding  once  more  that  our  distributions  of  redshift,
$\beta_{\rm X}$  and $\beta_{\rm OX}$ are consistent  with the largest
possible sample of previous literature values.\par

\subsection{RATIR SED fitting}
\label{sec:sed_fit}

In order  to understand  why GRBs within  the RATIR-\textit{Swift}/XRT
sample might  appear under-luminous in  the optical regime,  we fitted
the RATIR optical and NIR SEDs. We did so for all of the bursts listed
in  Table \ref{tab:fast_sample} that  had photometry  in a  minimum of
four filters, with  a minimum of three detections.  This allowed us to
model 19 GRBs.\par

The expected shape  of an optical and NIR GRB SED  is a power-law with
potential perturbations from  either absorption from the intergalactic
medium (IGM) or dust within the host galaxy of the burst.  To maximise
signal-to-noise ratio,  we therefore used SEDs  obtained when coadding
all  observations made  during the  first night  of  observations. The
RATIR photometry  obtained during the first night  of observations for
those  bursts  in  Table   \ref{tab:fast_sample}  is  shown  in  Table
\ref{tab:first_night_phot}.\par

\begin{table*}
  \centering
    \caption{Stacked  photometry obtained  during the  first  night of
      RATIR observations of those  bursts responded to within 10 hours
      of the initial  \textit{Swift}/BAT trigger. These magnitudes are
      in the AB  system and are not corrected  for Galactic extinction
      in the direction of the GRB.}
    \label{tab:first_night_phot}
    \begin{tabular}{ccccccc}
    \hline \hline
    GRB & \textit{r} & \textit{i} & \textit{Z} & \textit{Y} & \textit{J} 
    & \textit{H} \\
    \hline
    130215A & 17.47$\pm$0.06 & 17.22$\pm$0.05 & 16.94$\pm$0.06 
    & 16.80$\pm$0.14 & 16.72$\pm$0.06 & ... \\
    130327A & 21.01$\pm$0.10 & 20.66$\pm$0.09 & 20.20$\pm$0.12 
    & 20.02$\pm$0.15 & 19.89$\pm$0.18 & 20.56$\pm$0.41 \\
    130418A & 19.02$\pm$0.01 & 18.77$\pm$0.01 & 18.49$\pm$0.02 
    & 18.28$\pm$0.02 & 18.17$\pm$0.02 & 17.73$\pm$0.02 \\
    130420A & 19.76$\pm$0.02 & 19.56$\pm$0.02 & 19.39$\pm$0.05 
    & 19.10$\pm$0.09 & 19.04$\pm$0.06 & 18.81$\pm$0.12 \\
    130427A & ... & 13.88$\pm$0.04 & 13.78$\pm$0.04 & 13.67$\pm$0.04 
    & 13.69$\pm$0.04 & 13.73$\pm$0.04 \\
    130502A & $>$22.43 & $>$23.04 & $>$22.11 & $>$21.35 & $>$21.04 & $>$20.64 \\
    130504A & ... & ... & $>$22.59 & ... & $>$20.49 & ... \\
    130514A & $>$23.17 & $>$22.87 & $>$22.46 & $>$21.89 & $>$21.67 & $>$21.22 \\
    130606A & 24.49$\pm$0.34 & 21.83$\pm$0.28 & 19.32$\pm$0.04 
    & 19.06$\pm$0.03 & 18.97$\pm$0.03 & 18.58$\pm$0.03 \\
    130609A & $>$23.65 & $>$23.37 & $>$22.54 & $>$21.88 & $>$21.54 & $>$21.07 \\
    130610A & 20.48$\pm$0.09 & 21.01$\pm$0.18 & 20.24$\pm$0.13 
    & 20.61$\pm$0.17 & ... & $>$15.99 \\
    130612A & 22.41$\pm$0.08 & 22.05$\pm$0.08 & ... & ... & ... & ... \\
    130701A & 19.61$\pm$0.50 & 20.45$\pm$1.08 & 19.91$\pm$0.09 
    & 19.36$\pm$0.11 & ... & ... \\
    130907A & 20.01$\pm$0.03 & 19.31$\pm$0.02 & 18.82$\pm$0.05 
    & 18.45$\pm$0.06 & 18.16$\pm$0.06 & 17.62$\pm$0.06 \\
    130925A & 20.92$\pm$0.17 & 21.31$\pm$0.18 & 20.67$\pm$0.11 
    & 20.80$\pm$0.17 & 20.02$\pm$0.09 & 19.74$\pm$0.11 \\
    131004A & $>$23.90 & $>$23.37 & $>$22.61 & $>$21.97 & $>$21.73 & $>$21.22 \\
    131030A & 19.15$\pm$0.05 & 18.92$\pm$0.04 & 18.77$\pm$0.04 
    & 18.55$\pm$0.04 & 18.62$\pm$0.04 & 18.37$\pm$0.04 \\
    140114A & 21.81$\pm$0.10 & 21.24$\pm$0.07 & ... & ... & ... & ... \\
    140129A & 19.11$\pm$0.02 & 18.98$\pm$0.02 & 18.87$\pm$0.05 
    & 18.69$\pm$0.04 & 18.80$\pm$0.06 & 18.57$\pm$0.07 \\
    140215A & 17.92$\pm$0.28 & 17.56$\pm$0.21 & 17.24$\pm$0.16 
    & 16.99$\pm$0.13 & 16.80$\pm$0.12 & 16.55$\pm$0.09 \\
    140311A & 22.34$\pm$0.13 & 21.57$\pm$0.08 & 20.59$\pm$0.08 
    & 20.07$\pm$0.08 & ... & ... \\
    140318A & 21.94$\pm$0.18 & 21.48$\pm$0.14 & 20.83$\pm$0.16 
    & 20.80$\pm$0.21 & 20.91$\pm$0.21 & 20.25$\pm$0.13 \\
    140331A & $>$23.65 & $>$23.49 & $>$22.34 & $>$21.80 & $>$21.59 & $>$21.00 \\
    140419A & 17.65$\pm$0.17 & 17.30$\pm$0.12 & 16.91$\pm$0.09 
    & 16.76$\pm$0.08 & 17.19$\pm$0.11 & 16.57$\pm$0.07 \\
    140518A & 20.52$\pm$0.32 & 19.00$\pm$0.08 & 18.60$\pm$0.06 
    & 18.20$\pm$0.05 & 18.12$\pm$0.04 & 17.80$\pm$0.03 \\
    140614B & $>$22.71 & $>$22.56 & $>$21.66 & $>$21.09 & $>$21.05 & $>$20.52 \\
    140622A & $>$23.58 & $>$23.43 & $>$19.31 & $>$19.75 & ... & ... \\
    140703A & 20.32$\pm$0.04 & 19.72$\pm$0.03 & 18.53$\pm$0.05 
    & 18.31$\pm$0.05 & 19.60$\pm$0.17 & 19.98$\pm$0.09 \\
    140709A & $>$24.10 & $>$23.70 & $>$22.92 & $>$22.41 & $>$22.31 & $>$21.94 \\
    140710A & 21.35$\pm$0.08 & 21.09$\pm$0.07 & 20.76$\pm$0.14 
    & 20.41$\pm$0.16 & 20.29$\pm$0.14 & 19.74$\pm$0.13 \\
    \hline
    \end{tabular}
\end{table*}

We    used   the    SED   template-fitting    routine    outlined   in
\citet{2014AJ....148....2L} to estimate the amount of host galaxy dust
extinction.  This  algorithm accounts for the  intrinsic GRB spectrum,
Galactic  dust extinction,  the absorption  from  the IGM  due to  the
redshift  of the host  galaxy and  the dust  absorption from  the host
galaxy along the GRB sight line by fitting the optical spectral index,
$\beta_{\rm opt}$, redshift, $z$,  and dust extinction $A_{\rm V}$. As
the exact  nature of the  dust extinction law  is not known,  we apply
templates of  Milky Way (MW),  Large Magellanic Cloud (LMC)  and Small
Magellanic       Cloud      (SMC)      dust       extinction      laws
\citep{1992ApJ...395..130P}, allowing the  algorithm to choose between
the model  that best describes any  dust present in the  SED. While we
note   that   these   three   templates  are   likely   not   accurate
representations of  all galaxies at  all redshifts, they do  allow for
the best comparisons with other studies.\par

\begin{figure*}
  \begin{center}
    \includegraphics[width=7.5cm,clip,angle=0]{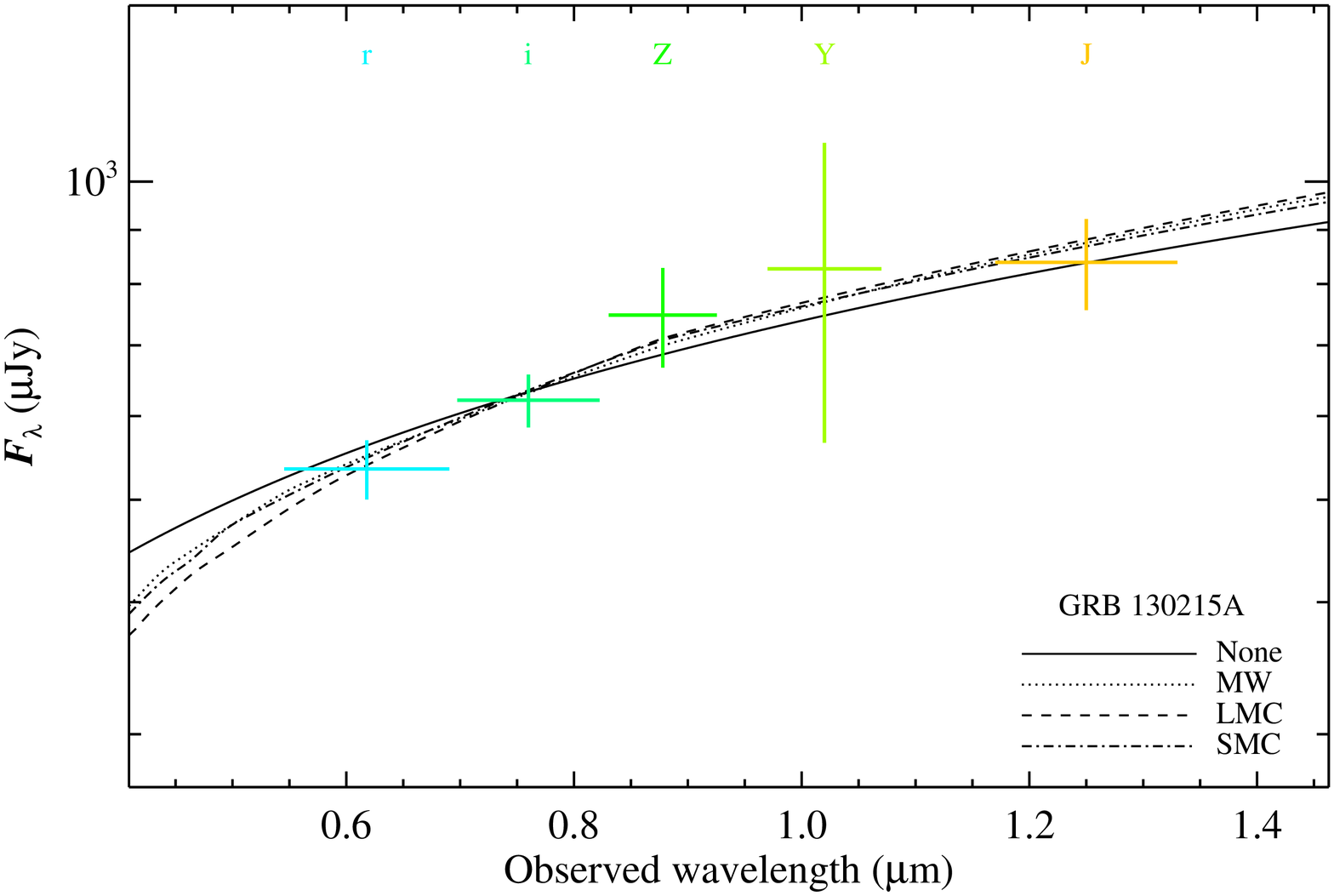}
    \quad
    \includegraphics[width=7.5cm,clip,angle=0]{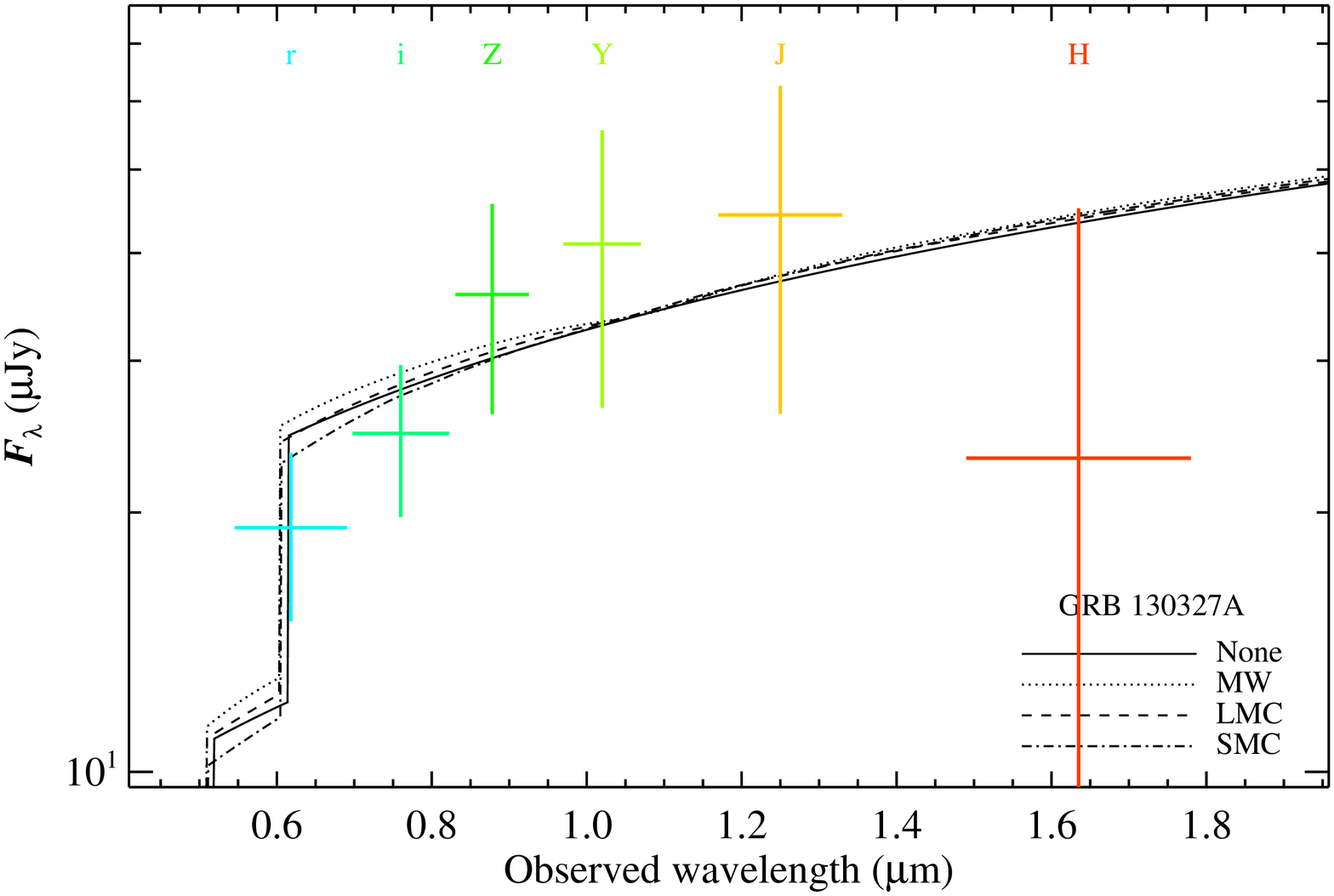}
    \\
    \includegraphics[width=7.5cm,clip,angle=0]{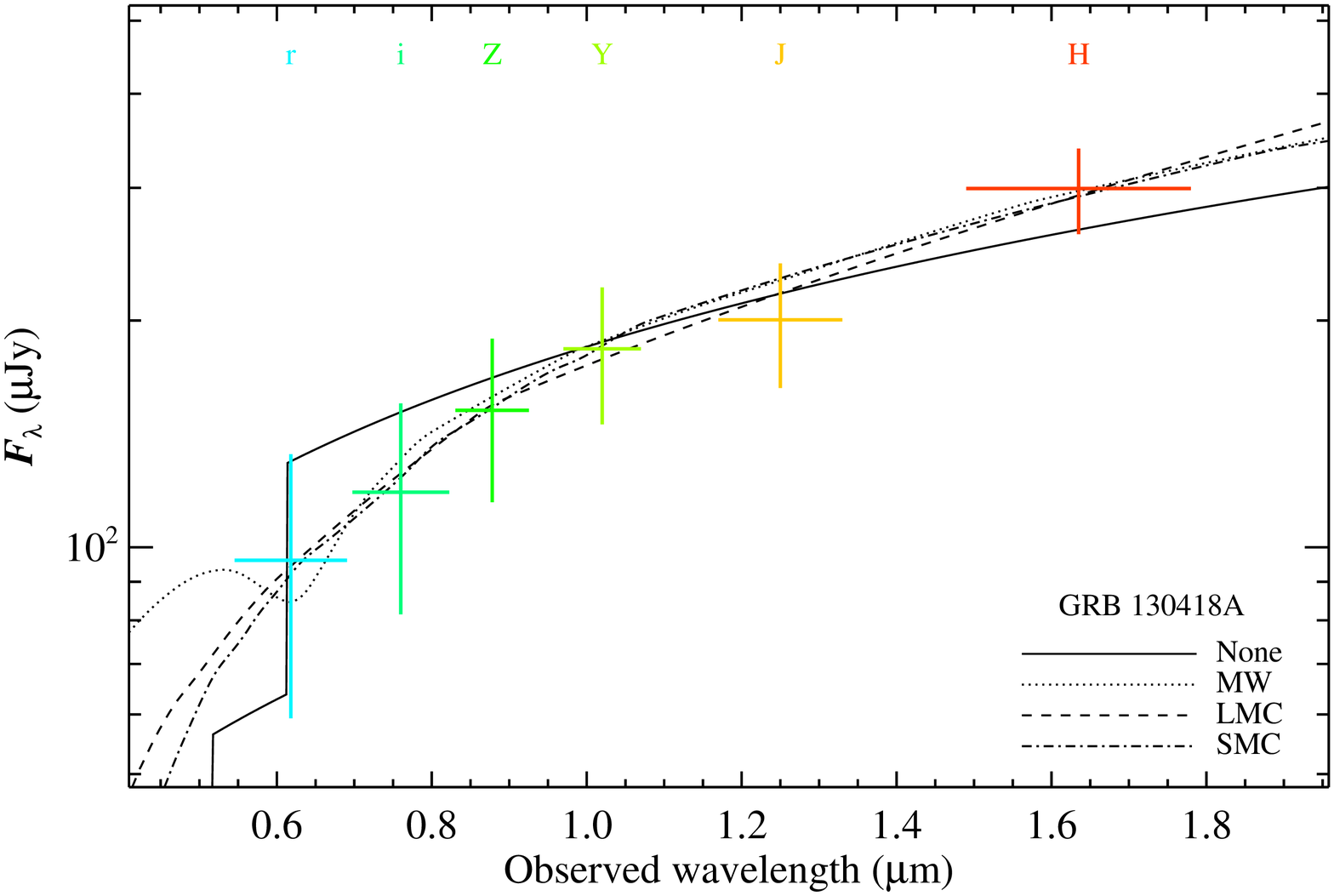}
    \quad
    \includegraphics[width=7.5cm,clip,angle=0]{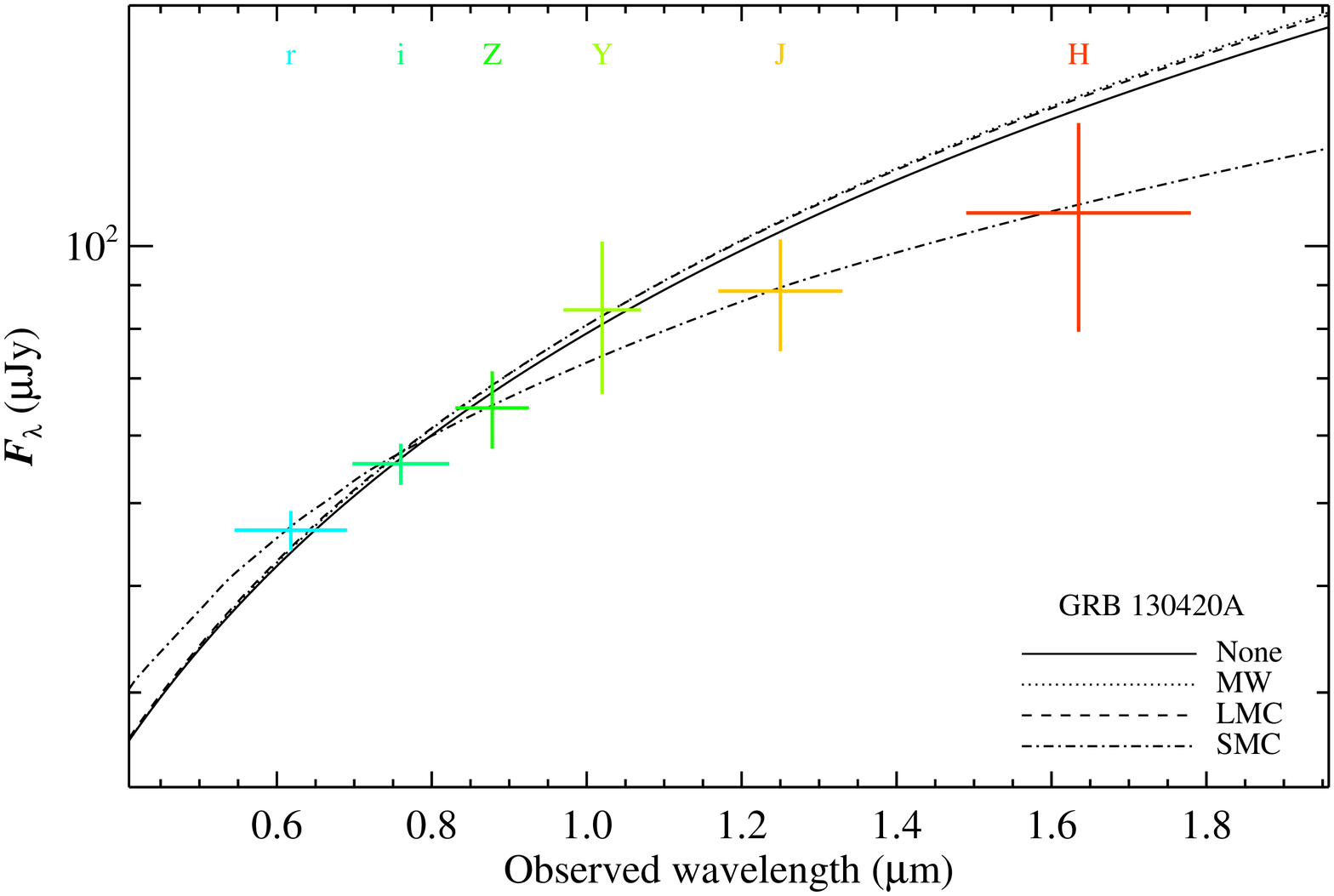}
    \\
    \includegraphics[width=7.5cm,clip,angle=0]{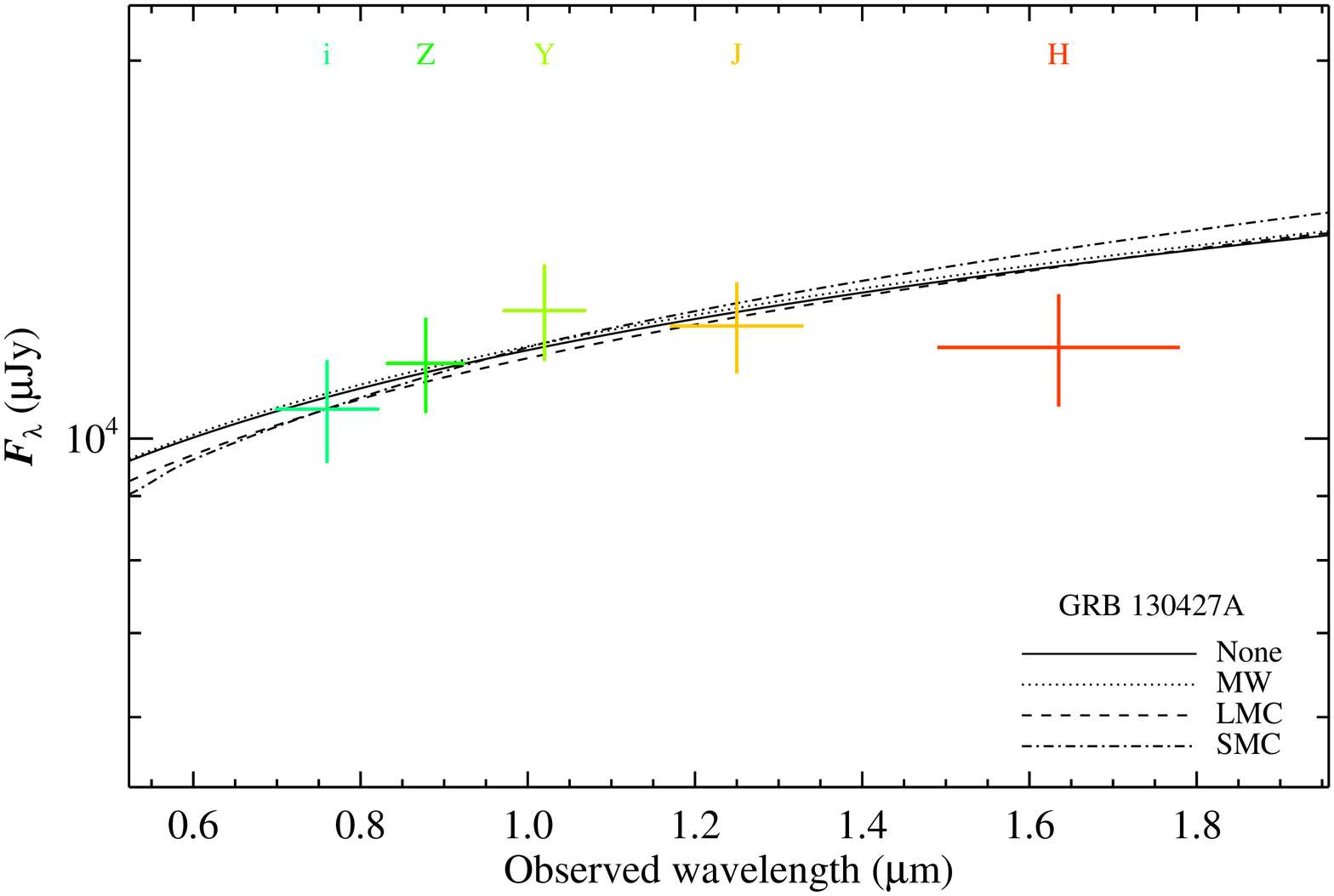}
    \quad
    \includegraphics[width=7.5cm,clip,angle=0]{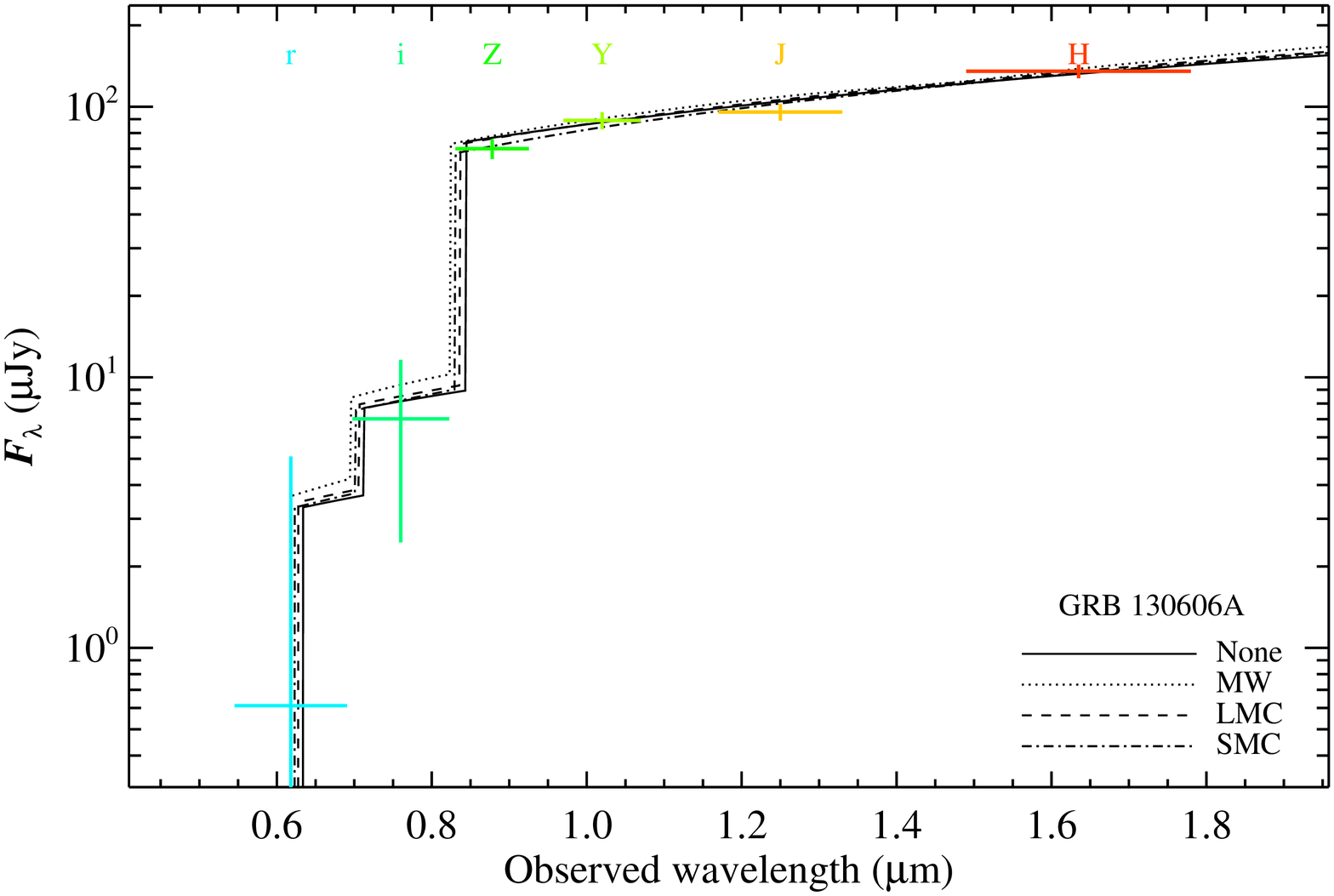}
    \\
    \includegraphics[width=7.5cm,clip,angle=0]{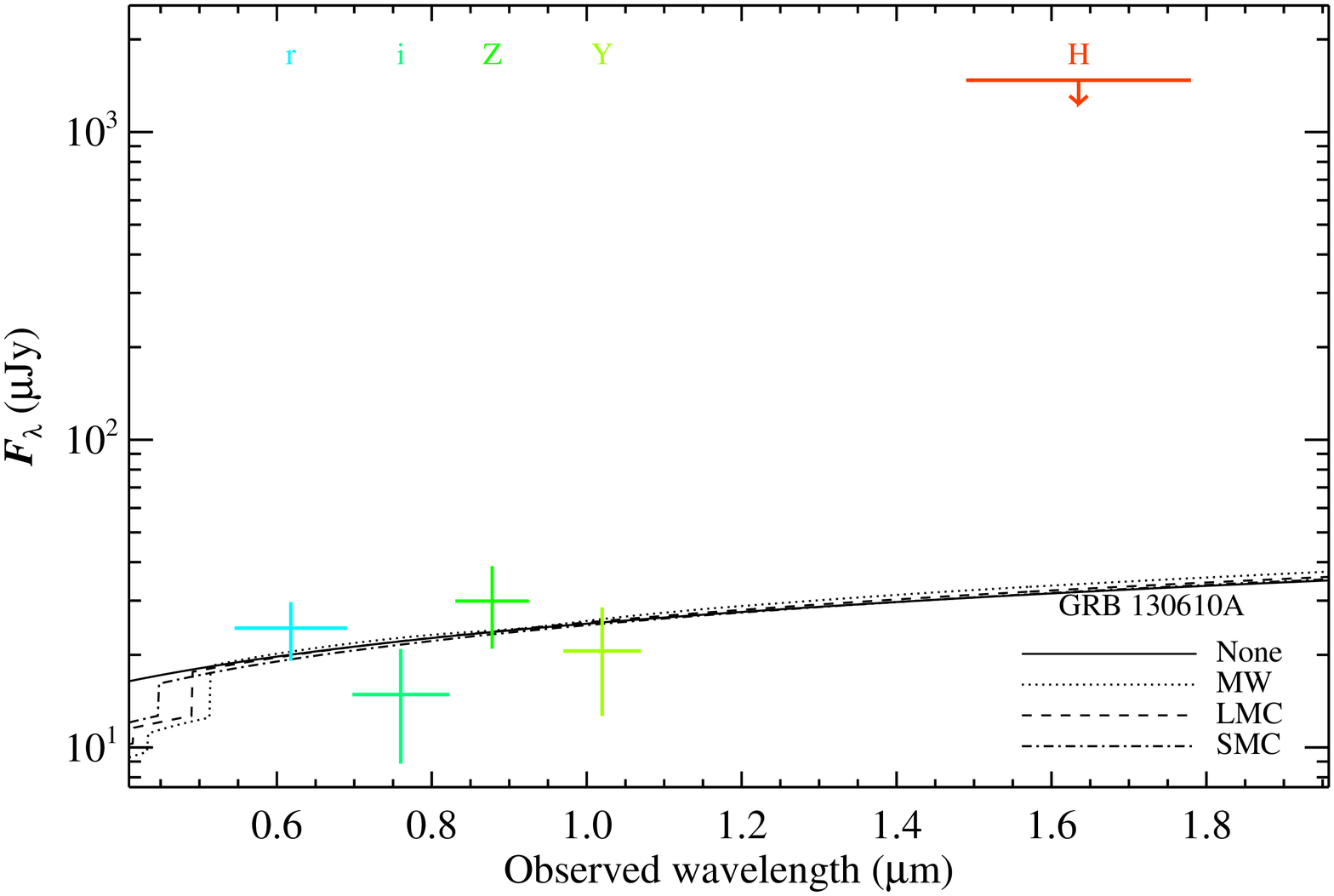}
    \quad
    \includegraphics[width=7.5cm,clip,angle=0]{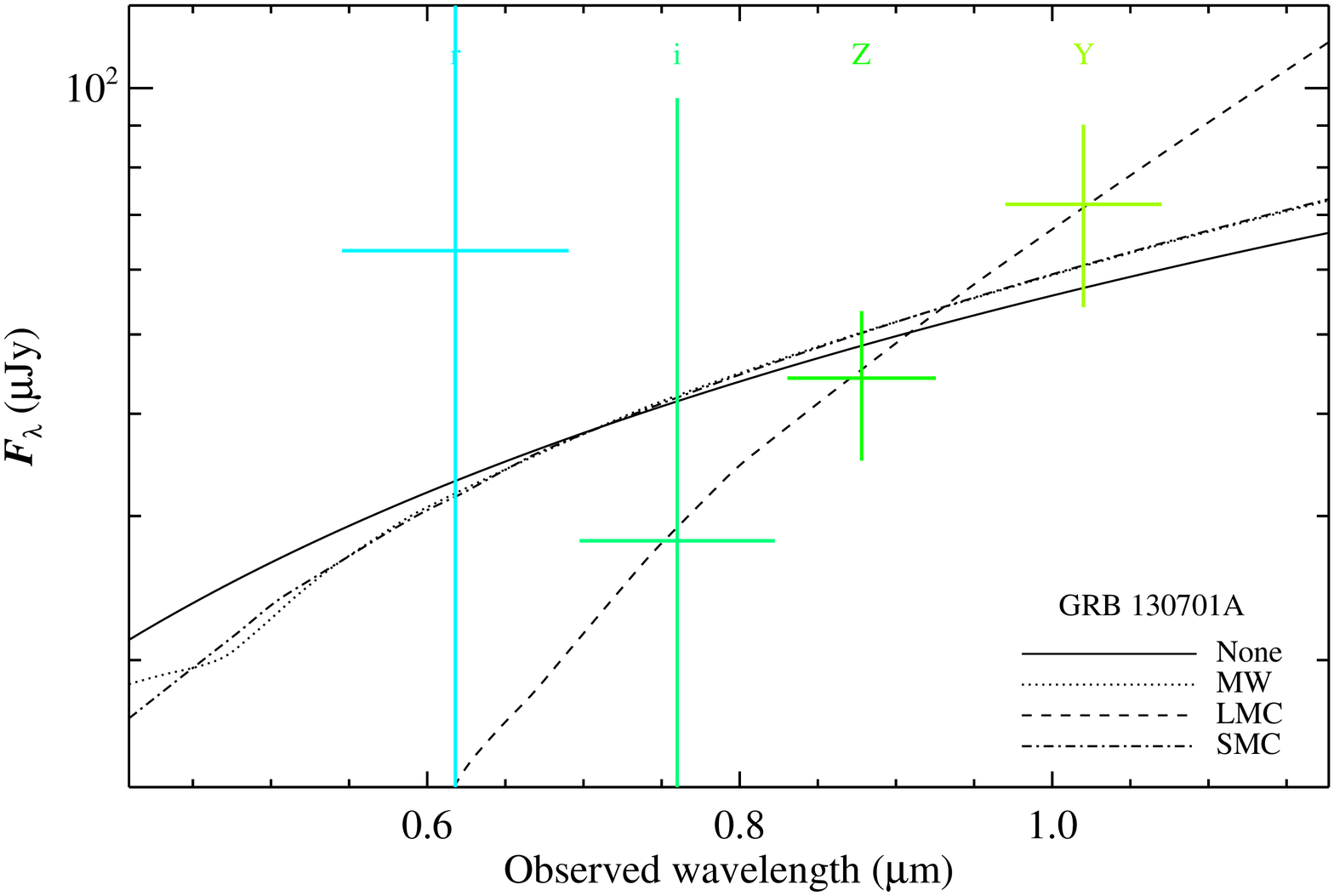}
    \\
  \end{center}
  \caption{Fitted   SED   templates   for  GRB~130215A,   GRB~130327A,
    GRB~130418A,  GRB~130420A,  GRB~130427A, GRB~130606A,  GRB~130610A
    and GRB~130701A.   The coloured points correspond  to the measured
    RATIR  photometry,  with  the   filter  being  marked  above  each
    measurement. The black lines indicate the best fits obtained using
    the extinction laws detailed in the key.}
  \label{fig:seds_fig1}
\end{figure*}

\begin{figure*}
  \begin{center}
    \includegraphics[width=7.5cm,clip,angle=0]{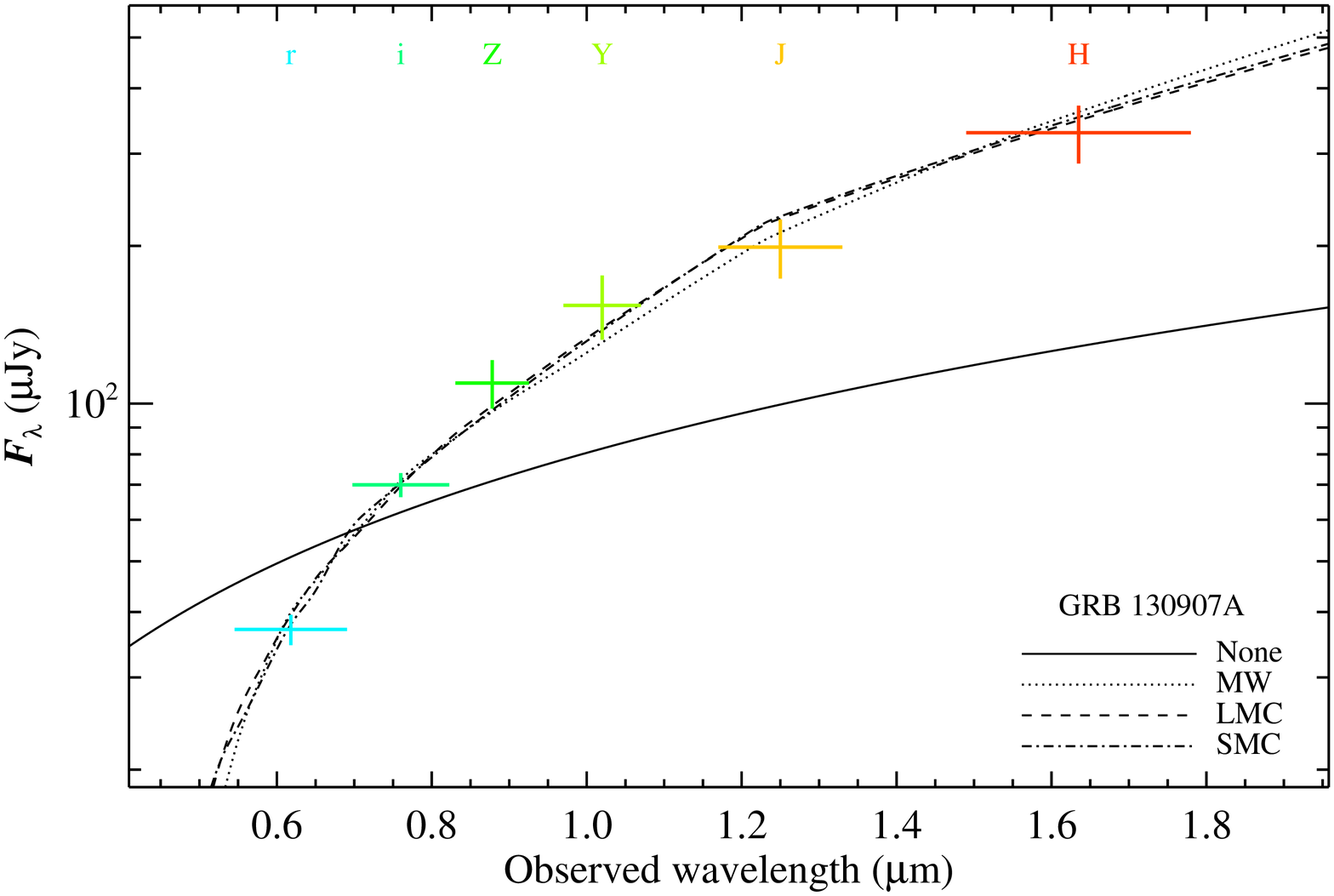}
    \quad
    \includegraphics[width=7.5cm,clip,angle=0]{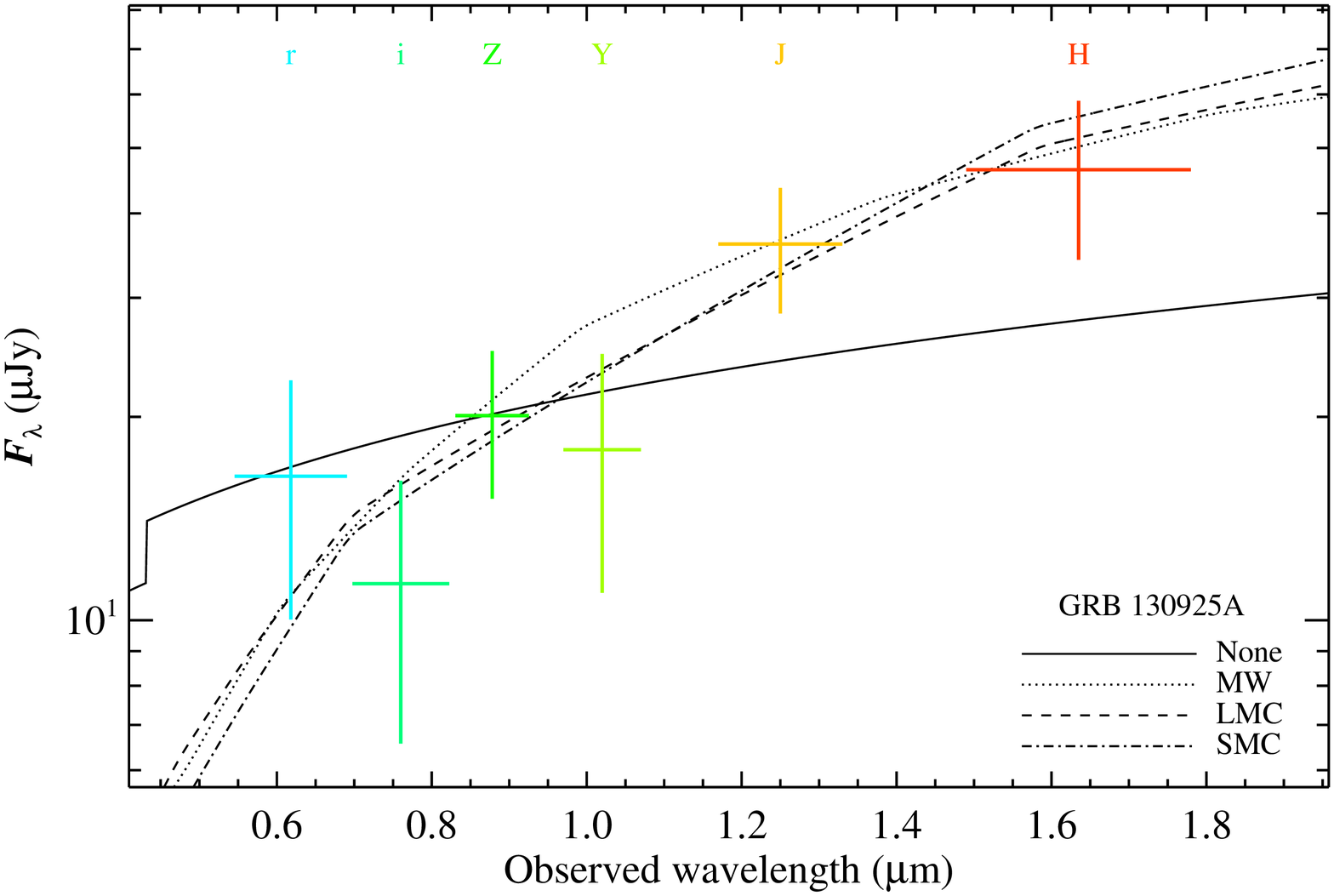}
    \\
    \includegraphics[width=7.5cm,clip,angle=0]{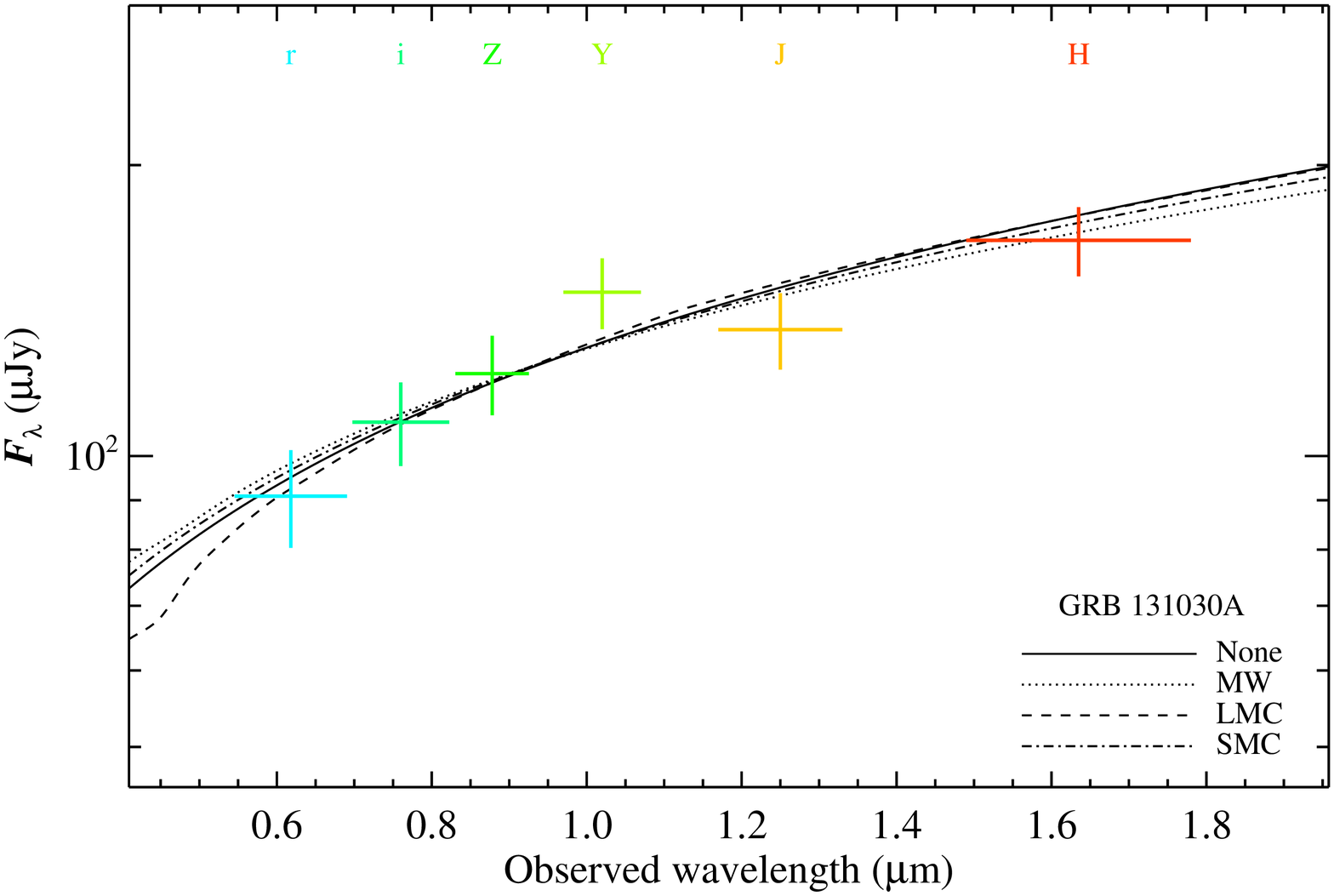}
    \quad
    \includegraphics[width=7.5cm,clip,angle=0]{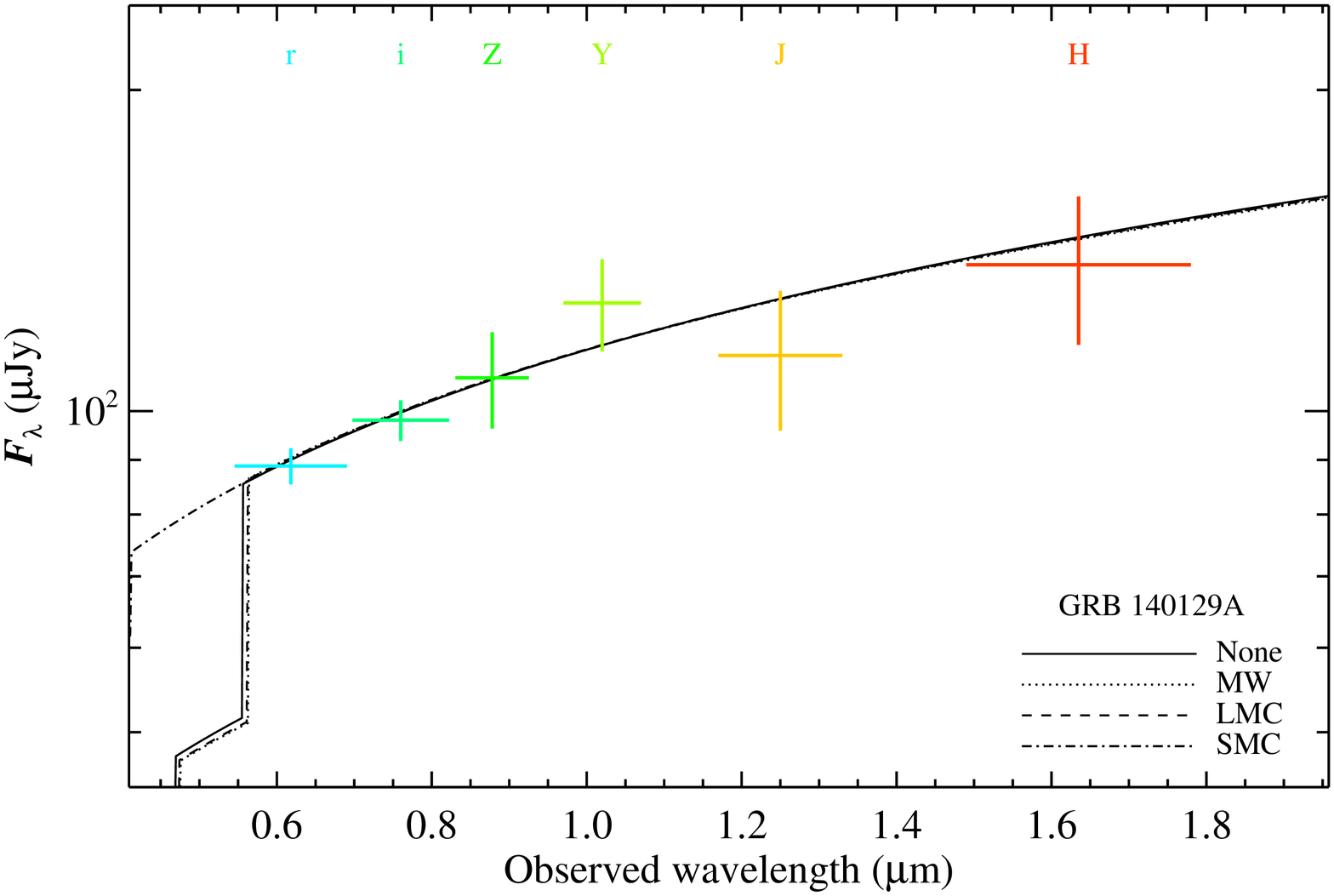}
    \\
    \includegraphics[width=7.5cm,clip,angle=0]{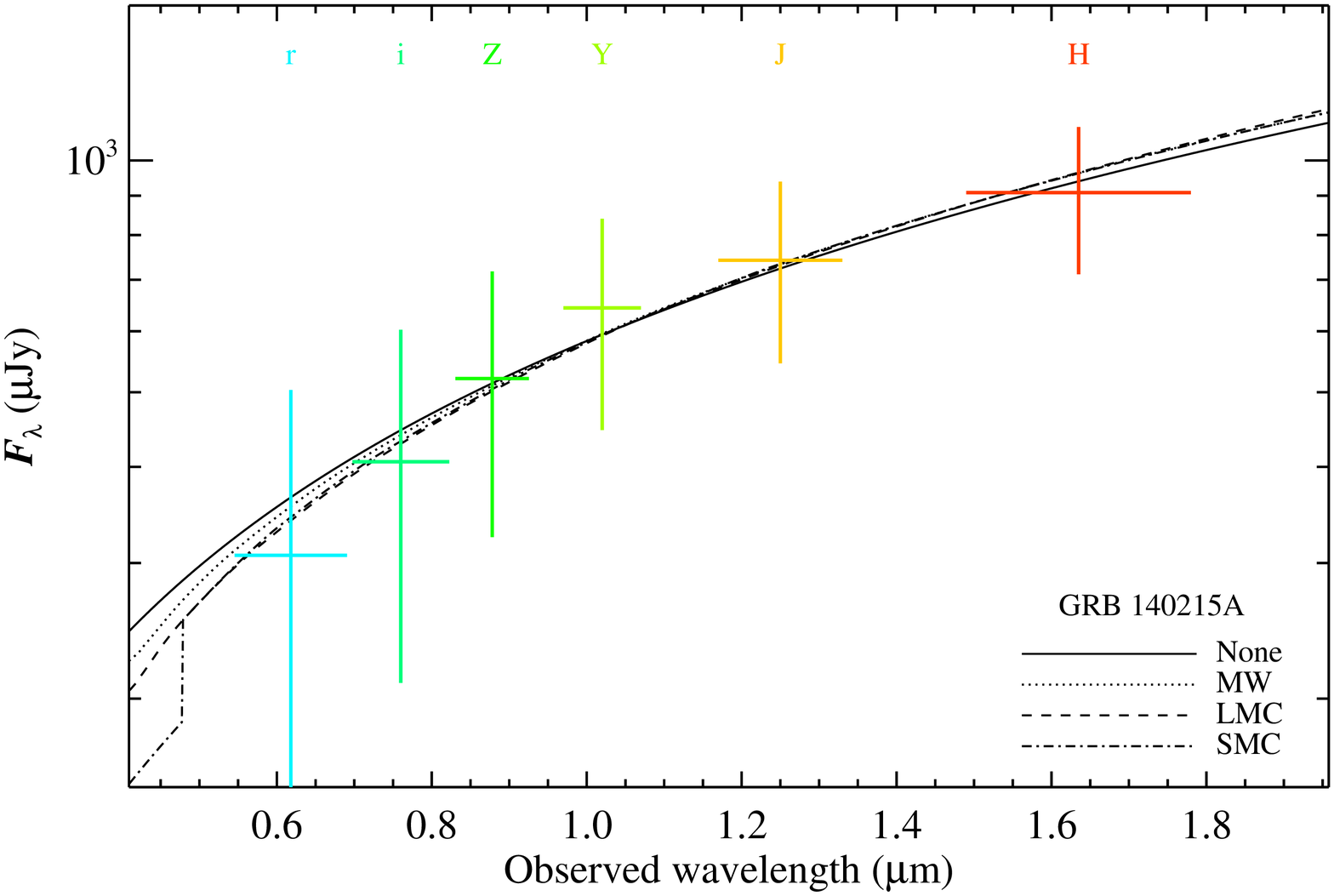}
    \quad
    \includegraphics[width=7.5cm,clip,angle=0]{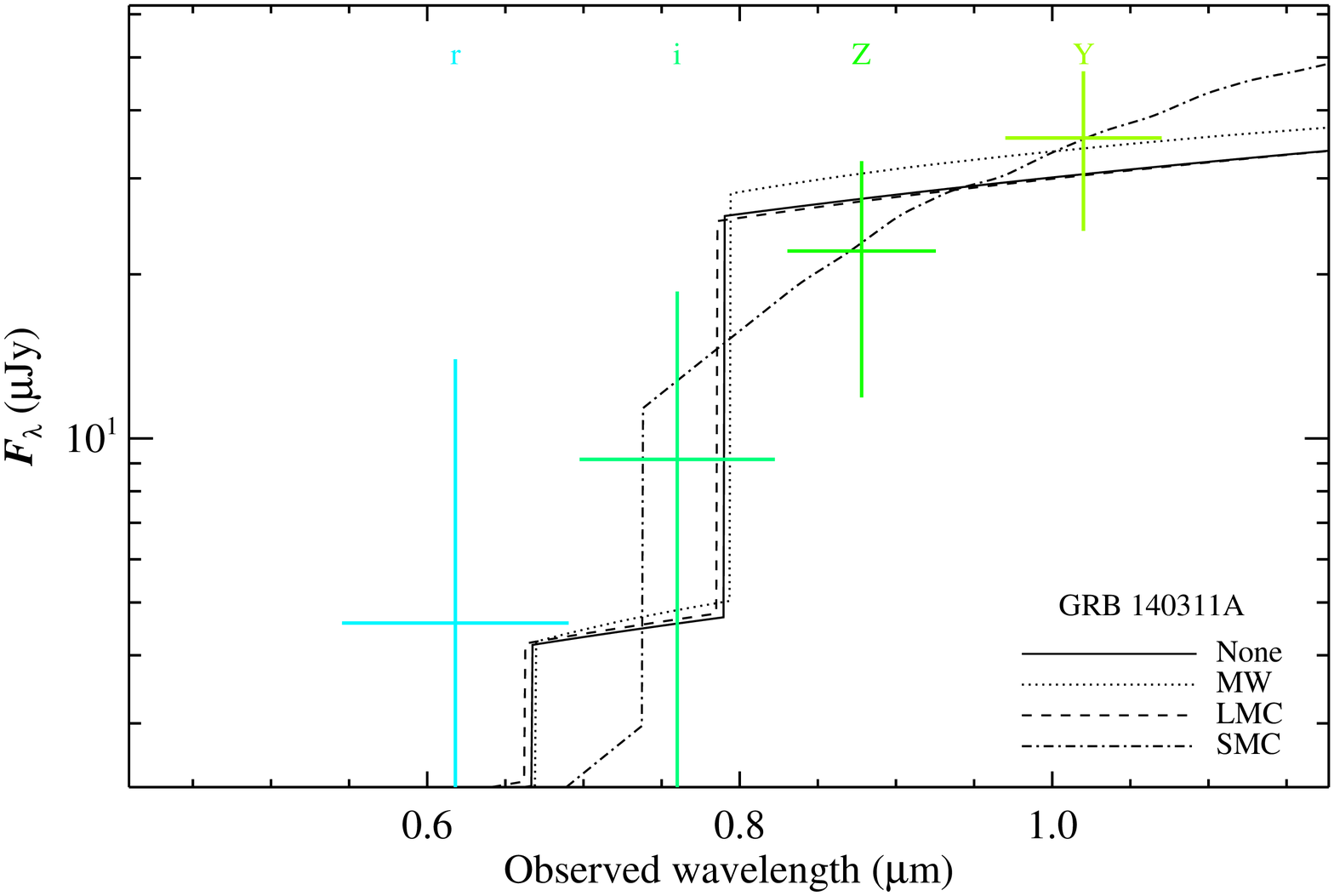}
    \\
    \includegraphics[width=7.5cm,clip,angle=0]{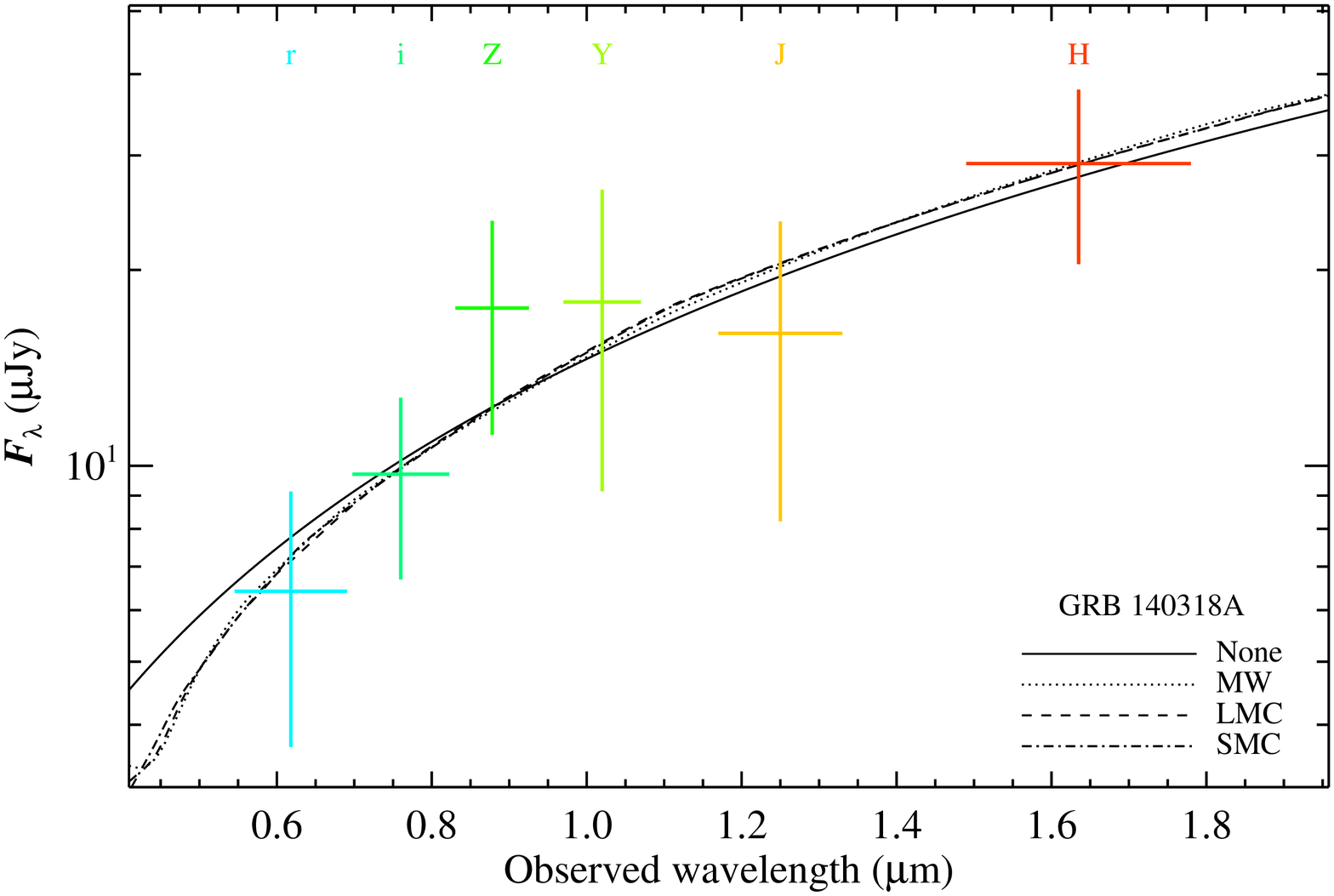}
    \quad
    \includegraphics[width=7.5cm,clip,angle=0]{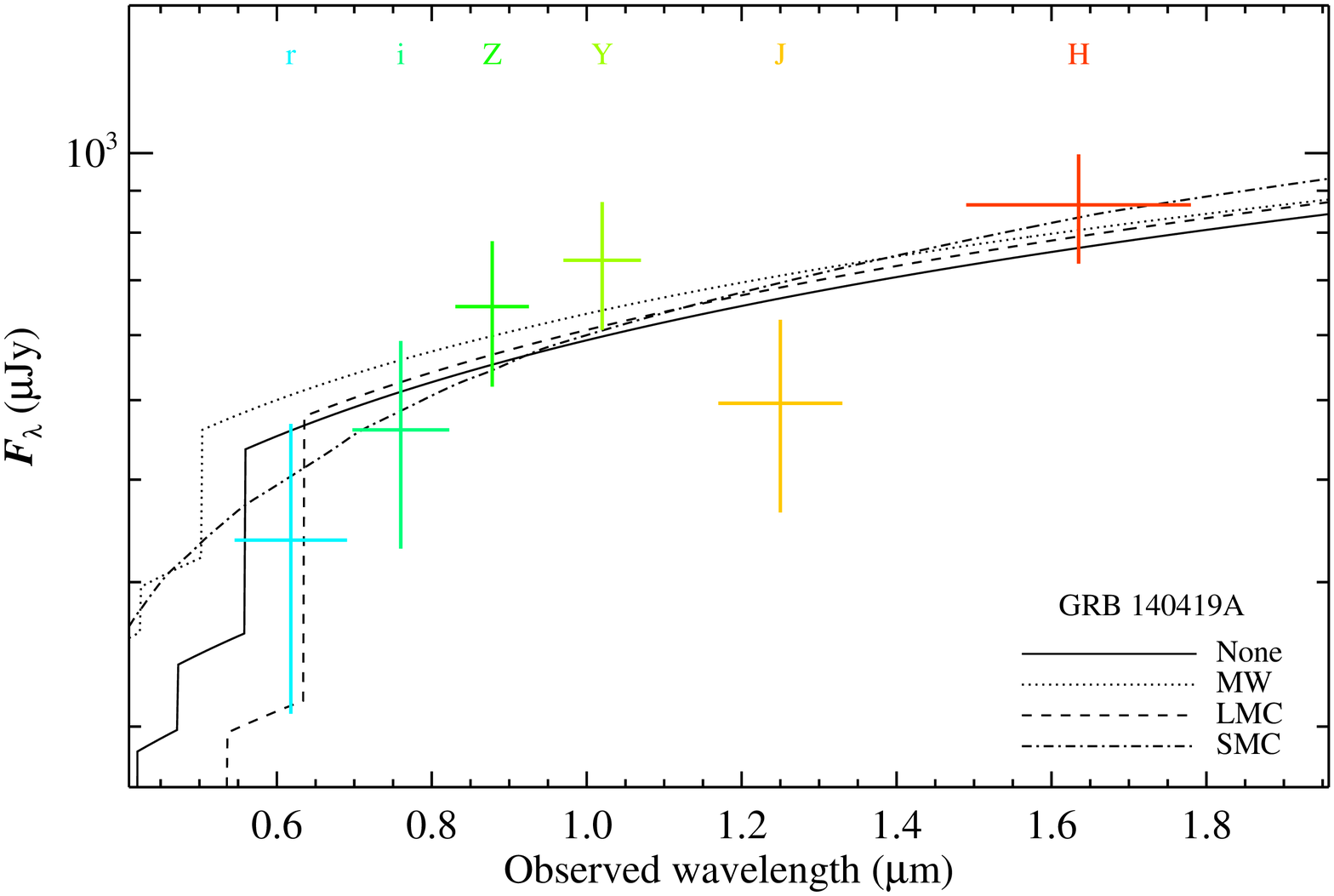}
    \\
  \end{center}
  \caption{Fitted   SED   templates   for  GRB~130907A,   GRB~130925A,
    GRB~131030A,  GRB~140129A, GRB~140215A,  GRB~1404311A, GRB~140318A
    and GRB~140419A.   The coloured points correspond  to the measured
    RATIR  photometry,  with  the   filter  being  marked  above  each
    measurement. The black lines indicate the best fits obtained using
    the extinction laws detailed in the key.}
  \label{fig:seds_fig2}
\end{figure*}

\begin{figure*}
  \begin{center}
    \includegraphics[width=7.5cm,clip,angle=0]{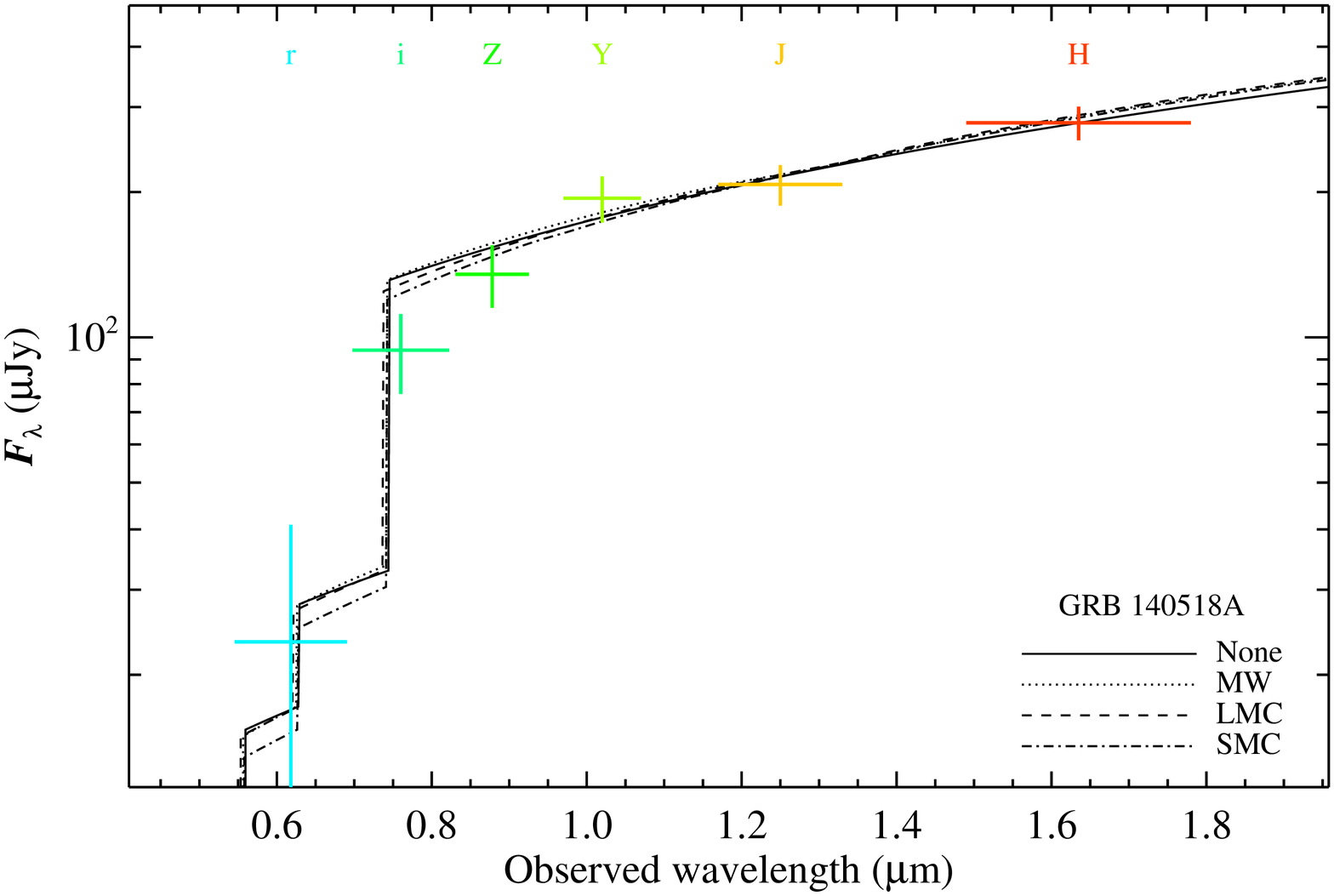}
    \quad
    \includegraphics[width=7.5cm,clip,angle=0]{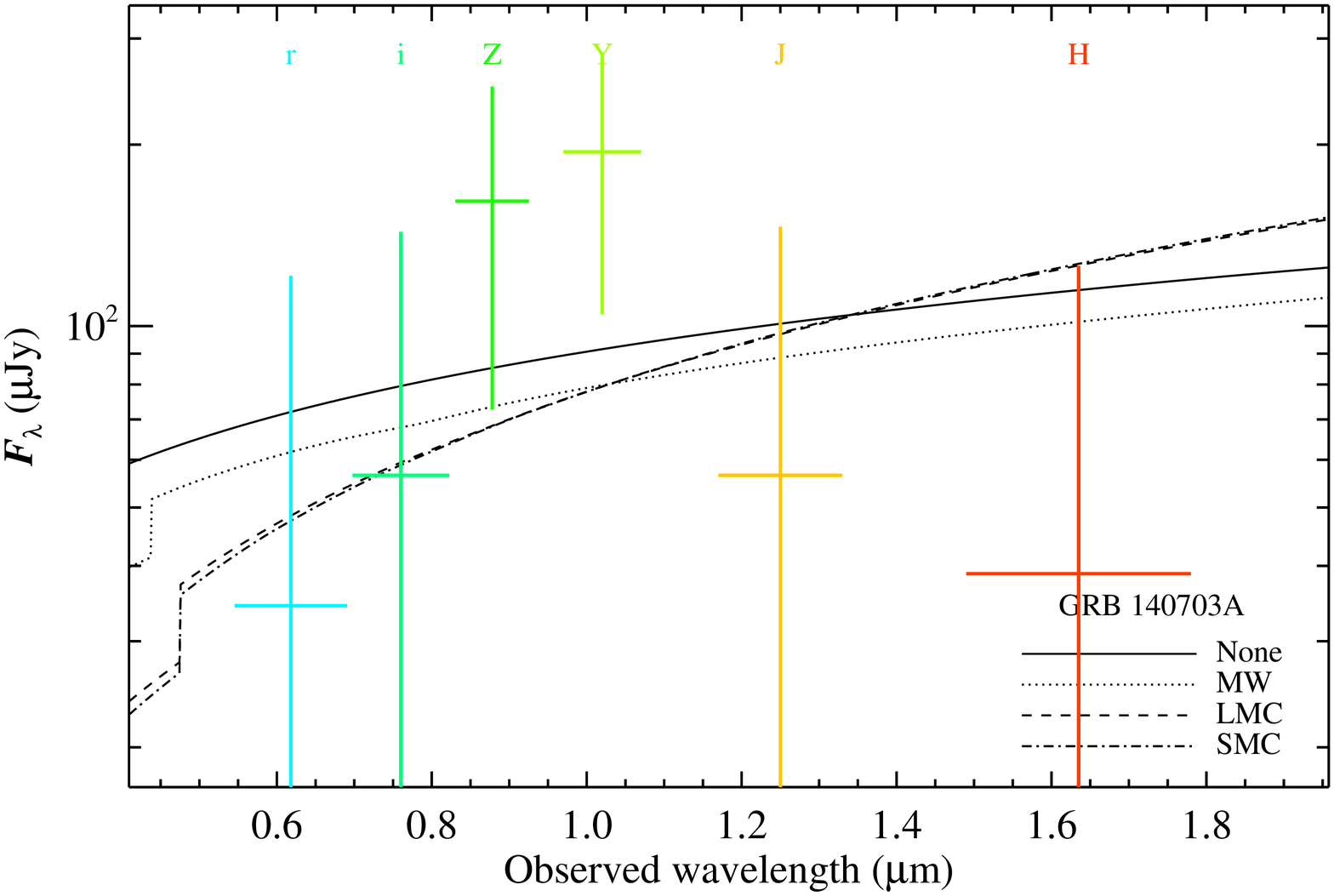}
    \\
    \includegraphics[width=7.5cm,clip,angle=0]{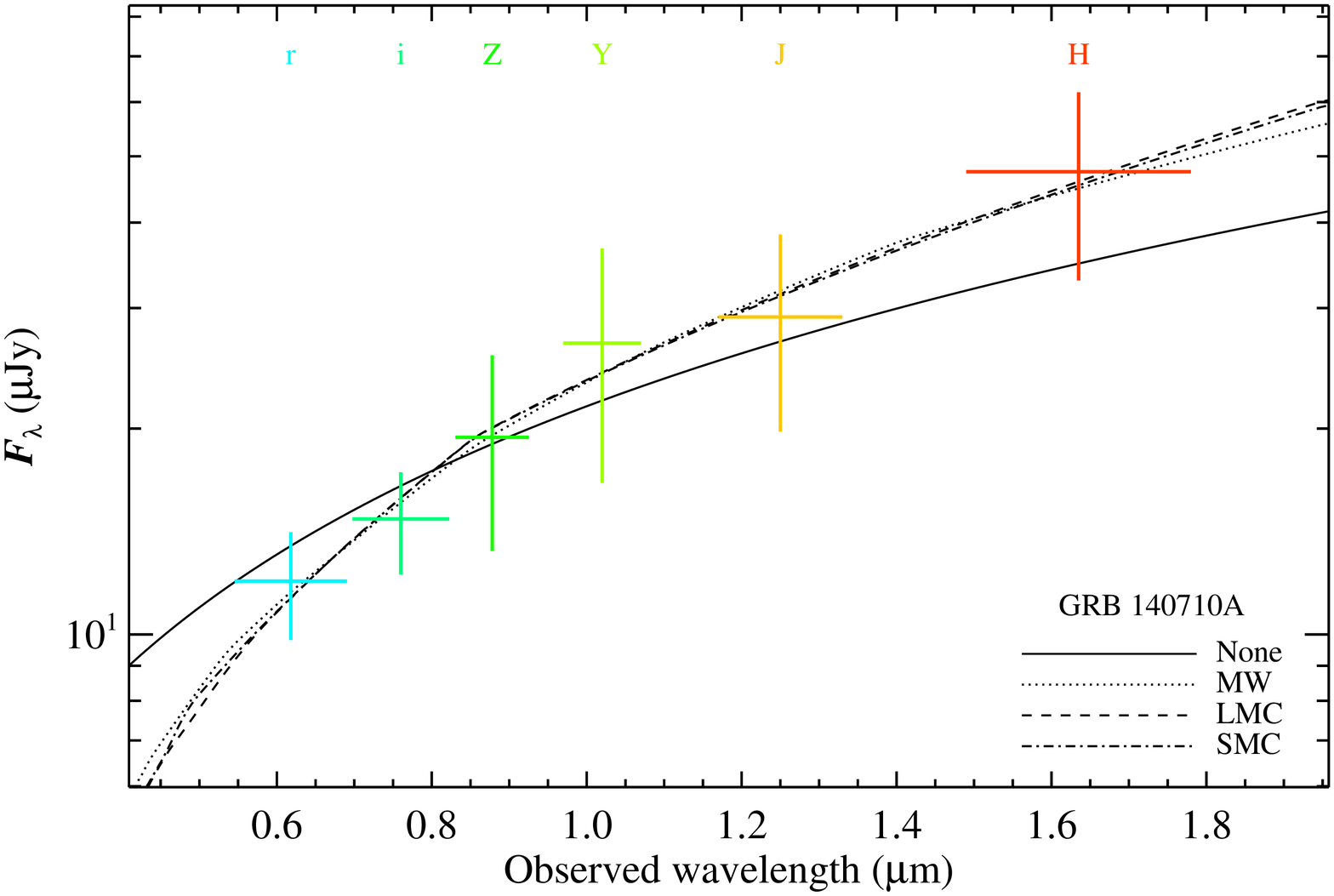}
    \\
  \end{center}
  \caption{Fitted  SED  templates  for  GRB~140518A,  GRB~140703A  and
    GRB~140710A.  The coloured points correspond to the measured RATIR
    photometry,   with   the    filter   being   marked   above   each
    measurement. The black lines indicate the best fits obtained using
    the extinction laws detailed in the key.}
  \label{fig:seds_fig3}
\end{figure*}

For  GRBs  with  a measured  redshift  we  fixed  $z$ in  our  fitting
algorithm to  the value  reported in Table  \ref{tab:fast_sample}. The
SEDs  with   the  resulting  fitted   models  are  shown   in  Figures
\ref{fig:seds_fig1}, \ref{fig:seds_fig2} and \ref{fig:seds_fig3}.  The
values of  $A_{\rm V}$ obtained from  each of the  three template dust
laws  are  reported  in   Table  \ref{tab:av_model},  along  with  the
$\chi^{2}$ fit statistic and  degrees of freedom associated with these
models.  In  each instance,  the preferred dust  model is  quoted.  We
report two measures of $\chi^{2}$  for each fitted model, the first is
the traditional  value, which  is a measure  of goodness of  fit.  The
second, $\chi_{\rm  eff}^{2}$, is the prior weighted  fit statistic as
described in \citet{2014AJ....148....2L}.  This value also includes an
additional  Bayesian prior  that compares  the local  optical  and NIR
spectral  index, $\beta_{\rm  opt}$,  to that  measured  in the  X-ray
regime, $\beta_{\rm  X}$.  Assuming a  synchrotron emission mechanism,
we expect the  optical SED to have either  the same intrinsic spectral
index as  measured with in  the X-ray spectrum,  or to have  a cooling
break between the two regimes.   If a cooling break is present between
the optical  and X-ray regimes, the  spectral index changes  by a well
defined  amount, dependent  on the  nature of  the  circumburst medium
\citep{2002ApJ...568..820G}.\par

In  the three  instances  where a  previous  spectroscopic measure  of
redshift  was  not   available  from  another  facility  (GRB~130327A,
GRB~140129A  and  GRB~140215A),  the  redshift  was  left  as  a  free
parameter  using the  \citet{2014AJ....148....2L}  algorithm. For  all
three  cases  we  were  only  able  to provide  upper  limits  on  the
photometric redshift, that $z_{\rm phot}\lesssim 4$.\par

\begin{table*}
  \centering
  \caption{Fitted dust models for GRBs with photometry in a minimum of
    four  RATIR bands,  modelled with  the \citet{2014AJ....148....2L}
    template fitting  algorithm. In each instance  the preferred model
    is stated.  Note  that in some instances the  preferred model does
    not have the lowest $\chi^{2}$,  however it was selected using the
    prior  weighted effective  $\chi_{\rm eff}^{2}$,  as  described in
    \citet{2014AJ....148....2L}. The  Bayesian prior in  this instance
    considers the  fitted value of optical spectral  index in relation
    to the  measured X-ray  spectral index. The  bottom row  shows the
    average value of  fitted host galaxy $A_{\rm V}$  across the whole
    sample for  each dust  extinction law. Also  shown are  the summed
    $\chi^{2}$  and $\chi_{\rm  eff}^{2}$  values to  show which  dust
    extinction law provides the best global fit to the entire sample.}
  \label{tab:av_model}
  \begin{tabular}{ccccccccccc}
    \hline \hline
    GRB & Preferred & $A_{\rm V, MW}$ & $\chi_{\rm MW}^{2}/\nu$ 
    & $\chi_{\rm eff, MW}^{2}/\nu$ &  $A_{\rm V, LMC}$ & $\chi_{\rm LMC}^{2}/\nu$ 
    & $\chi_{\rm eff, LMC}^{2}/\nu$ & $A_{\rm V, SMC}$ & $\chi_{\rm SMC}^{2}/\nu$ 
    & $\chi_{\rm eff,SMC}^{2}/\nu$ \\
     & Model & & & & & & & & & \\
    \hline
    130215A & LMC & 0.19$_{-0.04}^{+0.03}$ & 0.65/2 & $-$0.57/2 
    & 0.27$_{-0.04}^{+0.03}$ & 0.61/2 & $-$0.74/2 & 0.21$_{-0.03}^{+0.04}$ & 0.59/2 
    & $-$0.73/2 \\
    130327A & MW & 0.12$_{-0.43}^{+0.19}$ & 2.52/2 & 1.26/2 & 0.20$_{-0.81}^{+0.41}$ 
    & 2.54/2 & 2.20/2 & 0.01$_{-0.92}^{+0.89}$ & 2.78/2 & 1.45/2 \\
    130418A & LMC & 0.37$_{-0.05}^{+0.07}$ & 0.48/3 & $-$0.84/3 
    & 0.68$_{-0.11}^{+0.11}$ & 0.32/3 & $-$1.00/3 & 0.28$_{-0.04}^{+0.06}$ & 0.58/3 
    & $-$0.72/3 \\
    130420A & SMC & 0.00$_{-0.02}^{+0.02}$ & 4.74/3 & 4.89/3 & 0.00$_{-0.03}^{+0.02}$ 
    & 4.51/3 & 4.89/3 & 0.09$_{-0.02}^{+0.02}$ & 0.50/3 & 0.49/3 \\
    130427A & MW & 0.00$_{-0.04}^{+0.03}$ & 3.15/2 & 3.23/2 & 0.03$_{-0.10}^{+0.04}$ 
    & 3.40/2 & 3.40/2 & 0.07$_{-0.04}^{+0.05}$ & 4.15/2 & 4.16/2 \\
    130606A & LMC & 0.05$_{-0.01}^{+0.01}$ & 2.51/3 & 2.60/3 & 0.02$_{-0.01}^{+0.01}$ 
    & 2.19/3 & 2.22/3 & 0.03$_{-0.01}^{+0.01}$ & 2.25/3 & 2.28/3 \\
    130610A & SMC & 0.04$_{-0.18}^{+0.05}$ & 3.27/2 & 2.09/2 & 0.01$_{-0.07}^{+0.05}$ 
    & 3.07/2 & 1.84/2 & 0.01$_{-0.07}^{+0.05}$ & 3.05/3 & 1.80/2 \\
    130701A & LMC & 0.15$_{-0.10}^{+0.16}$ & 1.08/1 & 1.10/1 & 1.60$_{-0.15}^{+0.11}$ 
    & 0.54/1 & 0.50/1 & 0.15$_{-0.10}^{+0.16}$ & 1.08/1 & 1.08/1 \\
    130907A & SMC & 1.07$_{-0.00}^{+0.00}$ & 5.49/3 & 3.58/3 & 1.10$_{-0.03}^{+0.02}$ 
    & 3.10/3 & 3.15/3 & 1.10$_{-0.02}^{+0.02}$ & 3.00/3 & 3.00/3 \\
    130925A & MW & 1.46$_{-0.22}^{+0.23}$ & 2.10/3 & 2.10/3 & 1.32$_{-0.13}^{+0.26}$ 
    & 2.67/3 & 2.70/3 & 1.47$_{-0.16}^{+0.23}$ & 2.77/3 & 2.77/3 \\
    131030A & MW & 0.01$_{-0.05}^{+0.03}$ & 3.41/3 & 3.40/3 & 0.17$_{-0.04}^{+0.03}$ 
    & 3.86/3 & 3.79/3 & 0.00$_{-0.04}^{+0.04}$ & 3.50/3 & 3.62/3 \\
    140129A & LMC & 0.00$_{-0.04}^{+0.03}$ & 1.76/2 & 0.38/2 & 0.00$_{-0.01}^{+0.01}$ 
    & 1.76/2 & 0.38/2 & 0.00$_{-0.04}^{+0.04}$ & 1.77/2 & 0.39/2 \\
    140215A & SMC & 0.05$_{-0.25}^{+0.15}$ & 0.19/2 & 0.19/2 & 0.11$_{-0.36}^{+0.14}$ 
    & 0.18/2 & 0.20/2 & 0.04$_{-0.14}^{+0.07}$ & 0.17/2 & 0.18/2 \\
    140311A & SMC & 0.07$_{-0.06}^{+0.07}$ & 0.50/1 & 0.20/1 & 0.01$_{-0.09}^{+0.07}$ 
    & 0.56/1 & 0.06/1 & 0.45$_{-0.06}^{+0.08}$ & 0.17/1 & $-$0.33/1 \\
    140318A & LMC & 0.31$_{-0.14}^{+0.13}$ & 1.22/3 & $-$0.03/3 
    & 0.36$_{-0.13}^{+0.13}$ & 1.20/3 & $-$0.09/3 & 0.32$_{-0.12}^{+0.14}$ 
    & 1.20/3 & $-$0.06/3 \\
    140419A & LMC & 0.00$_{-0.09}^{+0.08}$ & 5.65/3 & 5.37/3 & 0.00$_{-0.03}^{+0.02}$ 
    & 4.12/3 & 3.03/3 & 0.11$_{-0.05}^{+0.06}$ & 4.42/3 & 3.33/3 \\
    140518A & MW & 0.03$_{-0.02}^{+0.02}$ & 1.51/3 & 1.52/3 & 0.04$_{-0.02}^{+0.02}$ 
    & 1.55/3 & 1.55/3 & 0.04$_{-0.02}^{+0.01}$ & 1.54/3 & 1.72/3 \\
    140703A & MW & 0.02$_{-0.23}^{+0.20}$ & 3.39/3 & 2.02/3 & 0.01$_{-0.23}^{+0.21}$ 
    & 3.95/3 & 4.01/3 & 0.02$_{-0.25}^{+0.21}$ & 3.97/3 & 4.04/3 \\
    140710A & MW & 0.49$_{-0.10}^{+0.08}$ & 0.47/3 & 0.24/3 & 0.60$_{-0.10}^{+0.09}$ 
    & 0.44/3 & 0.31/3 & 0.56$_{-0.09}^{+0.10}$ & 0.42/3 & 0.30/3 \\
    Average & SMC & 0.23$_{-0.04}^{+0.02}$ & 44.09/47 & 32.73/47 
    & 0.36$_{-0.05}^{+0.03}$ & 40.57/47 & 32.40/47 & 0.26$_{-0.05}^{+0.05}$ 
    & 37.91/47 & 28.77/47 \\
    \hline
  \end{tabular}
\end{table*}

Of    the    13    bursts     identified    as    dark    using    the
\citet{2009ApJ...699.1087V} criterion,  we were able to  model the SED
of  seven.    This  yielded  two  GRBs  of   high  redshift  ($z>3.5$;
GRB~130606A  \&  GRB~140518A),  two  with  a  high  quantity  of  dust
extinction ($A_{\rm  V}>1$; GRB~130907A \&  GRB~130925A) and 2  with a
moderate quantity of  dust extinction ($0.25<A_{\rm V}<1$; GRB~140318A
\& GRB~140710A).   Curiously, whilst dark, GRB~130420A has  both a low
redshift ($z  = 1.297$) and a  low amount of  modelled dust extinction
($A_{\rm      V}      =      0.09\pm0.02$).\par

We  considered  two  alternative  scenarios to  explain  the  modelled
optical  attenuation  of  GRB   130420A.   First,  we  looked  at  the
\citet{2009ApJ...699.1087V} condition  of optical darkness,  for which
GRB~130420A has $\beta_{\rm  OX}-\beta_{\rm X}=-0.63\pm0.13$.  A value
of  $\beta_{\rm OX}-\beta_{\rm X}=-0.5$  would indicate  a GRB  with a
cooling  break  immediately below  the  measured  X-ray spectrum,  and
therefore consistent  with an  intrinsic synchrotron spectrum  with no
optical  attenuation. Such a  scenario is  at the  upper limit  of the
error  bounds of  $\beta_{\rm OX}-\beta_{\rm  X}$ for  GRB~130420A. An
alternative explanation  can be  considered by looking  at the  SED of
GRB~130420A shown in  Figure~\ref{fig:seds_fig1}. This reveals a large
error in the \textit{H} band flux measurement.  It is this filter that
is least well represented by the MW and LMC dust profile templates, as
the  large   error  weights  the   fitted  templates  away   from  the
\textit{H} band.  It  is possible that  this value is  accurate, while
not precise.  As such if the errors were smaller the template would be
constrained into a shallower local spectral index, and therefore would
require a larger  value of $A_{\rm V}$ to produce  the lower fluxes in
the \textit{r}, \textit{i} and \textit{Z} bands.\par

Six  optically dark  GRBs do  not have  sufficient photometry  for SED
modelling   (GRB  130502A,   GRB~130514A,   GRB~130609A,  GRB~140114A,
GRB~140331A \&  GRB~140710A).  GRB~130514A has  a photometric redshift
from  the  GROND instrument  of  $z  \sim  3.6$, suggesting  this  may
contribute  to  the  under-luminous  nature  of  the  \textit{r} band,
however RATIR  did not  detect the GRB  in any  of the six  filters in
which  it  was  observed.   GRB~140331A  has a  measured  redshift  of
$z=1.09$, ruling it  out as a high-redshift event.  The remaining four
GRBs do not have a measured redshift.\par

GRB~140311A  has  a   measured  spectroscopic  redshift  of  $z=4.954$
\citep{2014GCN..16301...1C},  at  which it  is  expected that  optical
attenuation should  be observed in the  \textit{r} band. Despite this,
GRB~140311A  is  not  classified  as  optical dark  using  either  the
\citet{2009ApJ...699.1087V}       or       \citet{2004ApJ...617L..21J}
criteria.  The  measured  X-ray   spectral  index  for  this  GRB  was
$\beta_{\rm X}  = 0.75_{-0.15}^{+0.21}$,  while the best  fit template
obtained  from  the  \citet{2014AJ....148....2L} SED  fitting  routine
found   the  local   optical  and   NIR  spectral   index  $\beta_{\rm
  opt}\approx0.68$. As such it is  possible that the optical and X-ray
regimes both  lie on the same  power-law segment of  the intrinsic GRB
synchrotron spectrum. If  this is the case, then  the attenuation from
high redshift  does not reduce the  optical flux to a  level below the
minimum allowable flux resulting from  the presence of a cooling break
between the two regimes.  Unfortunately, GRB~140311A was only observed
in the four bluest RATIR  filters, thus giving only a loose constraint
on $\beta_{\rm opt}$.\par

\begin{figure}
  \begin{center}
    \includegraphics[width=8.5cm,angle=0,clip]{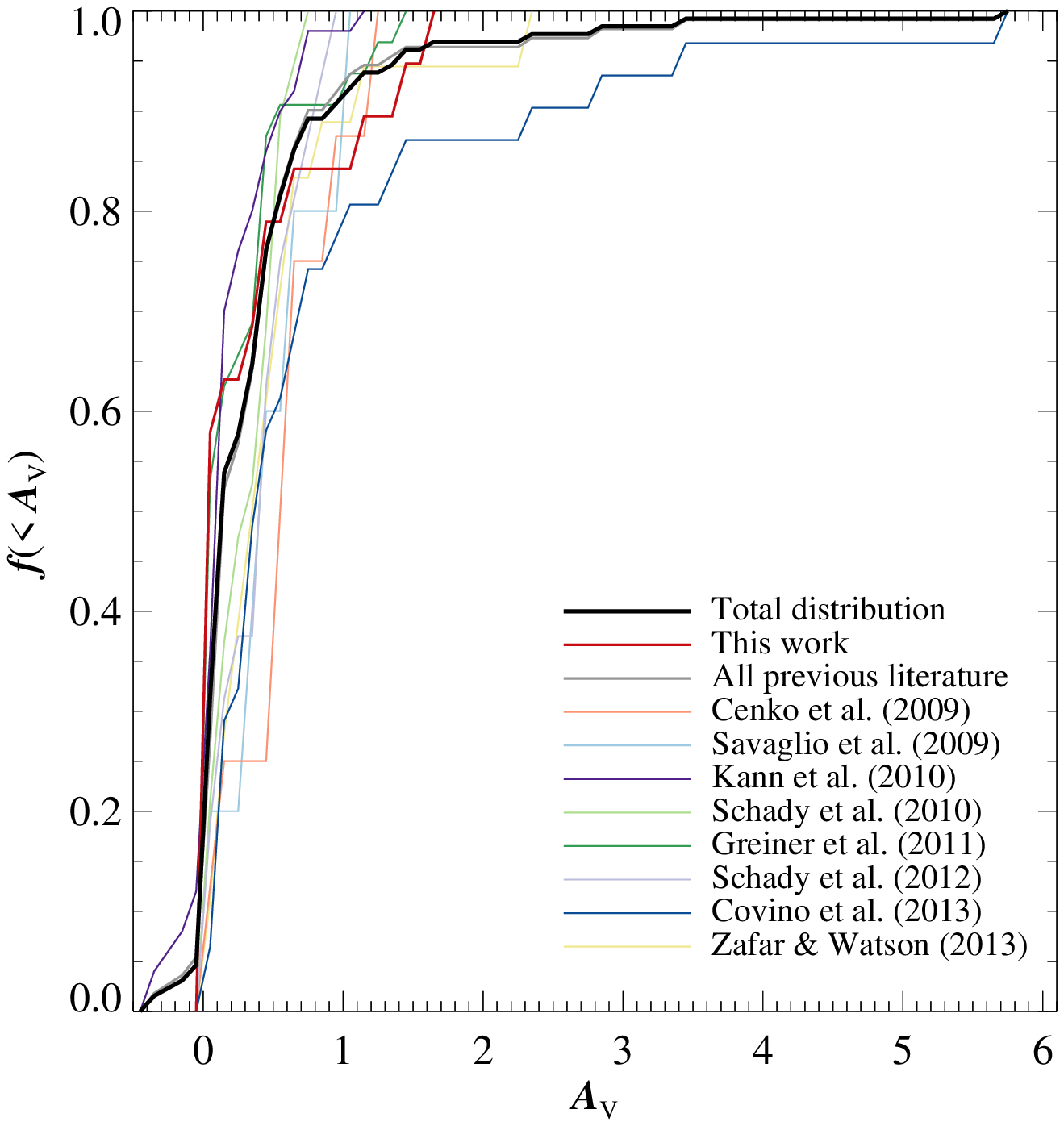}
    \caption{Normalised   cumulative  distributions  of   host  galaxy
      $A_{\rm  V}$  for  this  sample  and those  available  from  the
      literature.  As  well as  distributions  of individual  samples,
      those of the  sum of all previous literature,  and sum including
      our sample, are plotted.  All  samples are denoted in the key in
      the bottom right of the panel.}
    \label{fig:av_hist}
  \end{center}
\end{figure}

The normalised cumulative distribution  of fitted values of $A_{\rm V}$
for  the 19  GRBs  in  Table \ref{tab:av_model}  are  shown in  Figure
\ref{fig:av_hist},  along  with those  of  an  extensive selection  of
previous  samples  from the  literature.  We  also  display the  total
distribution of  all previous literature,  with and without  the RATIR
sample.  Care has been  taken to  ensure GRBs  that occur  in multiple
samples are  only included once  in the total literature  sample. From
Figure \ref{fig:av_hist}  it can be seen  that our sample,  denoted by the
red line,  appears consistent with  most of the previous  samples. The
one  distribution  that  appears   discrepant  is  that  presented  in
\citet{2013MNRAS.432.1231C}, which is the BAT6 sample.\par

\citet{2011A&A...526A..30G} found  25 per cent of  bursts with $A_{\rm
  V}  \sim 0.5$ (8/33).   In our  sample, this  fraction is  lower, at
11$\pm$7     per    cent     (2/19),    which     is     similar    to
\citet{2010ApJ...720.1513K} who  found approximately 12  per cent.  It
is important to note, however,  that the fundamental properties of our
GRB sample  are more similar to  those of \citet{2011A&A...526A..30G},
as  the  sample   of  \citet{2010ApJ...720.1513K}  is  biased  towards
optically      brighter      bursts.       In      another      study,
\citet{2013MNRAS.432.1231C} find  in a sample  of 53 GRBs that  50 per
cent  have an  extinction of  $A_{\rm  V} \lesssim  0.3$, whereas  our
sample  has  a  marginally  higher  fraction  of  63$\pm$11  per  cent
(12/19).\par

Those GRBs with values of  $A_{\rm V}$ inconsistent with zero are best
fitted by a variety  of dust laws. 27 per cent (3/11)  favour a MW dust
extinction law, with  45 per cent (5/11) favouring an  LMC dust law and
27 per  cent (3/11)  being best  fit by an  SMC dust  law.  Conversely,
\citet{2010MNRAS.401.2773S} find  that an  SMC extinction curve  to be
preferred in 56  per cent of their sample. In the  bottom row of Table
\ref{tab:av_model} we calculate the average required amount of dust if
the same  model was to be  assumed for the entire  population. We also
sum  the  $\chi^{2}$ for  all  these  models  to compare  which  model
provides the best  global fit across the entire  population should the
fitting  be  limited  to one  dust  extinction  law.  Due to  the  few
instances where  an SMC  dust extinction model  has a  markedly better
$\chi^{2}$, while  in cases where it  is not the  preferred model host
galaxy $A_{\rm  V}$ tends to be  low and so $\Delta  \chi^{2}$ is much
lower, an SMC  dust extinction law proves to be  the best when fitting
to the  entire sample.  This is  more consistent with  the findings of
\citet{2010MNRAS.401.2773S}.\par

From the  distribution of  best fitted $A_{\rm  V}$ values  plotted in
\ref{fig:av_hist}, we  find 16$\pm$8  per cent (3/19)  of GRBs  with a
fitted SED  have host  dust extinction $A_{\rm  V} > 1$,  which agrees
with \citet{2011A&A...526A..30G}.  25 per  cent (2/8) of the GRBs with
$A_{\rm  V}> 0.25$ favour  a MW  dust profile,  which has  a prominent
feature  at 2175~\AA.   \citet{2011A&A...526A..30G} find  a suggestion
that  GRBs with  a  larger dust  content  may favour  a  MW type  dust
profile. We  instead find that  50 per cent  (4/8) favour an  LMC type
dust extinction  law. It must be  noted that our sample  lacks a large
number of  high extinction  GRBs, and this  discrepancy may be  due to
small number statistics.\par

To compare  the distributions  of obtained host  $A_{\rm V}$,  we once
more employed  K-S tests between our  sample and an  extensive list of
previous      samples.     These      are      shown     in      Table
\ref{tab:ks_test_results}. Comparing  our distribution of  host galaxy
$A_{\rm  V}$  to   these  studies  shows  only  one   result  that  is
statistically significantly  different. Comparing our work  to that of
\citet{2013MNRAS.432.1231C} we find the two samples to be different at
a level of approximately  3$\sigma$. For completeness we also compared
the    results   of    \citet{2013MNRAS.432.1231C}    to   those    in
\citet{2010ApJ...720.1513K}  and \citet{2011A&A...526A..30G},  as well
as a composite  sample of all other existing  literature , finding the
BAT6 distribution  of host galaxy $A_{\rm V}$  values to significantly
differ from all three.\par

Looking  at  Figure \ref{fig:av_hist}  the  main  cause of  difference
between these samples arises from  a handful of extremely high $A_{\rm
  V}$ values in the BAT6 sample. The highest of these, for GRB~070306,
is  $A_{\rm  V}=5.74_{-1.45}^{+1.48}$.   As  shown  in  Table  A10  of
\citet{2013MNRAS.432.1231C},  however,  the fit  used  to derive  this
value  of dust  extinction has  zero degrees  of freedom,  which could
perhaps  lead to  the  very high  value  obtained. It  must be  noted,
however, that there are also several instances where $A_{\rm V}$ could
not be  constrained in the BAT6  sample, due to a  lack of photometric
detections in  a sufficient number  of optical and NIR  bands, however
lower  limits could  be derived  that are  indicative of  high $A_{\rm
  V}$. In  these cases,  as the  redshift is also  known for  the GRB,
high-redshift is precluded, thus  requiring higher dust content in the
host galaxy.   As the BAT6  sample membership is defined  by gamma-ray
fluence, rather than optical or NIR brightness, it is perhaps expected
that  it  might contain  a  larger  number  of highly  dust  extincted
GRBs.\par

\begin{table*}
  \centering
  \caption{K-S test results from  comparisons of $z$, $\beta_{\rm X}$,
    $\beta_{\rm OX}$,  $A_{\rm V}$ and  $N_{\rm H,rest}$ distributions
    between this work and values obtained from the literature. In each
    comparison, only values with  full detections of the parameter are
    considered, thus excluding upper  and lower limits. When comparing
    parameters from the  BAT6 samples to this work  and other previous
    literature,  $A_{\rm  V}$ and  $N_{\rm  H,rest}$  were taken  from
    \citet{2013MNRAS.432.1231C},   while  $z$,  $\beta_{\rm   X}$  and
    $\beta_{\rm  OX}$  were  taken  from  \citet{2012MNRAS.421.1265M}.
    Comparisons using  the BAT6 sample and previous  literature do not
    include this  RATIR sample.  $N_{\rm  1}$ and $N_{\rm 2}$  are the
    sizes of the samples in the first and second column, respectively.
    $D_{\rm KS}$  is the  Kolmogorov-Smirnov statistic found  for each
    test and $p_{\rm KS}$ is  the probability that both samples derive
    from a common parent population.}
  \label{tab:ks_test_results}
  \begin{tabular}{ccccccc}
    \hline
    \hline
    Sample 1 & Sample 2 & Parameter & $N_{\rm 1}$ & $N_{\rm 2}$ & $D_{\rm KS}$ 
    & $p_{\rm KS}$ \\
    \hline
    This work & Literature & $z$ & 16 & 156 & 0.20 & 6.22$\times10^{-1}$ \\
    This work & Fynbo et al. (2009) & $z$ & 16 & 85 & 0.28 
    & 2.02$\times10^{-1}$ \\
    This work & BAT6 & $z$ & 16 & 52 & 0.32 
    & 1.37$\times10^{-1}$ \\
    This work & Kann et al. (2010) & $z$ & 16 & 46 & 0.26
    & 3.54$\times10^{-1}$ \\
    This work & Greiner et al. (2011) & $z$ & 16 & 33 & 0.23 
    & 5.65$\times10^{-1}$ \\
    This work & Schady et al. (2010) & $z$ & 16 & 26 & 0.29 
    & 3.05$\times10^{-1}$ \\
    This work & Zafar \& Watson (2013) & $z$ & 16 & 25 & 0.52 
    & 5.65$\times10^{-3}$ \\
    This work & Cenko et al. (2009) & $z$ & 16 & 16 & 0.25 
    & 3.05$\times10^{-1}$ \\
    This work & Schady et al. (2012) & $z$ & 16 & 16 & 0.38 
    & 1.62$\times10^{-1}$ \\
    This work & Savaglio et al. (2009) & $z$ & 16 & 10 & 0.88 
    & 4.45$\times10^{-5}$ \\
    BAT6 & Literature & $z$ & 52 & 116 & 0.20 
    & 9.24$\times10^{-2}$ \\
    \hline
    This work & Literature & $\beta_{\rm X}$ & 28 & 228 & 0.11 
    & 9.32$\times10^{-1}$ \\
    This work & Fynbo et al. (2009) & $\beta_{\rm X}$ & 28 & 137 & 0.12 
    & 8.60$\times10^{-1}$ \\
    This work & BAT6 & $\beta_{\rm X}$ & 28 & 44 & 0.12 & 9.54$\times10^{-1}$ \\
    This work & van der Horst et al. (2009) & $\beta_{\rm X}$ & 28 & 40 & 0.16 
    & 7.51$\times10^{-1}$ \\
    This work & Jakobsson et al. (2004) & $\beta_{\rm X}$ & 28 & 37 & 0.32 
    & 6.44$\times10^{-2}$ \\
    This work & Greiner et al. (2011) & $\beta_{\rm X}$ & 28 & 34 & 0.15 
    & 8.55$\times10^{-1}$ \\
    This work & Cenko et al. (2009) & $\beta_{\rm X}$ & 28 & 28 & 0.14 
    & 9.17$\times10^{-1}$ \\
    This work & Melandri et al. (2008) & $\beta_{\rm X}$ & 28 & 22 & 0.16 
    & 8.87$\times10^{-1}$ \\
    This work & Galama \& Wijers (2001) & $\beta_{\rm X}$ & 28 & 5 & 0.36 
    & 5.48$\times10^{-1}$ \\
    BAT6 & Literature & $\beta_{\rm X}$ & 44 & 184 & 0.14 & 4.99$\times10^{-1}$ \\
    \hline
    This work & Literature & $\beta_{\rm OX}$ & 21 & 167 & 0.27 
    & 1.05$\times10^{-1}$ \\
    This work & Fynbo et al. (2009) & $\beta_{\rm OX}$ & 21 & 86 & 0.30 
    & 7.95$\times10^{-2}$ \\
    This work & van der Horst et al. (2009) & $\beta_{\rm OX}$ & 21 & 36 & 0.43 
    & 1.12$\times10^{-2}$\\
    This work & BAT6 & $\beta_{\rm OX}$ & 21 & 35 & 0.26 & 3.03$\times10^{-1}$ \\
    This work & Greiner et al. (2011) & $\beta_{\rm OX}$ & 21 & 34 & 0.30
    & 1.48$\times10^{-1}$ \\
    This work & Jakobsson et al. (2004) & $\beta_{\rm OX}$ & 21 & 25 & 0.36 
    & 8.46$\times10^{-2}$ \\
    This work & Cenko et al. (2009) & $\beta_{\rm OX}$ & 21 & 21 & 0.24 
    & 5.31$\times10^{-1}$ \\
    This work & Melandri et al. (2008) & $\beta_{\rm OX}$ & 21 & 9 & 0.37 
    & 2.96$\times10^{-1}$ \\
    BAT6 & Literature & $\beta_{\rm OX}$ & 35 & 134 & 0.12 
    & 8.07$\times10^{-1}$ \\
    \hline
    This work & Literature & $A_{\rm V}$ & 19 & 111 & 0.35 
    & 3.07$\times10^{-2}$ \\
    This work & Kann et al. (2010) & $A_{\rm V}$ & 19 & 50 & 0.29
    & 1.74$\times10^{-1}$ \\
    This work & Greiner et al. (2011) & $A_{\rm V}$ & 19 & 32 & 0.15
    & 9.27$\times10^{-1}$ \\
    This work & BAT6 & $A_{\rm V}$ & 19 & 31 & 0.54 
    & 1.29$\times10^{-3}$ \\
    This work & Schady et al. (2010) & $A_{\rm V}$ & 19 & 19 & 0.47
    & 1.81$\times10^{-2}$ \\
    This work & Zafar \& Watson (2013) & $A_{\rm V}$ & 19 & 18 & 0.53
    & 6.96$\times10^{-3}$ \\
    This work & Schady et al. (2012) & $A_{\rm V}$ & 19 & 16 & 0.46
    & 3.16$\times10^{-2}$ \\
    This work & Cenko et al. (2009) & $A_{\rm V}$ & 19 & 8 & 0.58
    & 2.66$\times10^{-2}$ \\
    This work & Savaglio et al. (2009) & $A_{\rm V}$ & 19 & 5 & 0.48
    & 2.22$\times10^{-1}$ \\
    BAT6 & Literature & $A_{\rm V}$ & 31 & 87 & 0.42
    & 3.79$\times10^{-4}$ \\
    BAT6 & Kann et al. (2010) & $A_{\rm V}$ & 31 & 50 & 0.52
    & 2.78$\times10^{-5}$ \\
    BAT6 & Greiner et al. (2011) & $A_{\rm V}$ & 31 & 32 & 0.53
    & 1.57$\times10^{-4}$ \\
    \hline
    This work & Literature & $N_{\rm H,rest}$ & 13 & 95 & 0.26 
    & 3.68$\times10^{-1}$ \\
    This work & BAT6 & $N_{\rm H,rest}$ & 13 & 52 & 0.25 & 4.73$\times10^{-1}$ \\
    This work & Greiner et al. (2011) & $N_{\rm H,rest}$ & 13 & 32 & 0.30
    & 3.15$\times10^{-1}$ \\
    This work & Schady et al. (2010) & $N_{\rm H,rest}$ & 13 & 26 & 0.15 
    & 9.77$\times10^{-1}$ \\
    This work & Schady et al. (2012) & $N_{\rm H,rest}$ & 13 & 16 & 0.26
    & 6.72$\times10^{-1}$ \\
    BAT6 & Literature & $N_{\rm H,rest}$ & 52 & 43 & 0.14 
    & 7.31$\times10^{-1}$ \\
    \hline
  \end{tabular}
\end{table*}

\begin{figure*}
  \begin{center}
    \includegraphics[width=7.5cm,clip,angle=0]{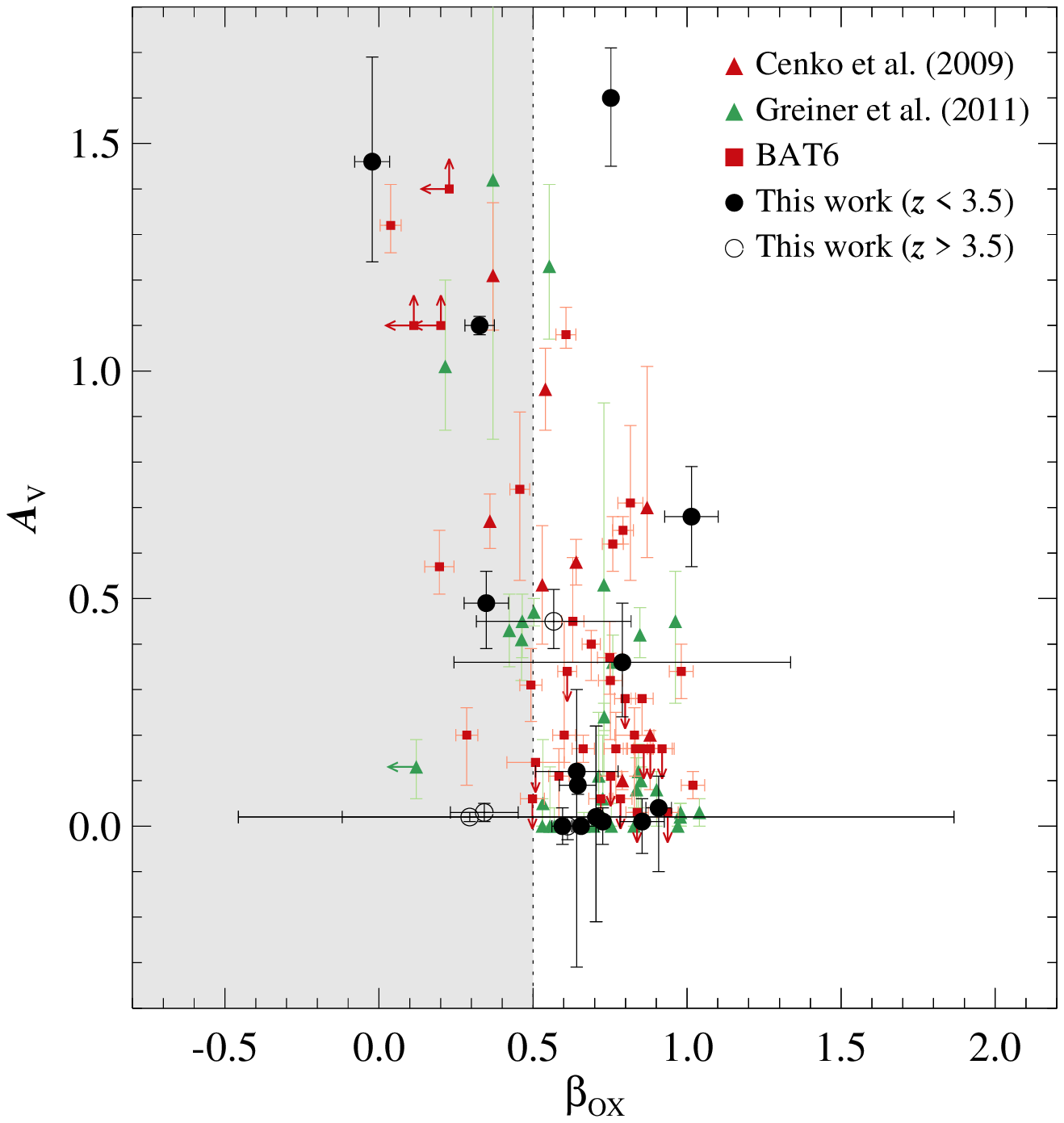}
    \quad
    \includegraphics[width=7.5cm,clip,angle=0]{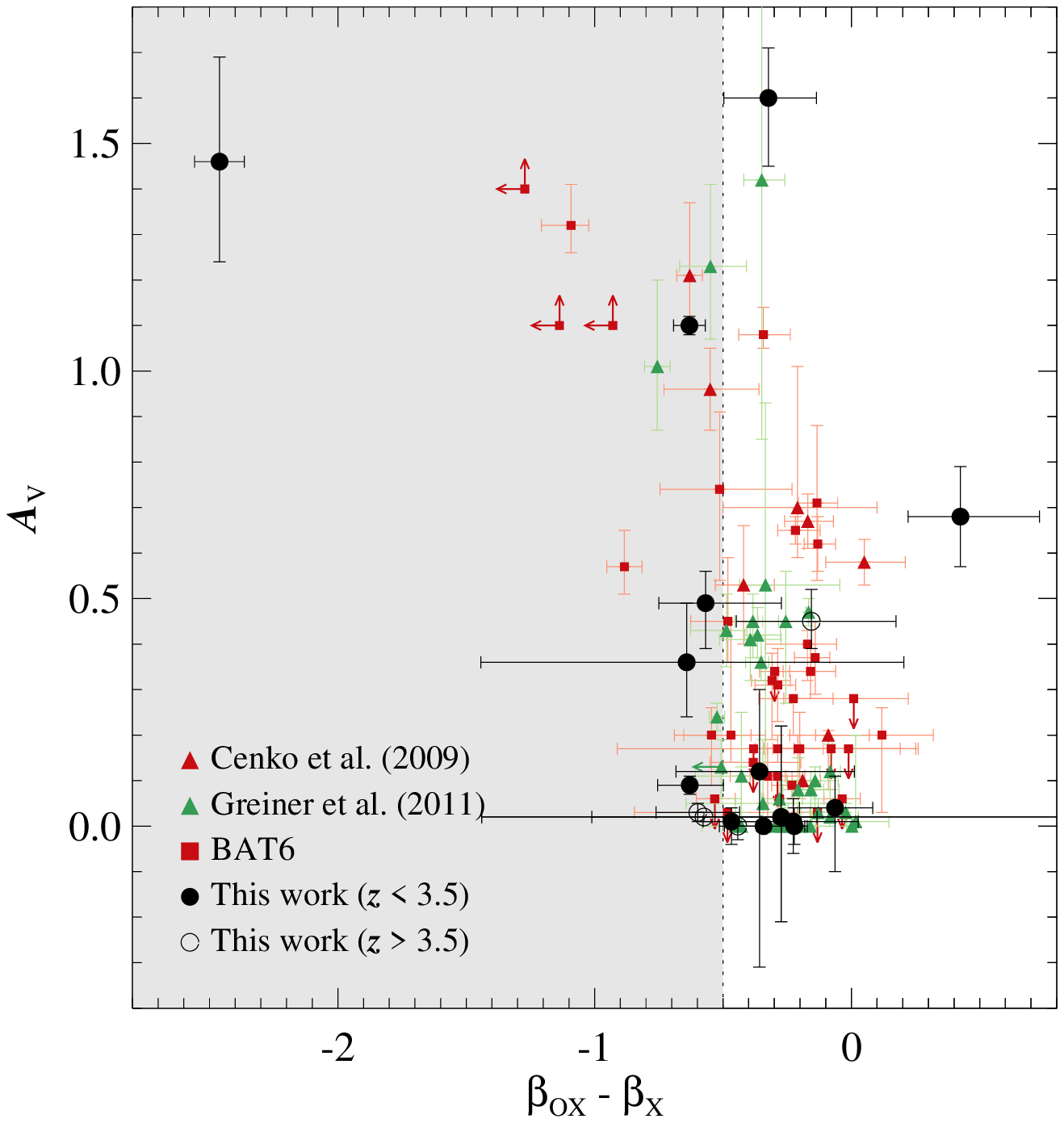}
    \\
  \end{center}
  \caption{Left panel: Fitted $A_{\rm V}$ as a function of $\beta_{\rm
      OX}$.  The  dotted line denotes  the \citet{2004ApJ...617L..21J}
    dark criterion,  with all  bursts to the  left being  dark.  Right
    panel:  $A_{\rm V}$  as a  function of  $\beta_{\rm OX}-\beta_{\rm
      X}$.   The dotted  line denotes  the \citet{2009ApJ...699.1087V}
    dark criterion.  All  bursts to the left of this  line are dark by
    this criterion. In both panels, black circles denote bursts in our
    sample with $z<3.5$, while empty circles denote bursts with in our
    sample  $z>3.5$.   The light  grey  regions  denote the  parameter
    spaces  where bursts are  considered to  be optically  dark. Other
    values available from the literature are also plotted, and denoted
    in  the  legend of  each  panel. The  BAT6  sample  is plotted  by
    cross-referencing values  of $\beta_{\rm OX}$  and $\beta_{\rm X}$
    from \citet{2012MNRAS.421.1265M} with  values for $A_{\rm V}$ from
    \citet{2013MNRAS.432.1231C}.}
  \label{fig:av_vs_darkness}
\end{figure*}

In  Figure  \ref{fig:av_vs_darkness}  we  show the  fitted  values  of
$A_{\rm V}$ as a function of  the two metrics for optical darkness. In
each case,  optically dark bursts are  in the grey  parameter space to
the left of the dotted vertical line.\par

Both  measures of optical  darkness reveal  a trend  where, typically,
bursts that are optically dark  have either high-redshift or modest to
high levels  of dust extinction in  their host galaxy.   There are two
GRB  that have  large  modelled  $A_{\rm V}$  value,  but that  aren't
consisent  with  being optically  dark  (GRB~130418A \&  GRB~130701A).
Figure \ref{fig:seds_fig1}  reveals that  the SED for  GRB~130701A was
only observed in four  filters (\textit{r}, \textit{i}, \textit{Z} and
\textit{Y}).  Furthermore,  the relative error in the  optical data is
large,  with two  possible types  of  solution.  The  first, with  the
smallest  $\chi^{2}$  fit  statistic,   for  an  LMC  type  extinction
law. Alternatively, the MW and SMC dust model templates prefer a lower
quantity    of    dust   in    the    host    galaxy   with    $A_{\rm
  V}=0.15_{-0.10}^{+0.16}$ in both  cases.  GRB~130701A was also found
to  be 5$^{\prime\prime}$  of  an $r=19.5$  magnitude source,  meaning
contaminating light  from this nearby source  may artificially enhance
the reported brightness of GRB~130701A.\par

GRB~130418A  has the softest  measured X-ray  spectrum in  our sample,
which  also has  a high  reported measurement  error. As  discussed in
\S~\ref{sec:rest_frame}, it  is possible that the true  value of X-ray
spectral index lies at the harder end of the 1$\sigma$ error bound and
that the optical  and X-ray regimes lie on  the same power-law segment
of  the  intrinsic GRB  synchrotron  spectrum.  In  such instances,  a
moderate amount of  dust could reduce the measured  optical flux by an
amount  less than  invoking a  cooling break  at 0.3~keV.  Thus  it is
possible for a  GRB host galaxy to contain  measurable amounts of dust
while not being reported as optically dark.\par

The optically darkest burst is GRB~130925A.  This burst was an unusual
event as the prompt high-energy emission was long-lived, making it one
of    the     few    ``ultra-long''    GRBs     observed    to    date
\citep{2014MNRAS.444..250E}.   Several  studies  of  this  event  have
suggested  that  the  central  engine  must occur  in  a  low  density
environment \citep{2014MNRAS.444..250E,2014ApJ...790L..15P}, such that
more emitted  shells have chance to interact  before being decelerated
by the circumburst medium. The SED for GRB~130925A, as shown in Figure
\ref{fig:seds_fig2},  implies   a  high  dust  content   in  the  host
galaxy. It  is perhaps possible that whilst  the immediate environment
of  the GRB  central engine  is low  density and  cleared by  a strong
stellar  wind,  outside  of this  the  host  galaxy  has a  high  dust
content.\par

\begin{figure}
  \begin{center}
    \includegraphics[width=8.5cm,clip,angle=0]{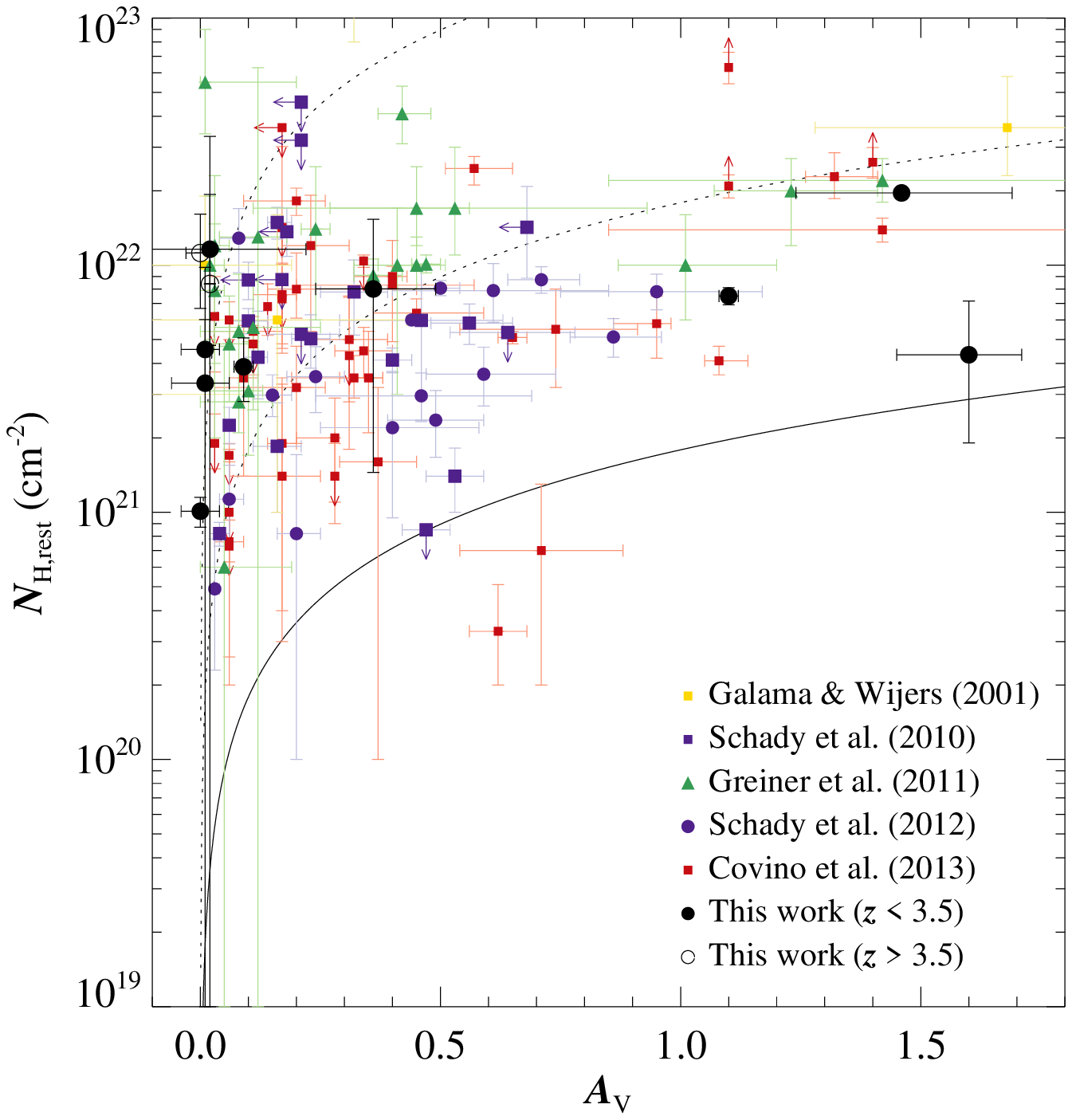}
  \caption{$N_{\rm H,rest}$ as a  function $A_{\rm V}$. The solid line
    is    the    Galactic    $A_{\rm    V}$-$N_{\rm    H}$    relation
    \citep{1995A&A...293..889P}.   The dash  lines  correspond to  the
    \citep{1995A&A...293..889P}   $A_{\rm  V}$-$N_{\rm   H}$  relation
    scaled according to a metal-to-dust ratio 10 and 100 times that of
    our Galaxy.  Black filled circles denote bursts in our sample with
    $z<3.5$,  while empty  circles denote  bursts in  our  sample with
    $z>3.5$. Values  available from  the literature are  also plotted,
    with corresponding plot symbols and  colours denoted in the key in
    the bottom right of the panel.}
  \label{fig:nh_vs_av}
  \end{center}
\end{figure}

We obtained  X-ray defined  measures of absorption  from the  GRB host
galaxy in  the form  of the neutral  hydrogen column  density, $N_{\rm
  H,rest}$
\citep{2006ApJ...637L..69W,2007ApJ...660L.101W,2010MNRAS.402.2429C,2011A&A...533A..16W},
using       the       spectral       fitting       algorithms       of
\citet{2007ApJ...663..407B}. The soft X-ray  spectra are fitted with a
power-law spectrum  and two absorption components,  corresponding to a
Galactic and extragalactic column.   In this pipeline solar abundances
are  assumed according  to \citet{1989GeCoA..53..197A}.   Those bursts
observed  by RATIR within  eight hours  of the  initial \textit{Swift}
trigger  with  a  measured  \textit{Swift}/XRT spectrum  have  $N_{\rm
  H,rest}$ reported in Table \ref{tab:fast_sample}. In total there are
15 GRBs with both a fitted value for $A_{\rm V}$ and $N_{\rm H,rest}$.
Of these  15, eleven  had a measurable  excess $N_{\rm  H,rest}$ above
that from  our own  Galaxy.  These eleven  GRBs are plotted  in Figure
\ref{fig:nh_vs_av},  once  more  with  previous values  obtained  from
samples in the  literature. We have also compared  our distribution of
$N_{\rm H,rest}$ to those available from the literature, using further
K-S  tests, and  find no  significant differences,  as shown  in Table
\ref{tab:ks_test_results}.\par

$A_{\rm V}$  is a measure of the  dust abundance of the  GRB host galaxy
along the sight line to  the burst. $N_{\rm H,rest}$, as measured from
soft X-ray spectra, is a probe  of the total metal content of the host
galaxy along the same line of sight, regardless of the phase it exists
in.  Figure  \ref{fig:nh_vs_av} confirms that the  sight-line from the
GRB central  engine probes  regions in the  host galaxy with  a higher
metal-to-dust ratio than our Galaxy.  This is in agreement with previous
studies  \citep{2001ApJ...549L.209G,2011A&A...526A..30G}  and suggests
that the host  galaxies of GRBs are similar  to molecular clouds, with
less dust than  our Galaxy.  \citet{2001ApJ...549L.209G} also consider
a scenario in which the central engine of the GRB photoionizes dust in
the  circumburst medium.  Such an  effect, however,  would  only occur
close to the GRB, depleting dust from  a region of order a few tens of
parsecs. This  is much less than  the distance of  host galaxy through
which the GRB emission must travel, and as such is less plausible than
the host galaxy having a lower bulk dust content.\par

It  must  also be  noted  that we  have  assumed  solar abundances  in
deriving  $N_{\rm H,rest}$.  The  curvature of  the X-ray  spectrum is
strongly  related to  absorption by  oxygen, and  as such  the derived
$N_{\rm H,rest}$  depends inversely on the  O/H ratio in  the GRB host
galaxy. As  the metal content in  GRB host galaxies  is actually lower
than                            that                           assumed
\citep{2012MNRAS.420..627S,2014MNRAS.441.3634C,2014arXiv1408.3578C},
for  a  given amount  of  X-ray  absorption,  a more  realistic  metal
abundance  would  reduce the  O/H  ratio  and  hence increase  $N_{\rm
  H,rest}$, therefore increasing the inferred metal-to-dust ratio.\par

We  also  obtained estimates  of  neutral  hydrogen column  densities,
$N_{\rm HI}$, derived from  optical spectroscopy as presented in Table
\ref{tab:optical_nh}.   Previous  studies  have shown  that  optically
derived  $N_{\rm  HI}$  are  usually significantly  lower  that  those
estimates               from               X-ray               spectra
\citep{2012A&A...537A..15S,2013A&A...560A..26Z}.  In contrast to X-ray
derived  $N_{\rm H,rest}$, $N_{\rm  HI}$ provides  an estimate  of the
quantity of  gas in the host  galaxy. Using this, we  can consider the
gas-to-dust ratio  of the  four host galaxies  for which  $N_{\rm HI}$
measurements are available.\par

\begin{table}
  \centering
  \caption{Optically  derived  estimates  of  $\log\left(  N_{\rm  HI}
    \right)$.}
  \label{tab:optical_nh}
  \begin{tabular}{ccc}
    \hline \hline
    GRB & $\log\left( N_{\rm HI} \right)$ & Reference \\
    \hline
    130606A & 19.93$\pm$0.2 & \citep{2013ApJ...774...26C} \\
     & & \citep{2014arXiv1409.4804H} \\
    140311A & 21.80$\pm$0.30 & \citep{2014arXiv1408.3578C} \\
    140419A & 19.3$\pm$0.2 & \citep{2014arXiv1408.3578C} \\
    140518A & 21.65$\pm$0.20 & \citep{2014arXiv1408.3578C} \\
    \hline
    \end{tabular}
\end{table}

The  host  $A_{\rm  V}$  for  three  of  these  GRBs  is  small,  with
GRB~140419A   having   $A_{\rm   V}   =   0$,  as   shown   in   Table
\ref{tab:av_model}.   However,   GRB~140311A  has  a   fitted  $A_{\rm
  V}=0.45_{-0.06}^{+0.08}$.    For    GRB~130606A,   GRB~140311A   and
GRB~140518A,   we    derive   $N_{\rm   HI}/A_{\rm    V}=4.26   \times
10^{21}$~cm$^{-2}$mag$^{-1}$,                               $1.40\times
10^{22}$~cm$^{-2}$mag$^{-1}$,              and             $1.49\times
10^{23}$~cm$^{-2}$mag$^{-1}$,  respectively.  In  comparison,  for the
LMC $N_{\rm HI}/A_{\rm  V}=8.3 \times 10^{21}$~cm$^{-2}$mag$^{-1}$ and
SMC  $N_{\rm  HI}/A_{\rm  V}=1.6  \times  10^{22}$~cm$^{-2}$mag$^{-1}$
\citep{2001ApJ...548..296W}.  This suggests that the gas-to-dust ratio
for GRB~140518A is  high in comparison to both the  LMC and SMC, while
that  of  GRB~130606A  is  most  consistent  with  that  of  the  LMC.
GRB~140311A appears to be  consistent with a gas-to-dust ratio similar
to that  of the  SMC.  These  latter two results  fit nicely  with the
preferred dust  extinction templates, albeit with a  very low quantity
of dust for  GRB~130606A. To reconcile the X-ray  $N_{\rm H,rest}$ and
optical values  of $N_{\rm  HI}$ requires either  a much  larger metal
abundance, specifically oxygen, in the  GRB host galaxy or for a large
fraction of  hydrogen gas in  the host to  be ionised.  The  latter is
perhaps more plausible as the  required oxygen abundance would have to
be at least an  order of magnitude higher. \citet{2013ApJ...768...23W}
consider  GRBs occurring  within  a star  forming  H~{\sc ii}  region,
attributing  the  X-ray absorption  to  a  He-dominated absorber.   In
another  study  \citet{2007ApJ...660L.101W}   discuss  the  sample  of
\citet{2006A&A...460L..13J}, in  which 17 GRBs with  optical and X-ray
measures of column  density are compared.  \citet{2007ApJ...660L.101W}
propose that the fundamental  difference between absorption in the two
regimes may  result from the  X-ray absorbing column  density sampling
the immediate  environment of the GRB, which  is substantially ionised
by the  burst.  The  H~{\sc i} column  density may, however,  probe an
environment  further from  the GRB  progenitor, and  thus one  that is
little affected by the GRB.\par

\subsection{Standardising $\beta_{\rm OX}$}
\label{sec:rest_frame}

\citet{2011A&A...526A..30G}  and  \citet{2012MNRAS.421.1265M}  observe
time evolution in the recovered  values of $\beta_{\rm OX}$.  With the
morphology  of  optical  and  X-ray  light  curves  not  always  being
correlated, the temporal power-law index with which both regimes decay
can  differ.   We  find  mean   values  of  temporal  decay  index  of
$\bar{\alpha}_{\rm      X}=1.37\pm0.08$     and     $\bar{\alpha}_{\rm
  opt}=0.97\pm0.11$, which  states that GRB light curves  in our RATIR
sample decay more slowly on average in the optical regime. As such, it
is more  likely that  GRBs with $\beta_{\rm  OX}$ measured  at earlier
times will be reported as  optically dark.  With bursts occurring at a
broad range  of redshifts, fixed observer frame  times and wavelengths
means that $\beta_{\rm OX}$ is not a standardised measure with regards
to the intrinsic  GRB emission.  To solve this,  a fiducial time fixed
in the rest  frame of the GRB can  be taken, $t_{\rm rest}=1.5$~hours.
With  our sample  having a  redshift range  of  $0.34<z<5.91$, $t_{\rm
  rest}$  corresponds to a  range of  observer frame  times from  2 to
11~hours.  At the lowest redshifts  a time of 2~hours should allow for
sampling  of  the  afterglow  at  a  time  after  the  plateau  phase.
Conversely, choosing $t_{\rm rest}$ to  be only slightly after the end
of the plateau  phase reduces any contamination of  the host galaxy to
the photometry of the GRB per se. At the highest redshift in the RATIR
sample $t_{\rm  rest}$ corresponds to approximately  10.5~hours in the
observer frame.\par

To  calculate a rest  frame measure  of optical  darkness, $\beta_{\rm
  OX,rest}$, one  should also define  the energies at which  the X-ray
and optical fluxes  are evaluated in the GRB  rest frame, with $E_{\rm
  X,rest}=3$~keV       and       $\lambda_{\rm       rest}=0.25~\mu$m,
respectively. $E_{\rm X,rest}$ was chosen such that at the peak of the
observed      GRB      redshift      distribution      ($z\approx2.2$;
\citealt{2012ApJ...752...62J}),     $E_{\rm    X,rest}\left(1+z\right)
\approx 1$~keV.  This  minimises the amount by which  the X-ray fluxes
must  be,  on  average,  extrapolated  from the  1~keV  light  curves.
$\lambda_{\rm  rest}=0.25~\mu$m  was  chosen  to  avoid  extrapolation
beyond the \textit{H} band.\par

The  two expected  mechanisms  for optical  darkness,  as measured  by
$\beta_{\rm OX}$, are  high dust content in the  host galaxy along the
GRB sight  line, or  high redshift.  By  selecting a fixed  rest frame
wavelength   in  the   ultraviolet  regime,   $\lambda_{\rm   rest}  =
0.25~\mu$m, $\beta_{\rm  OX,rest}$ always corresponds  to the observed
optical  (or indeed  NIR) flux  above the  observed Lyman  break, thus
removing  the effects of  high-redshift from  the measure.  All bursts
with  $\beta_{\rm OX,rest}-\beta_{\rm  X} <  -0.5$,  therefore, should
result from significant quantity of  dust in their host galaxy.  It is
also  worth   noting,  however,  that  a  rest   frame  wavelength  of
0.25~$\mu$m will  not sample the 2175~\AA  bump observed in  a MW type
dust  extinction law, although  this feature  only becomes  visible to
optical and NIR facilities at redshifts $z\gtrsim1.5$.\par

\begin{figure}
  \begin{center}
    \includegraphics[width=8.5cm,clip,angle=0]{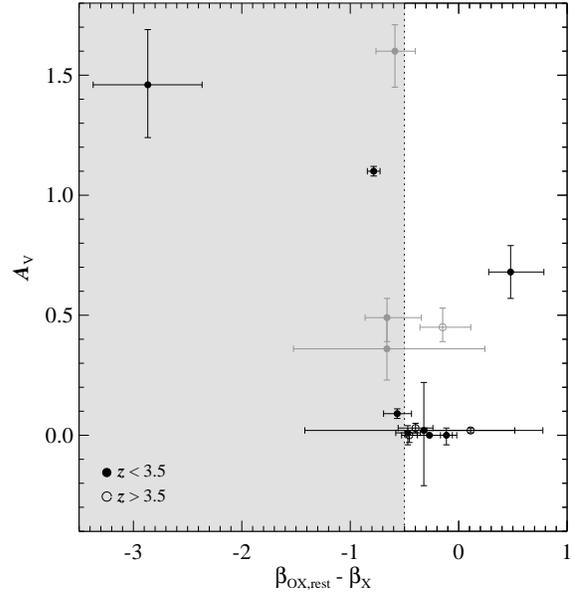}
  \end{center}
  \caption{$A_{\rm  V}$  as  a  function  of  $\beta_{\rm  OX,rest}  -
    \beta_{\rm X}$ at a fixed  rest frame time of $t_{\rm rest}$~=~1.5
    hours  after  the  high-energy  GRB  trigger. The  X-ray  flux  is
    evaluated at a fixed rest frame energy of 3~keV, while the optical
    flux  is   calculated  at  a   fixed  rest  frame   wavelength  of
    0.25~$\mu$m.  Black  circles  denote  bursts with  fitted  optical
    lights  curves,  while  grey  circles  have  detections  that  are
    extrapolated  to $t  = t_{\rm  rest}\left( 1  +  z\right)$. Filled
    circles are GRBs with $z<3.5$, while empty circles are bursts with
    $z>3.5$.}
  \label{fig:av_betadiff_rest}
\end{figure}

Looking at  Figure \ref{fig:av_betadiff_rest}, we see that  all of the
GRBs classified  as optically dark  in using $\beta_{\rm  OX,rest}$ do
indeed have a detectable amount of dust in their host galaxy along the
GRB sight  line. We note, however,  that our sample size  is small, as
only 15 GRBs  met the observational criteria required  to measure both
$A_{\rm V}$ and $\beta_{\rm  OX,rest}$.  It is perhaps surprising that
a  trend of increasing  $A_{\rm V}$  with increasing  optical darkness
(i.e. decreasing $\beta_{\rm OX,rest}-\beta_{\rm X}$) is not apparent,
however, the most notable  deviations from this trend are GRB~130418A,
GRB~130701A and GRB~140311A.\par

The  value of  optical flux  for GRB~130701A  was extrapolated  from a
single detection assuming $\bar{\alpha}_{\rm  opt}$ which may not best
represent  the actual  GRB  decay slope.   Additionally,  as noted  in
\S~\ref{sec:sed_fit}, GRB~130701A may suffer from contamination in the
optical  and NIR  regime from  a nearby  source. As  such,  this would
artificially   enhance  the  reported   optical  flux,   and  increase
$\beta_{\rm OX,rest}$ from its true value.\par

GRB~130418A  has the  softest measured  X-ray spectrum  in  our sample
($\beta_{\rm X}=0.59_{-0.18}^{+0.30}$),  thus increasing the  value of
$\beta_{\rm  OX,rest}-\beta_{\rm   X}$,  keeping  the   burst  in  the
optically    bright    region    of    the    parameter    space    in
Figure~\ref{fig:av_betadiff_rest},  despite  significant  fitted  dust
content ($A_{\rm V}=0.68\pm0.11$). As  measured in the observer frame,
GRB~130418A has $\beta_{\rm OX}=1.01\pm0.09$,  which is typical of the
expected value  of $\beta_{\rm  X}$.  It is  possible that  the actual
value of  $\beta_{\rm X}$ lies  closer to the  upper end of  the error
bound  in the measured  value. In  addition to  this, the  optical and
X-ray  regimes could  lie on  the same  segment of  the  intrinsic GRB
synchrotron  spectrum, thus  leading  to the  burst  remaining in  the
bright region  of the parameter space.\par

GRB~140311A  is the  final burst  with significant  amounts  of fitted
dust, $A_{\rm  V}=0.45_{-0.06}^{+0.08}$, that is  within the optically
bright  region of  Figure~\ref{fig:av_betadiff_rest}. As  discussed in
\S~\ref{sec:sed_fit},  this GRB  was  only observed  in four  filters,
giving poorer  constraints on the template  fitted to the  SED.  It is
also possible that, like GRB~130418A, GRB~140311A has both the optical
and X-ray regimes  on the same power-law segment  of the intrinsic GRB
spectrum. As  such, to  be reported as  optically dark, the  amount of
dust  along the line  of sight  in the  host galaxy  would have  to be
sufficient to  reduce the  optical flux below  the level  predicted by
having a cooling  break just below the X-ray  regime. Both GRB~130418A
and GRB~140311A  highlight the complicated role played  by the cooling
break in  identifying optically attenuated  GRBs.  As a  cooling break
could occur anywhere between the  optical and X-ray regimes, or indeed
not at  all, a simple linear  relation of increasing  $A_{\rm V}$ with
decreasing  $\beta_{\rm  OX,rest}-\beta_{\rm  X}$  is unlikely  to  be
observed.     However,     all    GRBs    denoted     as    dark    in
Figure~\ref{fig:av_betadiff_rest} are successfully explained by either
having a high  redshift or moderate to high amounts  of dust along the
GRB line of sight in the host galaxy.\par

A further  test of the results  from Figure \ref{fig:av_betadiff_rest}
would be to  consider a fixed rest frame wavelength  that is also less
affected by dust extinction. This could be achieved by considering the
rest-frame  \textit{i} band, which  would allow  a measure  of optical
darkness that should only consider  the intrinsic GRB spectrum. If the
GRB is intrinsically underluminous $\beta_{\rm OX,rest}$, as evaluated
in the  rest-frame \textit{i} band, will continue  to indicate optical
darkness independent of redshift  and dust content.  Conversely, a GRB
that is optically attenuated, due  to either being at high-redshift or
the  dust  content  of its  host  galaxy,  would  not be  reported  as
optically  dark   using  such  a  measure.  We   have  not  calculated
$\beta_{\rm OX,rest}$  at a rest-frame  \textit{i} band for  the RATIR
sample of GRBs as this would require an extrapolation further into the
NIR, outside of the RATIR coverage for $z\gtrsim1.9$.\par

\section{Conclusions}
\label{sec:conc}

In  this work we  present photometry  of 28  GRBs rapidly  observed by
RATIR.    Combining   these   data   with  those   obtained   by   the
\textit{Swift}/XRT  allows us  to quantify  optical darkness  in these
GRBs at a fiducial time of  11 hours after the high-energy trigger. To
account for  the expected synchrotron  emission mechanism, we  use the
\citet{2009ApJ...699.1087V}  definition of  darkness to  find 46$\pm$9
per cent (13/28) of our  sample of GRBs are considered optically dark,
or 50$\pm$10  per cent  (13/26) when only  including long  GRBs.  This
fraction    is    broadly    consistent    with    previous    studies
\citep{2008ApJ...686.1209M,2009ApJS..185..526F,2009ApJ...693.1484C,2011A&A...526A..30G,2012MNRAS.421.1265M}. The
optically dark  fraction of GRBs  in our sample also  remains constant
when calculated at an earlier epoch of 5.5 hours.\par

To investigate the  underlying causes of optical darkness,  we use the
template fitting algorithm presented in \citet{2014AJ....148....2L} to
model the  optical and  NIR SEDs for  the 19  GRBs in our  sample with
coverage  in a sufficient  number of  filters. We  were able  to model
seven of the bursts identified  as optically dark. Of these seven, two
have high  redshift, two  have $A_{\rm V}  > 1$  and two have  $0.25 <
A_{\rm  V} <  1$. GRB~130420A  has been  modelled, but  shown  to have
neither  high  redshift nor  high  dust  extinction.  We consider  two
alternative  explanations,  which   suggest  either  a  cooling  break
immediately below the X-ray regime  or that a more precise measurement
in  the \textit{H} band  would prefer  a template  with a  larger dust
content in the GRB host galaxy.\par

Optical  darkness is  indicative of  interesting GRB  events,  as they
either  occur  at  high-redshift   or  within  highly  dust  extincted
galaxies. Considering  optically attenuated GRBs, we find  that 23 per
cent (3/13)  are due to moderate  or high redshifts ($z  > 3.5$). Four
dark GRBs have  unknown redshift, and so this fraction  may in fact be
higher.\par

Of  the 19  GRBs with  modelled SEDs,  37$\pm$11 per  cent  (7/19) had
moderate  or high  amounts of  dust  extinction.  This  is in  general
agreement              with              previous              studies
\citep{2011A&A...526A..30G,2013MNRAS.432.1231C}, where the majority of
the  sample have  low dust  extinction.  Averaging  across  the entire
fitted sample using  a single type of dust extinction  law, we find an
average  best  fit  of  $A_{\rm  V}=0.26\pm0.05$ with  an  SMC  model.
Individually, only 5  GRBs prefer an SMC type  extinction law, but the
improvement in $\chi^{2}$ is much larger  in a few of these cases than
obtained in choosing  a different dust extinction law  in the other 14
SED fits.\par

We  perform  an extensive  array  of  K-S  tests comparing  the  RATIR
distributions of  $z$, $\beta_{\rm X}$, $\beta_{\rm  OX}$, $A_{\rm V}$
and  $N_{\rm  H,rest}$  to  samples provided  in  previous  literature
\citep{2008ApJ...686.1209M,2009ApJ...693.1484C,2009ApJS..185..526F,2009ApJ...699.1087V,2010ApJ...720.1513K,2011A&A...526A..30G,2012MNRAS.421.1265M,2013MNRAS.432.1231C}. These
tests confirm,  with one exception,  that our sample  is statistically
consistent  with these  previous studies,  both individually  and when
considered  as  a  single  sample.   We  do  however,  find  that  our
distribution  of host  galaxy $A_{\rm  V}$  along the  GRB sight  line
differs   significantly  from  that   of  \citet{2013MNRAS.432.1231C}.
Further K-S  tests reveal  that the BAT6  distribution of  host galaxy
$A_{\rm V}$  also differs from  other previous literature,  whilst the
RATIR sample is consistent with  these other studies.  The BAT6 sample
is notable  for a handful of  very high $A_{\rm  V}$ values, including
one in  excess of $A_{\rm  V}>5.5$, which is  derived from a  fit with
zero  degrees of  freedom. However,  it must  also be  noted  that the
selection criteria for the BAT6 sample differ significantly from other
studies.\par

Within  the sample  of optically  dark GRBs,  an X-ray  derived excess
$N_{\rm  H,rest}$ from  the  host  galaxy is  detected  in 11  bursts.
Figure  \ref{fig:nh_vs_av}  compares  $N_{\rm  H,rest}$  to  the  dust
content, as probed along the line  of sight.  As such, we find the GRB
host galaxies tend  to have a higher metal-to-dust  ratio, which is in
agreement          with         some          previous         studies
\citep{2001ApJ...549L.209G,2011A&A...526A..30G}.    Optically  derived
estimates of $N_{\rm HI}$ are  also presented for four bursts, showing
GRB~130606A  and  GRB~140311A  to  have a  gas-to-dust  ratio  broadly
consistent with that of the LMC and SMC, respectively.\par

Finally,  we present  a standardised  measurement of  optical darkness
$\beta_{\rm  OX,rest}$, which  corresponds  to optical  darkness in  a
fixed rest  frame time  of 1.5 hours,  considering flux at  fixed rest
frame   values    of   $E_{\rm   X,rest}=3$~keV    and   $\lambda_{\rm
  rest}=0.25~\mu$m. In  doing so, we reduce the  dependency of optical
darkness solely to the host galaxy  dust content along the GRB line of
sight. As such, we demonstrate  that optical darkness in our sample is
only due  to either  high-redshift or host  galaxy dust  content. This
statement  is limited  by the  small sample  size of  rapidly observed
RATIR  GRBs,  but  further   population  of  the  $A_{\rm  V}$  versus
$\beta_{\rm OX,rest}-\beta_{\rm X}$ parameter space should finally and
conclusively confirm this to be true.\par

\section*{Acknowledgements}

We thank  Pall Jakobsson  for useful comments  and suggestions  on the
manuscript.   We  also thank  Jochen  Greiner  for  supplying us  with
detailed  data related  to \citet{2011A&A...526A..30G}.  We  thank the
RATIR project  team and the staff of  the Observatorio Astron\'{o}mico
Nacional  on Sierra San  Pedro M\'{a}rtir.   RATIR is  a collaboration
between  the  University   of  California,  the  Universidad  Nacional
Auton\'{o}ma  de M\'{e}xico,  NASA  Goddard Space  Flight Center,  and
Arizona State University, benefiting from the loan of an H2RG detector
and  hardware  and  software  support  from  Teledyne  Scientific  and
Imaging.  RATIR, the automation of the Harold L.  Johnson Telescope of
the  Observatorio   Astron\'{o}mico  Nacional  on   Sierra  San  Pedro
M\'{a}rtir, and the  operation of both are funded  through NASA grants
NNX09AH71G,  NNX09AT02G, NNX10AI27G,  and  NNX12AE66G, CONACyT  grants
INFR-2009-01-122785  and CB-2008-101958,  UNAM PAPIIT  grant IN113810,
and UC MEXUS-CONACyT  grant CN 09-283. A.C.  is  supported by the NASA
Postdoctoral Program at the  Goddard Space Flight Center, administered
by  Oak   Ridge  Associated  Universities  through   a  contract  with
NASA. This work made use of data supplied by the UK Swift Science Data
Centre at the University of Leicester.\par

\begin{bibliography}{ms}
  \bibliographystyle{mn2e}
\end{bibliography}

\label{lastpage}
\end{document}